\documentclass[prd,notitlepage,preprintnumbers,superscriptaddress,nofootinbib,twocolumn]{revtex4-1}
\usepackage[utf8]{inputenc}

\usepackage{amsmath}
\usepackage{amsfonts}
\usepackage{amssymb}
\usepackage{physics}
\usepackage{slashed}
\usepackage{simplewick}
\usepackage{mathtools}
\usepackage{braket}
\usepackage{siunitx}
\usepackage{orcidlink}
\usepackage{placeins}
\usepackage{graphicx}
\usepackage{float}
\usepackage[caption=false]{subfig}

\usepackage{listings}

\usepackage{hyperref}
\hypersetup{ 
    pdfnewwindow=true,      
    colorlinks=true,       
    allcolors=[RGB]{31 119 180}
} 
\usepackage{tikz}
\usetikzlibrary{calc}
\usetikzlibrary{patterns}
\usetikzlibrary{patterns}
\usetikzlibrary{positioning,decorations.pathmorphing,decorations.markings,snakes}

\usepackage{xcolor}
\definecolor{jpac-blue}{rgb}{0.12,0.47,0.71}
\definecolor{jpac-orange}{rgb}{1,0.5,0.05}
\definecolor{jpac-green}{rgb}{0,0.62,0.45}
\definecolor{jpac-red}{rgb}{0.84,0.15,0.15}
\definecolor{jpac-purple}{rgb}{0.58,0.40,0.71}
\definecolor{jpac-brown}{rgb}{0.54,0.34,0.29}
\definecolor{jpac-pink}{rgb}{0.89,0.47,0.76}
\definecolor{jpac-grey}{rgb}{0.5,0.5,0.5}
\definecolor{jpac-gold}{rgb}{0.74,0.74,0.13}
\definecolor{jpac-aqua}{rgb}{0.09,0.75,0.81}
\definecolor{jpac-white}{rgb}{1,1,1}

\usepackage[capitalise]{cleveref}
\usepackage{pgfplots,pgfplotstable}
\usetikzlibrary{pgfplots.groupplots}
\usetikzlibrary{decorations.markings}
\usepgfplotslibrary{fillbetween}
\pgfplotsset{compat=1.14}
\usepackage{tikz}
\usepackage{tikz-3dplot}
\usetikzlibrary{positioning}
\usetikzlibrary{calc}
\tikzset{->-/.style={decoration={
  markings,
  mark=at position .5 with {\arrow{>}}},postaction={decorate}}}
\tikzset{>=stealth}
\makeatletter
\tikzoption{canvas is plane}[]{\@setOxy#1}
\def\@setOxy O(#1,#2,#3)x(#4,#5,#6)y(#7,#8,#9)%
  {\def\tikz@plane@origin{\pgfpointxyz{#1}{#2}{#3}}%
   \def\tikz@plane@x{\pgfpointxyz{#4}{#5}{#6}}%
   \def\tikz@plane@y{\pgfpointxyz{#7}{#8}{#9}}%
   \tikz@canvas@is@plane
  }
\makeatother

\usepackage{feynmp-auto}
\usepackage{multirow}

\newcommand{\p}{\ensuremath{\prime}}

\newcommand{\re}[1]{\ensuremath{\text{Re}\{#1\}}}

\newcommand{\comment}[1]{}

\newcommand{\HISKP}{Helmholtz-Institut f\"{u}r Strahlen- und Kernphysik (Theorie) and Bethe Center for Theoretical Physics, Universit\"{a}t Bonn, D-53115 Bonn, Germany}
\newcommand{\AGH}{AGH University of Krakow, Faculty of Physics and Applied Computer Science, PL-30-059 Krak\'ow, Poland}
\newcommand{\catania}{INFN Sezione di Catania, I-95123 Catania, Italy}
\newcommand{\ceem}{Center for  Exploration  of  Energy  and  Matter, Indiana  University, Bloomington,  IN  47403,  USA}

\newcommand{\indiana}{Department of Physics,
Indiana  University, Bloomington,  IN  47405,  USA}
\newcommand{\jlab}{Theory Center, Thomas  Jefferson  National  Accelerator  Facility, Newport  News,  VA  23606,  USA}
\newcommand{\messina}{Dipartimento di Scienze Matematiche e Informatiche, Scienze Fisiche e Scienze della Terra,
Universit\`a degli Studi di Messina, I-98166 Messina, Italy}
\newcommand{\ub}{Departament de F\'isica Qu\`antica i Astrof\'isica and Institut de Ci\`encies del Cosmos, Universitat de Barcelona, E-08028 Barcelona, Spain}

\newcommand{\uned}{Departamento de F\'isica Interdisciplinar, Universidad Nacional de Educaci\'on a Distancia (UNED), E-28040 Madrid, Spain}

\newcommand{\gwu}{Department of Physics,
The George Washington University, Washington, DC 20052, USA
}
\newcommand{\ucb}{Department of Physics, University of California, Berkeley, CA 94720, USA}
\newcommand{\lbnl}{Nuclear Science Division, Lawrence Berkeley National Laboratory, Berkeley, CA 94720, USA}
\newcommand{\odu}{Department of Physics, Old Dominion University, Norfolk, Virginia 23529, USA
}
\begin{document}
    \preprint{JLAB-THY-24-4078}
\begin{fmffile}{feynDiags}

\begin{abstract}
Forward photoproduction of $\pi^+\pi^-$ pairs with invariant mass of the order of $m_\rho\sim 770~\si{MeV}$ is traditionally understood to be produced via Pomeron exchange. Based on a detailed analysis of the CLAS photoproduction data, it is shown that the dynamics of two-pion photoproduction for $|t|\gtrsim 0.5~\si{GeV^2}$ cannot be explained by Pomeron exchange alone. This motivates the development of a new theoretical model of two-pion photoproduction which incorporates both two-pion and pion-nucleon resonant contributions. After fitting free parameters, the model provides an excellent description of the low moments of the angular distribution measured at CLAS, and enables an assessment of the relative contributions of particular production mechanisms and an interpretation of the various features of the data in terms of these mechanisms.
\end{abstract}

\title{Studying $\pi^+\pi^-$ photoproduction beyond Pomeron exchange}

\author{{\L}ukasz Bibrzycki\orcidlink{0000-0002-6117-4894}}
\email{lukasz.bibrzycki@agh.edu.pl}
\affiliation{\AGH}
\author{Nadine Hammoud\orcidlink{0000-0002-8395-0647}}
\email{nadine.hammoud.28@gmail.com}
\affiliation{Institute of Nuclear Physics, Polish Academy of Sciences, PL-31-342 Krak\'ow, Poland}
\author{Vincent Mathieu\orcidlink{0000-0003-4955-3311}}
\affiliation{\ub}
\author{Robert~J.~Perry\orcidlink{0000-0002-2954-5050}}
\email{perryrobertjames@gmail.com}
\affiliation{\ub}

\author{Alex Akridge}
\affiliation{\ceem}
\affiliation{\indiana}
\author{C\'esar~Fern\'andez-Ram\'irez\orcidlink{0000-0001-8979-5660}}
\affiliation{\uned}
\author{Gloria~\surname{Monta\~na}\orcidlink{0000-0001-8093-6682}}
\affiliation{\jlab}
\author{Alessandro~\surname{Pilloni}\orcidlink{0000-0003-4257-0928}}
\affiliation{\messina}
\affiliation{\catania}
\author{Arkaitz~\surname{Rodas}\orcidlink{0000-0003-2702-5286}}
\affiliation{\jlab}
\affiliation{\odu}
\author{Vanamali~\surname{Shastry}\orcidlink{0000-0003-1296-8468}}
\affiliation{\ceem}
\affiliation{\indiana}
\author{Wyatt~A.~\surname{Smith}\orcidlink{0009-0001-3244-6889}}
\affiliation{\gwu}
\affiliation{\ucb}
\affiliation{\lbnl}
\author{Daniel~Winney\orcidlink{0000-0002-8076-243X}}
\affiliation{\HISKP}
\author{Adam P. Szczepaniak\orcidlink{0000-0002-4156-5492}}
\affiliation{\jlab}
\affiliation{\indiana}
\affiliation{\ceem}

\collaboration{Joint Physics Analysis Center}

\maketitle

\section{Introduction}
\label{sec:introduction}
Two-pion photoproduction has long been a reaction of interest for studies of hadron spectroscopy. Since free pion targets are difficult to obtain, multipion hadro- and photoproduction measurements are necessary to understand the spectrum of light meson resonances. In recent years, the field of hadron spectroscopy has experienced a revolution due to the observation of a number of resonances in the heavy sector which do not fit into the conventional quark model (for reviews, see~\cite{Esposito:2016noz,Guo:2017jvc,Olsen:2017bmm,Brambilla:2019esw}). The existence of such exotic states, although long heralded~\cite{Gell-Mann:1964ewy}, has nonetheless recently motivated further experimental studies of light-meson spectroscopy~\cite{GlueX:2020fam,GlueX:2023fcq}. This modern high-precision data necessitates sophisticated amplitude analysis methods to extract the full physics content. 

The study of the two-pion final state has significant theoretical value because the current understanding of this reaction is primarily limited to the region of small momentum transfer between the impinging photon and the produced $\pi^+\pi^-$ system ($|t|\lesssim 0.4~\si{GeV^2}$). In the limit where the total squared center-of-mass energy, $s$ becomes large while the momentum transfer, $t$ remains small, it is natural to describe the data using Regge theory~\cite{Regge:1959mz,Irving:1977ea}. This theory predicts that in this kinematical region the scattering amplitude is dominated by Regge exchanges, leading to asymptotic energy dependence of the form $s^{\alpha(t)}$, where $\alpha(t)$ is known as the Regge trajectory. These trajectories may be calculated from low-energy properties of resonances in the related $t$-channel process. The asymptotic behaviour of the scattering amplitude, and via the optical theorem, the total cross section, is determined by the Regge trajectory with the largest $t=0$ intercept. A fit to high-energy data for the total cross section for a range of hadronic reactions yields a value $\alpha(t=0)\approx1$~\cite{Donnachie:1992ny}. This trajectory is known as the Pomeron ($\mathbb{P}$), and provides an explanation within Regge theory of the approximately constant hadronic total cross sections at large center-of-mass energies. 

Several features of two-pion photoproduction measurements at small momentum transfer are easily understood in terms of Pomeron exchange. Firstly, in this kinematic region ($|t|\lesssim 0.4~\si{GeV^2}$), the process is known to be $P$-wave dominated with a prominent $\rho(770)$ peak. The line-shape of the $\rho(770)$ in the two-pion spectrum appears deformed with respect to pion-initiated $\rho(770)$ production~\cite{Crouch1966,Aachen-Berlin-Bonn-Hamburg-Heidelberg-Munich:1968rzt} or to the electromagnetic pion form factor~\cite{Belle:2008xpe,BaBar:2012bdw}. This feature is effectively described by the \mbox{Drell--S\"oding} or Deck models~\cite{Drell:1961zz,Soding:1965nh,Deck:1964hm,Pumplin:1970kp,PhysRevLett.25.485}.

\begin{figure}
\centering
\includegraphics[scale=0.5]{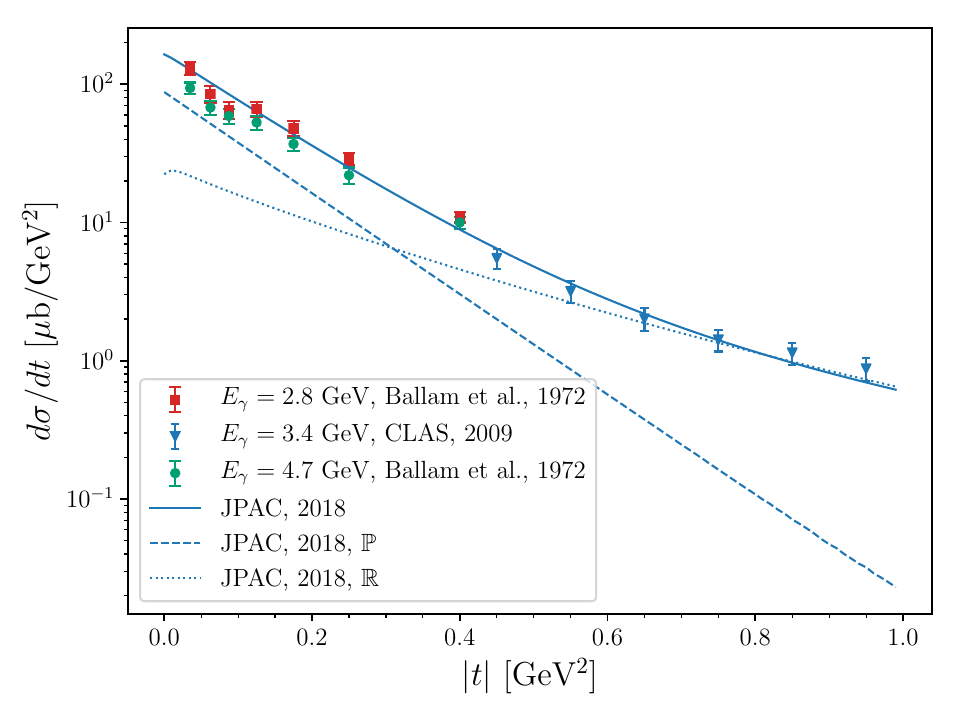}
\caption{ 
Comparison of $\rho(770)$ photoproduction measurements from Refs.~\cite{Ballam:1972eq,CLAS:2009ngd} with the Regge model from Ref.~\cite{Mathieu:2018xyc}. The model is calculated at a photon energy of $E_\gamma=3.4~\si{GeV}$. Note that the Pomeron ($\mathbb{P}$) exchange reproduces the differential cross section slope at small-$t$, while contributions from other Reggeons ($\mathbb{R}$) become increasingly important at larger $|t|$.
}
\label{fig:pomeron_model}
\end{figure}

Secondly, from a study of the spin-density matrix elements (SDMEs) for \mbox{$|t|\lesssim0.4~\si{GeV^2}$} and \mbox{$E_\gamma=2.8,4.7~\si{GeV}$}, it has been shown that the produced $\rho(770)$ resonance inherits the helicity of the incoming photon in the $s$-channel center-of-mass frame~\cite{Ballam:1970qn}. This phenomenon is known as $s$-channel helicity conservation (SCHC). Regge theory provides a natural explanation for SCHC because it predicts the factorization of the Regge pole residue into two vertices, which are proportional to $(-t)^{n/2}$ near the forward limit, where in this case $n$ is the net helicity flip between the photon and the $\rho(770)$~\cite{Irving:1977ea}. 
Thus, Regge theory provides a natural explanation of the dominance of $n=0$ helicity-conserving processes in the forward ($t\to0$) limit. Recently, high-precision measurements at higher photon energies ($E_\gamma\approx8.5~\si{GeV}$) show that SCHC approximately holds in the forward limit~\cite{GlueX:2023fcq}, with increasing violations at larger momentum transfer.

Finally, from this reaction, it is possible to extract the cross section for the process $\gamma  p \to \rho^0  p$. Regge models in which $\rho(770)$ photoproduction is assumed to be saturated by Pomeron exchange alone reproduce the slope of the differential cross section for $|t|\lesssim 0.4~\si{GeV^2}$, while data for $|t|\gtrsim 0.4~\si{GeV^2}$ suggest that additional Reggeon exchanges are also required. This is shown in~\cref{fig:pomeron_model}, where the Regge model employed in Ref.~\cite{Mathieu:2018xyc} which incorporates both Pomeron- and Reggeon-exchange contributions is compared with experimental data. 

Thus at small momentum transfer ($|t|\lesssim 0.4~\si{GeV^2}$), a relatively simple picture of the process emerges, in which at least three of the prominent features  of the data (the $\rho(770)$ lineshape, SCHC and the $t$-dependence of the $\rho(770)$ photoproduction differential cross section) can be explained by a model which incorporates both a Pomeron-induced $\rho(770)$ resonant amplitude and a nonresonant Deck background.

Less is known about the reaction at larger momentum transfers ($|t|\gtrsim0.4~\si{GeV^2}$). Theoretically speaking, the validity of the Pomeron exchange picture at larger $t$ is questionable, since this is no longer in the kinematical limit where Regge theory is applicable. In addition, while a good description of the $\rho(770)$ is essential to reproduce the two-pion lineshape, the Particle Data Group (PDG)~\cite{ParticleDataGroup:2022pth} also lists several other light meson resonances with quantum numbers and masses which can be expected to contribute to the experimental cross section. While their influence on the cross section is subleading with respect to the $\rho(770)$, precise data in this region offer the prospect of further study of these resonances and their associated production mechanisms. 

Rather than working directly with the multidimensional differential cross sections, it is possible to use the set of moments of the angular distribution, $\expval{Y^L_M}$. These moments are rigorously defined observables and are bilinear in the partial wave amplitudes. 

Results for the experimentally measured low angular moments for two-pion photoproduction were presented in Ref.~\cite{PhysRevD.5.545} at photon energies of $E_\gamma =2.8$ and $4.7~\si{GeV}$ and a momentum transfer range of $0.02 < |t| < 0.4~\si{GeV^2}$. The moments $\braket{Y_0^0}$, $\braket{Y^2_0}$, and $\braket{Y_2^2}$ revealed a prominent peak in the two-pion invariant mass distribution corresponding to the $P$-wave contribution, primarily associated with the $\rho(770)$ meson. However, the angular moment $\braket{Y_0^1}$ was relatively small, indicating that other resonance contributions are small in this kinematic region. A reasonable description of these low angular moments was obtained by employing a \mbox{Drell--S\"oding} model~\cite{PhysRevD.5.545}. 

\begin{figure*}
\centering
\includegraphics[scale=0.45]{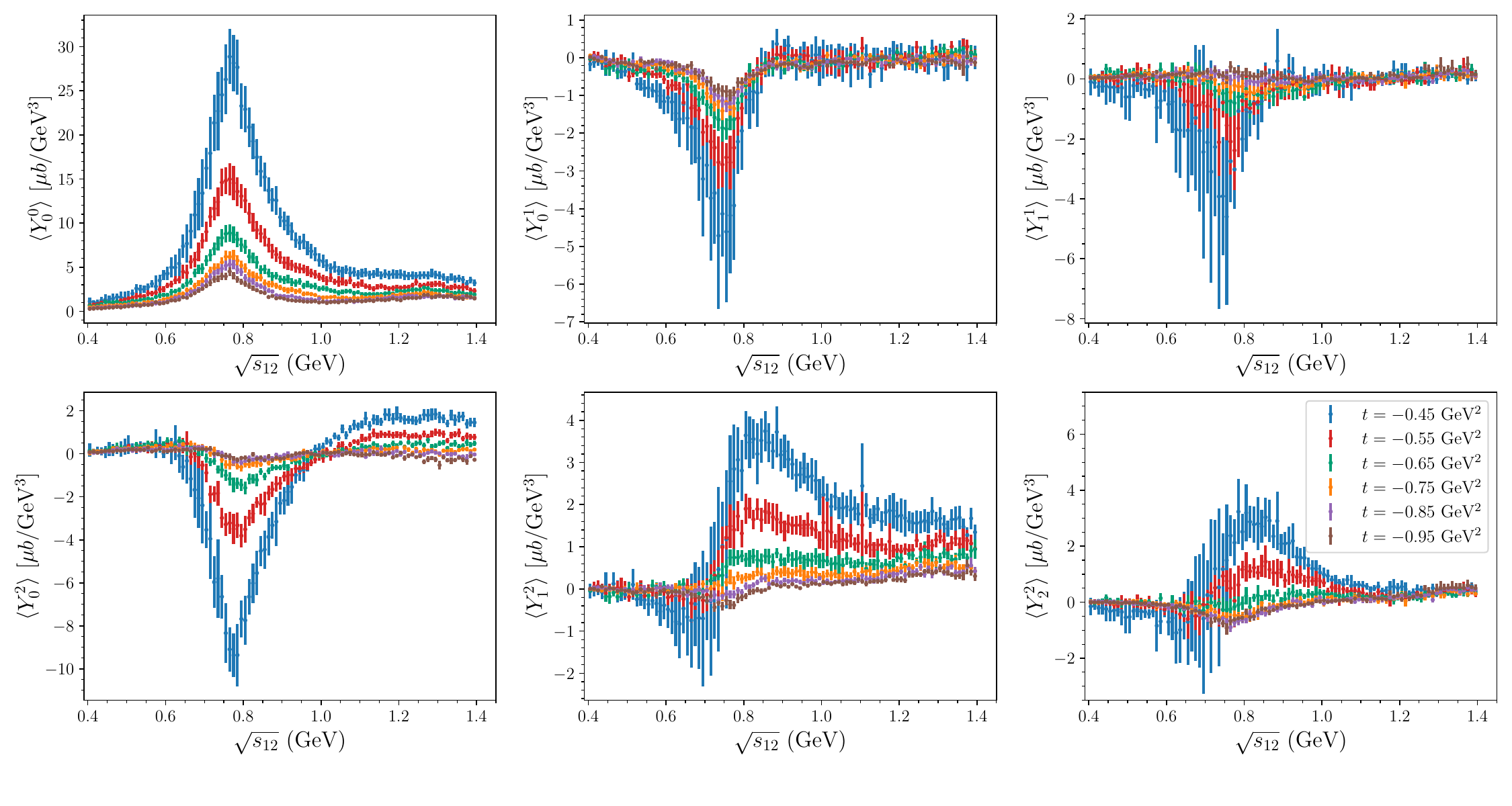}
\caption{Moments of $\pi^+\pi^-$ angular distribution measured by CLAS at $E_\gamma=3.7~\si{GeV}$ and across all $t$-bins~\cite{CLAS:2009ngd}.}
\label{fig:clas_data}
\end{figure*}

The more recent CLAS dataset~\cite{CLAS:2009ngd} analyzed in this work covers a range of larger momentum transfers, of  $0.4 < |t| < 1.0~\si{GeV^2}$, for similar photon energies and two-pion invariant masses. A characteristic subset of the data is shown in Fig.~\ref{fig:clas_data}. The presence of the $\rho(770)$ resonance is again clear in $\braket{Y_0^0}$, $\braket{Y^2_0}$, and $\braket{Y_2^2}$; however, the more precise data suggests some evidence that a broad enhancement is present at higher two-pion invariant masses ($\sqrt{s_{12}}\sim 1.2~\si{GeV}$). There are at least two relevant known resonances in this mass region: the $f_2(1270)$ and $f_0(1370)$. In contrast to the older data of Ref.~\cite{PhysRevD.5.545}, a detailed analysis of the angular moments $\braket{Y^1_0}$ and $\braket{Y^1_1}$ around $1~\si{GeV}$ has been interpreted as evidence for the presence of the $f_0(980)$~\cite{CLAS:2008ycy,CLAS:2009ngd} resonance. The complex interference patterns exhibited by this precise modern data present challenges for models of two-pion photoproduction, including the Regge models of the type mentioned above. In particular, it will be shown that a model which incorporates only a Pomeron-induced resonant amplitude and a non-resonant $P$-wave amplitude cannot reproduce the higher precision angular moments measured at larger momentum transfer ($|t|\gtrsim0.4~\si{GeV^2}$).

The failure of this simple model to describe the angular moments motivates the development of a more detailed description of two (charged) pion photoproduction, which is valid at intermediate momentum transfers ($0.4~\si{GeV^2}<|t|<1.0~\si{GeV^2}$). 
The model parameters are determined from a global fit of the available experimental data for angular moments up to $L=2$ and $M=2$. The angular moments fulfill $\braket{Y^L_M}=(-1)^M\braket{Y^L_{-M}}$, thus only moments for $M\geq0$ are considered. The resulting model provides a good description of the angular moments for all $t$-bins studied. By examining the fitted model in more detail, it is observed that the three features mentioned above --- the deformation of the $\rho(770)$ lineshape, SCHC and the $t$-dependence of the $\rho(770)$ photoproduction differential cross section --- cannot be explained by simple models which incorporate only Pomeron exchange. Having fit the model to the low angular moments, the model is validated by comparison to the higher angular moments in the two-pion channel. 

This paper is organized as follows: in~\cref{sec:kinematics}, kinematical variables and key observables are defined, in~\cref{sec:model} the theoretical model is developed, and in~\cref{sec:results} the fits to the data and the model predictions are presented. Finally, in~\cref{sec:conclusion} the conclusions are discussed, and possible further studies are suggested.

\section{Kinematics and angular moments}
\label{sec:kinematics}
\begin{figure}
\tdplotsetmaincoords{70}{20}
\begin{tikzpicture}[scale=0.9,tdplot_main_coords]
\coordinate (origin) at (0,0,0);
\pgfmathsetmacro{\lx}{6}
\pgfmathsetmacro{\ly}{4}
\pgfmathsetmacro{\lz}{5.5}
\pgfmathsetmacro{\nlx}{-1.5}
\pgfmathsetmacro{\nly}{-0.7}
\pgfmathsetmacro{\nlz}{-1.5}
\pgfmathsetmacro{\ax}{5.0}
\pgfmathsetmacro{\az}{4.5}
\pgfmathsetmacro{\nax}{-2.5}
\pgfmathsetmacro{\naz}{-2.5}
\pgfmathsetmacro\pe{ 5}
\pgfmathsetmacro\posi{0.05}
\pgfmathsetmacro\eD{-1}
\fill[jpac-blue, opacity=0.15] (\az,\ax,0) -- (\naz,\ax,0) -- (\naz,\nax,0) -- (\az,\nax,0) -- cycle; 
    \draw[->] (0,0,0) -- (\lz,0,0) node[anchor=north east]{$z$};
    \draw[->] (0,0,0) -- (0,\lx,0) node[anchor=north west]{$x$};
    \draw[->] (0,0,0) -- (0,0,\ly) node[anchor=south]{$y$};
    \draw[dashed] (0,0,0) -- (\nlz,0,0) node[anchor=north east]{};
    \draw[dashed] (0,0,0) -- (0,\nlx,0) node[anchor=north west]{};
    \draw[dashed] (0,0,0) -- (0,0,\nly) node[anchor=south]{};
\pgfmathsetmacro\th{40}
\pgfmathsetmacro\ph{60}
\pgfmathsetmacro\fpe{0.5}
\pgfmathsetmacro{\peZu}{cos(\th)};
\pgfmathsetmacro{\peXu}{sin(\th)*sin(\ph)};
\pgfmathsetmacro{\peYu}{sin(\th)*cos(\ph)};
\pgfmathsetmacro{\peXYu}{sin(abs(\th))};
\pgfmathsetmacro{\peZ}{\pe*\peZu};
\pgfmathsetmacro{\peX}{\pe*\peXu};
\pgfmathsetmacro{\peY}{\pe*\peYu};
\pgfmathsetmacro{\peXY}{\pe*sqrt(\peXu*\peXu + \peYu*\peYu)};
\coordinate (eta)  at ({+1.*\peZ},{+1.*\peX},{+1.*\peY});
\coordinate (pi)   at ({-1.*\peZ},{-1.*\peX},{-1.*\peY});
\coordinate (etaY)   at (\peZ,\peX,0);
\coordinate (etaX)   at (\peZ,0,\peY);
\coordinate (etaZ)   at (0,\peX,\peY);
\coordinate (etaZZ)   at (\peZ,0,0);
\coordinate (etaYY)   at (0,0,\peY);
\coordinate (etaXX)   at (0,\peX,0);
\pgfmathsetmacro\pg{ 4}
\pgfmathsetmacro\prix{4}
\pgfmathsetmacro\priy{-2}
\coordinate (gamma) at (2,-2,0);
\coordinate (gammaprol) at (-2,+2,0);
\coordinate (pro)   at (-2,0,0);
\coordinate (pri)   at ($(pro) - (gamma)$);
\draw[very thick,color=jpac-blue,->-] (origin) -- (gamma) node[midway,above right] {}; 
\node[color=jpac-blue, right] at (gamma){$\gamma$};
\draw[very thick,color=jpac-blue,->-] (gamma) -- (pro)    node[midway,below]  {$p_1$};
\draw[very thick,color=jpac-blue,->-] (origin) -- (pro)   node[midway,above left] {$p_2$};
\draw[very thick,jpac-red,->]  (origin) -- (eta)   node[anchor=south] {$\pi^+$};    
\begin{scope}[canvas is xy plane at z=0.0]
\draw[jpac-red,fill] (etaY)+(\posi,0) arc (0:360:\posi) {};
\end{scope}
\draw[thin,color=jpac-red,densely dashed]  (eta) -- (etaY)   node[below] {};    
\begin{scope}[canvas is yz plane at x=0.0]
\draw[jpac-red,fill] (etaZ)+(\posi,0) arc (0:360:\posi) {};
\end{scope}
\draw[thin,color=jpac-red,densely dashed]  (eta) -- (etaZ)   node[below] {};    
\draw[thin,color=jpac-red,densely dashed]  (origin) -- (etaZ)   node[below] {};    
\begin{scope}[canvas is yz plane at x=0.0]
\draw[jpac-red] ({\fpe*\peXY},0.0) arc (0:90-\ph:{\fpe*\peXY}) node[midway,above right] {$\phi^\text{H}$};
\end{scope}
\begin{scope}[canvas is plane={O(0,0,0)x(1,0,0)y(0,\peXu/\peXYu,\peYu/\peXYu)}]
\draw[jpac-red] ({\fpe*\pe},0.0) arc (0:\th:{\fpe*\pe}) node[midway,right] {$\theta^\text{H}$};
\end{scope}
\pgfmathsetmacro\phPol{40}
\pgfmathsetmacro\fpPol{0.5}
\pgfmathsetmacro\pPol{3}
\pgfmathsetmacro{\pepsXu}{cos(\phPol)};
\pgfmathsetmacro{\pepsYu}{sin(\phPol)};
\pgfmathsetmacro{\pepsX}{\pPol*\pepsXu};
\pgfmathsetmacro{\pepsY}{\pPol*\pepsYu};
\coordinate (eps)   at (0,\pepsX,\pepsY);
\end{tikzpicture}
\caption{The $\pi^+\pi^-$ helicity frame. Three-vectors in blue lie in the $x$-$z$ plane, while three-vectors in red make angles $(\theta^\text{H},\phi^\text{H})$ to the plane. In this frame, the two charged pions decay back-to-back. For clarity, only the three-momentum of the positive charged pion is shown. 
}\label{fig:helicity_frame}
\end{figure}
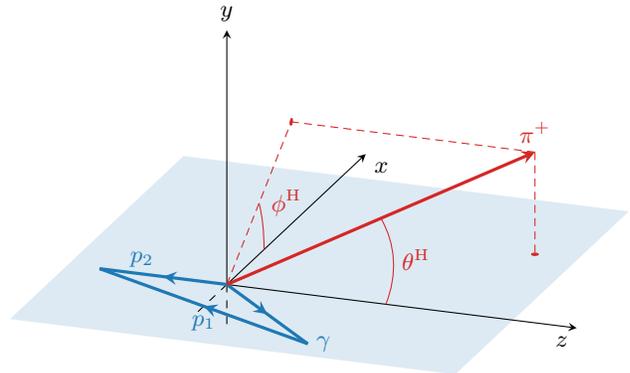

The following process is considered:
\begin{equation}
{\gamma}(q,\lambda_\gamma)+p(p_1,\lambda_1)\to \pi^+(k_1)+\pi^-(k_2)+p(p_2,\lambda_2).
\end{equation}
The helicities of the particles are defined in the $\pi^+\pi^-$ helicity frame, where the two-pion three-momenta are related via 
\mbox{$\mathbf{k}_1^\text{H}=-\mathbf{k}_2^\text{H}$}
and the recoiling proton ($\mathbf{p}_2^\text{H}$) defines the negative $z$-axis. The reaction ($x$-$z$) plane is defined by the three-vectors of the photon, target and recoiling proton. Then the unit vector normal to this plane defines the $y$-axis, that is, \mbox{$\mathbf y \parallel \mathbf{p}_2^\text{H} \times \mathbf{q}^\text{H}$}. With this choice of axes, \mbox{$\Omega^\text{H}=(\theta^\text{H},\phi^\text{H})$} define the angles of the $\pi^+$. The orientation of the helicity frame with respect to relevant kinematic variables is shown in~\cref{fig:helicity_frame}.

The helicity amplitudes for a $2\to3$ process are maximally described by five independent kinematic variables. Thus in addition to the two angles for the $\pi^+$, the following kinematic invariants shall be used:
\begin{subequations}
\begin{align}
s&=(p_1+q)^2=(p_2+k_1+k_2)^2,
\\
t&=(p_1-p_2)^2=(k_1+k_2-q)^2,
\\
s_{12}&=(k_1+k_2)^2=(p_1-p_2+q)^2.
\end{align}
\end{subequations}
The set of kinematic variables, $(s,t,s_{12},\Omega^\text{H})$ is complete, in the sense that all other kinematic invariants may be computed by knowing these five. Further details on the kinematics and the definitions of additional kinematic invariants are given in~\cref{app:kinematics}. The differential cross section is given by
\begin{equation}
\frac{d\sigma}{dt\, d\sqrt{s_{12}}\, d\Omega^\text{H}}= \kappa \sum_{\lambda_1\lambda_\gamma\lambda_2}|\mathcal{M}_{\lambda_\gamma \lambda_1 \lambda_2}(s,t,s_{12},\Omega^\text{H})|^2,
\end{equation}
where $\kappa$ contains the kinematical factors and is given by\footnote{The normalization of the $\kappa$ defined here is related to the version defined in Ref.~\cite{Mathieu:2019fts}, $\kappa^\text{prev}$ by $\kappa=(2\pi)\frac{\kappa^\text{prev}}{2}$. This choice is made to keep the notation for the unpolarized moments simple.}
\begin{equation}
\kappa=\frac{1}{8}\frac{1}{(2\pi)^4} \frac{\lambda^{1/2}(s_{12},m_\pi^2,m_\pi^2)}{16\sqrt{s_{12}}(s-m_N^2)^2},
\label{kappa}
\end{equation}
where $\lambda(a,b,c)=a^2+b^2+c^2-2(ab+ac+bc)$ is the K\"all\'en function. The angular moments of this distribution are defined as
\begin{equation}
\braket{Y^L_M}=\sqrt{4\pi}\int d\Omega^\text{H} \frac{d\sigma}{dt\, d\sqrt{s_{12}}\, d\Omega^\text{H}} \re{Y^L_M(\Omega^\text{H})},
\end{equation}
where the normalization is chosen to  ensure that the moment $\braket{Y_0^0}$ is equal to the integrated cross section, i.e.,\footnote{These moments may be related to the $H_{LM}^0$ defined in Ref.~\cite{Mathieu:2019fts} via $\expval{Y_M^L} = 2\pi \sqrt{2L+1}H_{LM}^0$. Only the real part of the spherical harmonics is retained in the definition, as the imaginary part identically vanishes for parity-conserving interactions.}

\begin{equation}
\label{eq:y00}
\expval{Y_0^0}=\frac{d\sigma}{dt\, d\sqrt{s_{12}}}.
\end{equation} 
By decomposing the amplitude into partial-waves,
\begin{equation}
\label{eq:partial_waves}
\mathcal{M}_{\lambda_\gamma\lambda_1\lambda_2}(s,t,s_{12},\Omega^\text{H})=\sum_{lm}\mathcal{M}_{\lambda_\gamma\lambda_1\lambda_2 m}^l(s,t,s_{12})Y^l_m(\Omega^\text{H}),
\end{equation}
it is possible to see that the angular moments are bilinear in the partial waves:
\begin{equation}
\begin{split}
\braket{Y^L_M}&=\sqrt{4\pi}\kappa \sum_{lml^\prime  m^\prime }A^{Lll^\prime}_{Mmm^\prime} 
\\
&\times \sum_{\lambda_\gamma\lambda_1\lambda_2} \mathcal{M}_{\lambda_\gamma\lambda_1 m^\prime \lambda_2 }^{l^\prime *}(s,t,s_{12})\mathcal{M}_{\lambda_\gamma\lambda_1 m \lambda_2 }^l(s,t,s_{12}),
\end{split}
\label{angular-moments}
\end{equation}
where
\begin{equation}
\begin{split}
A^{Lll^\prime}_{Mmm^\prime}&=\int d\Omega^\text{H} Y^l_m(\Omega^\text{H})Y^{l^\prime*}_{m^\prime}(\Omega^\text{H})\, \re{Y^L_M(\Omega^\text{H})}.
\end{split}
\label{ang.mom}
\end{equation}
Parity imposes relations on the helicity amplitudes. In particular, the amplitudes should obey
\begin{equation}
\begin{split}
\label{eq:parity_relation}
\mathcal{M}_{\lambda_\gamma\lambda_1 M\lambda_2 }^{l}(s,t,s_{12})&
\\
=(-1)^{\lambda_2-\lambda_1+\lambda_\gamma-M}&\mathcal{M}_{-\lambda_\gamma-\lambda_1-M -\lambda_2 }^{l}(s,t,s_{12}).
\end{split}
\end{equation}
This relation can be used to reduce by half the number of helicity amplitudes one must consider. In this work, the above symmetry will be used to relate the positive and negative photon polarizations. 
It will sometimes be advantageous to use spectroscopic notation for the partial-waves. 
To this end, the notation
\begin{equation}
[l]_{m}(s,t,s_{12})\equiv \mathcal{M}_{+1\lambda_1 m\lambda_2 }^l(s,t,s_{12}),
\end{equation}
where $[l]=S,P,D,\dots$ for $l=0,1,2,\dots$ will be used. Note that the photon helicity is fixed to be $\lambda_\gamma=+1$ and  the dependence on the nucleon helicities in the $[l]_{m}$'s is left implicit. The negative photon helicity $\lambda_\gamma=-1$ can be obtained using Eq.~\eqref{eq:parity_relation}.
The advantage of the $\pi^+\pi^-$ helicity frame lies in the fact that the helicity of the two-pion system coincides with the $m$ quantum number of the partial waves. Full expressions for these angular moments in terms of the partial waves are given in Appendix~\ref{app:angular_moments_partial_waves}.

\section{The Model}
\label{sec:model}

\begin{figure}
\vspace{10pt}
\subfloat[\label{subfig:resonant_production}]{
\begin{fmfgraph*}(50,60)
\fmfleft{l1,l2}
\fmfright{r1,r2,r3}
\fmf{fermion}{l1,v1,r1}
\fmf{photon}{l2,v2a}
\fmf{dbl_zigzag}{v1,v2a}
\fmf{dbl_plain}{v2a,v2b}
\fmf{dashes}{v2b,r2}
\fmf{dashes}{v2b,r3}
\fmflabel{$p(p_1,\lambda_1)$}{l1}
\fmflabel{$\gamma(q,\lambda_q)$}{l2}
\fmflabel{$p(p_2,\lambda_2)$}{r1}
\fmflabel{$\pi^+(k_1)$}{r3}
\fmflabel{$\pi^-(k_2)$}{r2}
\end{fmfgraph*}
\vspace{40pt}
}
\hspace{70pt}
\subfloat[\label{subfig:nonresonant_production}]{
\begin{fmfgraph*}(50,60)
\fmfleft{l1,l2}
\fmfright{r1,r2,r3}
\fmf{fermion}{l1,v1,r1}
\fmf{photon}{l2,v2}
\fmf{dashes}{v1,r2}
\fmf{dashes}{v1,v2,r3}
\fmfblob{0.2w}{v1}
\fmflabel{$p(p_1,\lambda_1)$}{l1}
\fmflabel{$\gamma(q,\lambda_q)$}{l2}
\fmflabel{$p(p_2,\lambda_2)$}{r1}
\fmflabel{$\pi^-(k_2)$}{r2}
\fmflabel{$\pi^+(k_1)$}{r3}
\end{fmfgraph*}
\vspace{40pt}
}
\caption{
The scattering amplitude for two pion photoproduction can be written as a sum of terms as shown in~\cref{eq:full-amp}. Dominant contributions at small momentum transfer can be separated into two categories based on their Feynman diagram topology. Diagrams of the type shown in (a) represent resonant, $\mathcal{R}$ contributions, while diagrams of the type (b) represent the Deck contribution. Note that there exists another diagram with the topology of (b) where the charged pions are exchanged. With this approximation, it is possible to relate this $2\to3$ process to $2\to2$ processes.
}\label{fig:model}
\end{figure}
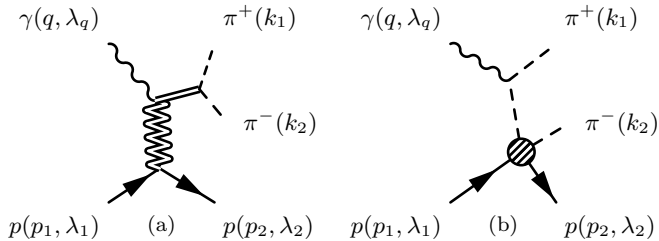

Since the focus of this work is the $\pi^+ \pi^-$ angular moments, particular emphasis is placed on the known low-energy resonances which decay to $\pi^+\pi^-$. These contributions are modeled via a two-step process, whereby a resonance is produced from $t$-channel scattering between the target nucleon and the photon beam, which then decays to form the two-pion final state. This generic process is depicted in~\cref{subfig:resonant_production}. The most obvious of these resonances is the large enhancement at a two-pion invariant mass of $\sqrt{s_{12}}\sim 0.77~\si{GeV}$ which may be attributed to the presence of the $\rho(770)$. However, a closer inspection of the two-pion angular moments suggests the presence of several other resonance-like structures. 
In this work, the resonances $f_0(500)$, $f_0(980)$, $f_0(1370)$ and $f_2(1270)$ are considered.\footnote{Some evidence exists for the presence of a radial excitation of the $\rho$ in the kinematic region studied here~\cite{Hammoud:2020aqi}, although at present, this state is not included in the PDG, and is not included in this analysis.}

In addition to these direct resonance contributions, the leading background~\cite{Bibrzycki:2018pgu} arises from the so-called Deck or \mbox{Drell--S\"oding} mechanism~\cite{Drell:1961zz,Deck:1964hm,Soding:1965nh}. Here a photon diffractively dissociates into a pion pair, with one of the pions being off-shell and brought on-shell by a further elastic scattering off the target proton.
As a result, one anticipates the dominance of one-pion exchange. The resulting amplitude may be factorized into an electromagnetic $\gamma \pi\pi$ vertex and a $2\to2$ subprocess related to elastic $\pi N$ scattering, which is well-known. This process is represented diagrammatically in~\cref{subfig:nonresonant_production}.

The full amplitude for this model may be written
\begin{equation}
\begin{split}
\mathcal{M}_{\lambda_\gamma\lambda_1 \lambda_2}(s,t,s_{12},\Omega^\text{H})&=\mathcal{M}_{\lambda_\gamma\lambda_1 \lambda_2}^\text{R}(s,t,s_{12},\Omega^\text{H})
\\
&+\, \mathcal{M}_{\lambda_\gamma \lambda_1\lambda_2}^\text{NR}(s,t,s_{12},\Omega^\text{H}),
\end{split}
\label{eq:full-amp}
\end{equation}
where the first term describes the resonant component of the model, while the second term describes the nonresonant (Deck) component. In the following sections, these two components are described in detail.

\subsection{Resonant Amplitude}
The resonant contribution to the full amplitude may be written as a sum of individual lineshapes:
\begin{table}
\caption{Resonance parameters employed in this work. Values are taken from Ref.~\cite{ParticleDataGroup:2022pth}.}
\label{tab:resonance_parameters}
\renewcommand{\arraystretch}{1.2}
\begin{ruledtabular}
\begin{tabular}{c|c c c}
$\mathcal{R}$ & $J$ & $m_\mathcal{R}$ [\si{GeV}] & $\Gamma_\mathcal{R}$ [\si{GeV}] \\ \hline
$f_0(500)$ & 0 & 0.500\phantom{0} & 0.450\phantom{0} \\
$\rho(770)$ & 1 & 0.775\phantom{0} & 0.149\phantom{0} \\
$f_0(980)$ & 0 & 0.990\phantom{0} & 0.055\phantom{0} \\
$f_2(1270)$ & 2 & 1.2755 & 0.1867 \phantom{0}\\
$f_0(1370)$ & 0 & 1.370\phantom{0} & 0.350\phantom{0}\\
\end{tabular}
\end{ruledtabular}
\end{table}
\begin{equation}
\mathcal{M}_{\lambda_\gamma\lambda_1 \lambda_2}^\text{R}(s,t,s_{12},\Omega^\text{H})=\sum_{\mathcal{R}} \mathcal{M}_{\lambda_\gamma\lambda_1 \lambda_2 }^{\mathcal{R}}(s,t,s_{12},\Omega^\text{H}).
\end{equation}
where the sum over $\mathcal{R}$ runs over all resonances considered and listed in Table~\ref{tab:resonance_parameters}. Each of these amplitudes is assumed to be decomposable into the product of a production amplitude, which describes the formation of an approximately stable resonance $\mathcal{R}$ with spin $J$, and a decay amplitude, which describes the decay of the resonance to two pions:
\begin{equation}
\begin{split}
\mathcal{M}_{\lambda_\gamma\lambda_1 \lambda_2 }^{\mathcal{R}}&(s,t,s_{12},\Omega^\text{H})
\\
&=\sum_{M=-J}^J \mathcal{M}_{\lambda_\gamma\lambda_1 M\lambda_2 }^{\gamma p \to \mathcal{R} p}(s,t) A^{\mathcal{R}}(s_{12})Y^J_M(\Omega^\text{H}).
\label{eq:resonant_amplitude}
\end{split}
\end{equation}
In principle, it is possible for the production amplitude, $\mathcal{M}_{\lambda_1\lambda_\gamma \lambda_2 M}^{\gamma p \to \mathcal{R} p}(s,t)$, to depend on the two-pion invariant mass. However, in this work, all two-pion mass dependence is assumed to originate from $A^{\mathcal{R}}(s_{12})$, which 
is an unconstrained dynamical function describing the two-pion invariant mass spectrum and is taken to be a \mbox{Breit--Wigner} distribution. More sophisticated approaches to the modelling of these resonances have been discussed extensively in the literature. Generally, modern spectroscopy analyses of the two-pion spectrum eschew the use of Breit-Wigner distributions in favor of a more theoretically sound approach. This point has been emphasized in particular in the study of the broad $f_0(500)$ resonance~\cite{Pelaez:2015qba,Garcia-Martin:2011iqs,Pelaez:2019eqa,Caprini:2005zr,Garcia-Martin:2011nna,Moussallam:2011zg}. In this work, the focus is on a global description of the angular moments, rather than a precise study of two-pion resonances. It is therefore not necessary to employ these more sophisticated modelling techniques. 

To complete the definition of the model, one must propose a form for the production amplitudes, $\mathcal{M}_{\lambda_1\lambda_\gamma \lambda_2 M}^{\gamma p\to \mathcal{R} p}$. In the kinematic region considered here, it is necessary to go beyond effective field theoretic techniques which are appropriate near threshold~\cite{Bernard:1994ds}. Thus, a combination of Regge theory and effective Lagrangians are employed to inspire the forms of these amplitudes. In particular, a version of the formalism employed in Ref.~\cite{Albaladejo:2020tzt,Winney:2022tky} is used here. In order to fix the notation, the essential results are restated.

 A generic $t$-channel exchange may be written
\begin{equation}
\mathcal{M}_{\lambda_\gamma \lambda_1 M \lambda_2  }^{\text{E}}(s,t)=\sum_j \mathcal{T}_{\lambda_\gamma M}^{\alpha_1\cdots \alpha_j} \mathcal{P}_{\alpha_1\cdots \alpha_j;\beta_1\cdot \beta_j}^{\text{E}} \mathcal{B}_{\lambda_1\lambda_2}^{\beta_1\cdots \beta_j}, 
\label{eq:J} 
\end{equation}
where $\mathcal{T}$, $\mathcal{B}$, and $\mathcal{P}$ are the top vertex, bottom vertex, and propagator, respectively. 
At high energies, and low momentum transfer, $t$ these amplitudes may be matched to Born-term $t$-channel diagrams where a hadron $\text{E}$ with spin $j$ is exchanged (not to be confused with the spin $J$ of the produced resonance). This fact is used to fix the form of the vertices $\mathcal{T}$ and $\mathcal{B}$. 
Summation of all allowed exchanged spins  in~\cref{eq:J}  results in a Regge pole amplitude.
In this study, the dominance of leading Regge pole amplitudes is assumed, corresponding to the exchange of $\rho$ and $\omega$ trajectories in production of the two-pion system with $J^P=0^+,2^+$, and  $\mathbb{P}$ and $a_2$, $f_2$ trajectories in production of  $J^P=1^-$. Explicitly, the $\gamma p\to \mathcal{R} p$ production amplitude is then written 
\begin{align}
\mathcal{M}_{\lambda_\gamma \lambda_1 M\lambda_2 }^{\gamma p\to f_{0,2} \, p}(s,t)&= \mathcal{M}_{\lambda_\gamma\lambda_1 M\lambda_2  }^{\rho/\omega}(s,t),
\\
\mathcal{M}_{\lambda_\gamma\lambda_1 M\lambda_2 }^{\gamma p\to \rho \, p}(s,t)&= \mathcal{M}_{\lambda_\gamma \lambda_1 M\lambda_2  }^{\mathbb{P}}(s,t)+\mathcal{M}_{\lambda_\gamma\lambda_1 M\lambda_2  }^{a_2/f_2}(s,t).
\end{align}
A Regge pole dominated amplitude has a generic factorized form, which is analogous to~\cref{eq:J} with  
\begin{equation}
\mathcal{P}^{\text{E}} \to 
 R^\text{E}(s,t)  =\frac{1}{s_0}
\frac{\alpha^E(t)}{\alpha^\text{E}(0)}\frac{1+\tau^\text{E} e^{-i\pi \alpha^\text{E}(t)}}{\sin\pi\alpha^\text{E}(t)}\bigg(\frac{s}{s_0}\bigg)^{\alpha^\text{E}(t)-1},
\end{equation}
being the Regge pole propagator. Here $\tau^\text{E}$ is the signature factor, $\alpha^\text{E}(t)=\alpha_0^\text{E}+t\, \alpha_1^\text{E}$, is the Regge trajectory and $s_0$ is a scale parameter conventionally chosen to be $s_0=1~\si{GeV^2}$ (the factor of $1/s_0$ has been included to preserve the dimensions of the amplitude).  Numerical values for these parameters and the Regge trajectories for the Reggeons employed in this work are given in Table~\ref{tab:regge_parameters}. For the product of the  helicity-dependent  couplings of the Reggeon to the photon (top vertex) and to the nucleon (bottom vertex) the form 
\begin{equation} 
 \mathcal{T} \times  
 \mathcal{B} 
 \to \mathcal{T}^\alpha_{\lambda_\gamma,M} \overline{u}_{\lambda_2}(p_2)\gamma_{\alpha} u_{\lambda_1}(p_1),
 \label{eq:BT} 
\end{equation}
is chosen,\footnote{This bottom vertex is sometimes referred to as the `Vector Pomeron Model'. Despite the name, it does not assume that the Pomeron carries vector quantum numbers, but is rather a model for the helicity structure that fulfills SCHC at high energies.} where  the top vertex, consistent with gauge invariance is given by  
\begin{subequations}
\begin{align}
{\mathcal{T}}_{\lambda_\gamma\hphantom{M}}^{\alpha}&=a^{\text{E},\mathcal{R}}(t)
[q^{\alpha} \epsilon_{\lambda_\gamma}^\sigma(q)-q^\sigma \epsilon_{\lambda_\gamma}^{\alpha}(q)]{k}_\sigma,
\\
{\mathcal{T}}_{\lambda_\gamma M}^{\alpha}&=a_{\lambda_\gamma M}^{\text{E},\mathcal{R}}(t)
[q^{\alpha} \epsilon_{\lambda_\gamma}^\sigma(q)-q^\sigma \epsilon_{\lambda_\gamma}^{\alpha}(q)]\epsilon_{M\sigma}^*(k),
\\
{\mathcal{T}}_{\lambda_\gamma M}^{\alpha}&=a_{\lambda_\gamma M}^{\text{E},\mathcal{R}}(t)
\{[q^\mu \epsilon_{\lambda_\gamma}^\rho(q) - q^\rho\epsilon_{\lambda_\gamma}^\mu(q)] (k-q)_\rho \epsilon_{M\mu}^{\alpha*}(k) \nonumber
\\
&-[q^\mu \epsilon_{\lambda_\gamma}^{\alpha}(q)-q^{\alpha}\epsilon_{\lambda_\gamma}^\mu(q)](k-q)^\nu \epsilon_{M\mu\nu}^*(k)\},
\end{align}
\end{subequations} 
for the production of a two-pion system in the $J=0,1,2$ partial waves respectively. 
Here $\epsilon_{\lambda_\gamma}^{\alpha}(q)$ is the polarization vector for the incoming photon, $\epsilon_{M\sigma}^*(k)$ is the polarization for the outgoing spin-1 particle with momentum $k=k_1+k_2$, while $\epsilon_{M \sigma_1\sigma_2}^*(k)$ is the polarization vector for the outgoing spin-2 particle. It may be written as
\begin{equation}
\epsilon_{M\mu\nu}(k) = 
\sum_{m_1,m_2} C^{JM}_{1m_1,1m_2}
\epsilon_{m_1\mu}(k)\epsilon_{m_2\nu}(k),
\end{equation}
where \mbox{$C^{JM}_{l_1 m_1,l_2m_2}=\braket{ J,M| l_1,m_1; l_2,m_2 }$} is the \mbox{Clebsch-Gordan} coefficient.
The choice of the 
helicity structure in the nucleon vertex is motivated by the expected dominance of the helicity conserving amplitude, although given that the data lacks polarization information this assumption cannot be verified. The helicity structure of the top vertex is similarly motivated. However, since the angular moments provide additional information about the relative weights of the helicity amplitudes, the effective couplings $a_{\lambda_\gamma M}^{\text{E},\mathcal{R}}(t)$ are fitted to the data. 

Since parity relates the positive and negative photon helicity amplitudes via Eq.~\eqref{eq:parity_relation}, this implies relations between the positive and negative photon helicity components of the effective couplings, $a_{\lambda_\gamma M}^{\text{E},\mathcal{R}}$. In order to keep notation for these effective couplings relatively simple, couplings are labelled for fixed $\lambda_\gamma=+1$. Then, the negative photon helicity amplitudes are defined by Eq.~\eqref{eq:parity_relation}. As a result, the explicit photon helicity label is suppressed in the following expressions.

The effective couplings depend on the two-pion final state and the Reggeon exchange, $\text{E}$. In the analysis, the parameters
\begin{align}
a_{M}^{\mathbb{P},\rho}(t)=g_{\gamma \mathbb{P}\rho}g_{\rho\pi\pi}\beta_\mathbb{P}(t),~~~~M=-1,0,1
\end{align}
are fixed with \mbox{$g_{\gamma \mathbb{P}\rho}=5.96$}, \mbox{$g_{\rho\pi\pi}= 2.506$} and \mbox{$\beta_\mathbb{P}(t)=\exp(bt)$} with $b=3.6~\si{GeV^{-2}}$ are taken from Ref.~\cite{Mathieu:2018xyc}, while the parameters
\begin{equation}
\begin{split}
&a^{\rho/\omega,f_0(500)}(t),~~
a^{\rho/\omega,f_0(980)}(t),~~
a^{\rho/\omega,f_0(1370)}(t),~~
\\
&a_{M}^{a_2/f_2,\rho}(t),~~
a_{M}^{\rho/\omega,f_2}(t),
\end{split}
\end{equation}
are fitted to experimental data.
Values of all the parameters (fixed and fitted) can be found in Appendix~\ref{parameter_values}. Finally, note 
that since the bottom vertex in~\cref{eq:BT} is $O(s)$ for large $s$, a factor of $1/s$ is included in the definition of the Regge propagator to ensure proper asymptotic behavior of the full amplitude. 

\begin{table}
\caption{Regge parameters used in this study.}
\label{tab:regge_parameters}
\renewcommand{\arraystretch}{1.2}
\begin{ruledtabular}
\begin{tabular}{c|c c c c}
$\text{E}$ & $\tau^\text{E}$ & $\alpha_0^\text{E}$ & $\alpha_1^\text{E}$ [\si{GeV^{-2}}]\\ \hline
$\mathbb{P}$ &  $+1$ & 1.08 & 0.2\\
$a_2/f_2$ & $+1$ & 0.5 & 0.9\\
$\rho/\omega$ & $-1$ & 0.55 & 0.8
\end{tabular}
\end{ruledtabular}
\end{table}

In the fits performed below, the $a_{ M}^{\text{E},\mathcal{R}}$ are allowed to be complex. Strictly speaking, in Regge theory one would expect these parameters to be real. However, complex phases are found to be essential for accounting for nonresonant contributions and interference effects arising from various production mechanisms which are not explicitly accounted for in the model construction. In this sense, the complex parameters can be viewed as an effective parametrization of features of the amplitude not explicitly implemented. It is important to note that the use of such complex couplings is not essential, and it is expected that a more sophisticaled model which incorporates additional corrections from these subleading effects (such as coupled-channels, daughters and absorption) can also reproduce the data. 

\subsection{Nonresonant Amplitude}
\label{sec:model:continuum}
In addition to modeling the leading resonant contributions, the model incorporates the expected leading background from the Deck process. To improve the description of the angular moments, empirically-motivated polynomial backgrounds are added to the low-lying partial waves. The nonresonant model may be written
\begin{equation}
\begin{split}
\mathcal{M}_{\lambda_\gamma\lambda_1 \lambda_2}^\text{NR} (s,t,s_{12},\Omega^\text{H})&=\mathcal{M}_{\lambda_\gamma\lambda_1 \lambda_2}^\text{Deck} (s,t,s_{12},\Omega^\text{H})
\\
&+\, \mathcal{M}_{\lambda_\gamma\lambda_1 \lambda_2}^\text{empr.} (s,t,s_{12},\Omega^\text{H}).
\end{split}
\end{equation}
In the following sections, these two background contributions are described in detail.

\subsubsection{Deck mechanism}
The Deck mechanism, which is expected to be the leading contribution to the non-resonant background, may be understood as a two-step process. One imagines the incoming photon first decaying into two pions, one of which is off-shell, (denoted here with an asterisk, i.e., $\pi^*$). Then, one of these pions recoils elastically against the nucleon target, producing the desired \mbox{$\pi^+\pi^-p$} final state. While the pion involved in the scattering with the nucleon target is off-shell, it is expected that this amplitude may be well approximated by the physical the $\pi N$ scattering amplitude, provided the virtuality of the pion is small. This allows one to relate the $2\to3$ process to $\pi N$ elastic scattering, which enables an essentially parameter-free description of the leading nonresonant background. It is useful to define the additional kinematic variables
\begin{align}
s_i&=(k_i+p_2)^2,\quad i=1,2,
\\
u_i&=(q-k_i)^2,\quad i=1,2,
\end{align}
which may be expressed in terms of the kinematic variables $(s,t,s_{12},\Omega^\text{H})$. Using these additional kinematic variables, the Deck mechanism for this process may be written as
\begin{widetext}
\begin{equation}
\begin{split}
\mathcal{M}_{\lambda_\gamma\lambda_1\lambda_2}^\text{Deck}(s,t,s_{12},\Omega^\text{H})&=e\frac{\epsilon_{\lambda_\gamma}(q)\cdot(2k_1-q)}{u_1-m_\pi^2}\beta(u_1){M}_{\lambda_1\lambda_2}^-(s_2,t;u_1)
-e\frac{\epsilon_{\lambda_\gamma}(q)\cdot(2k_2-q)}{u_2-m_\pi^2}\beta(u_2){M}_{\lambda_1\lambda_2}^+(s_1,t;u_2)
\\
&+\mathcal{M}_{\lambda_\gamma\lambda_1 \lambda_2}^\text{cont.}(s,t,s_{12},\Omega^\text{H}),
\end{split}
\end{equation}
\end{widetext}
where relative minus sign between the first two terms in the above expression arises because the photon couples to the charge of the exchanged pion. The four-point vertex, ${M}_{\lambda_1\lambda_2}^\pm$ represents the process \mbox{$\pi^{*\pm }p\to \pi^\pm p$}. In this study this vertex is approximated with a generalized form of the $\pi N$ scattering amplitude taken from Ref.~\cite{Mathieu:2015gxa}. The term $\mathcal{M}^\text{cont.}$ is a phenomenological contact term added to ensure the preservation of gauge invariance and 
\begin{equation}
\beta(u_i)=\exp(\frac{u_i-u_i^\text{min}}{\Lambda_\pi^2}),
\end{equation}
is a hadronic form factor introduced to suppress the Born term pion propagator for the one-pion exchange at large $u_i$. In this study, \mbox{$\Lambda_\pi=0.9~\si{GeV}$}. The transversity of the photon implies that \mbox{$\epsilon_{\lambda_q}(q)\cdot q=0$}, and so one may write the Deck amplitude in the equivalent form
\begin{equation}
\begin{split}
\mathcal{M}_{\lambda_\gamma\lambda_1\lambda_2}^\text{Deck}(s,t,s_{12},\Omega)&=e\bigg[\frac{\epsilon_{\lambda_\gamma}(q)\cdot k_2}{q\cdot k_2}\beta(u_2){M}_{\lambda_1\lambda_2}^+(s_1,t;u_2)
\\
&-\frac{\epsilon_{\lambda_\gamma}(q)\cdot k_1}{q\cdot k_1}\beta(u_1){M}_{\lambda_1\lambda_2}^-(s_2,t;u_1)\bigg]
\\
&
+\mathcal{M}_{\lambda_\gamma\lambda_1 \lambda_2}^\text{cont.}(s,t,s_{12},\Omega^\text{H}).
\end{split}
\end{equation}
Before proceeding any further, one must specify the form of the phenomenological contact interaction. This contact-term is assumed to have the functional form
\begin{equation}
\mathcal{M}_{\lambda_\gamma\lambda_1 \lambda_2}^\text{cont.}(s,t,s_{12},\Omega^\text{H})=\epsilon_{\lambda_\gamma}(q)\cdot V(s,t,s_{12},\Omega^\text{H}),
\end{equation}
where $V^\mu$ is an unknown four-vector. Recall that gauge invariance is expressed by the \mbox{Ward--Takahashi} identity, which in this case requires that under the substitution \mbox{$\epsilon_{\lambda_\gamma}^\mu(q)\to q^\mu$}, the amplitude is zero. This constraint fixes the form of the four-vector. Following the prescription of Pumplin~\cite{Pumplin:1970kp}, $V^\mu$ is chosen as $V^\mu=(p_1+p_2)^\mu v$, where $v$ is a scalar function of the five independent kinematic variables
\begin{equation}
v=\frac{\beta(u_1){M}_{\lambda_1\lambda_2}^-(s_2,t;u_1)-\beta(u_2){M}_{\lambda_1\lambda_2}^+(s_1,t;u_2)}{q\cdot (p_1+p_2)}.
\end{equation}
Thus the full Deck mechanism is
\begin{widetext}
\begin{equation}
\begin{split}
\mathcal{M}_{\lambda_\gamma\lambda_1\lambda_2}^\text{Deck}(s,t,s_{12},\Omega)&=\sqrt{4\pi\alpha}\bigg[\bigg(\frac{\epsilon_{\lambda_\gamma}(q)\cdot k_2}{q\cdot k_2}-\frac{\epsilon_{\lambda_\gamma}(q)\cdot(p_1+p_2)}{q\cdot(p_1+p_2)}\bigg)\beta(u_2){M}_{\lambda_1\lambda_2}^+(s_1,t;u_2)
\\
&- \bigg(\frac{\epsilon_{\lambda_\gamma}(q)\cdot k_1}{q\cdot k_1}-\frac{\epsilon_{\lambda_\gamma}(q)\cdot(p_1+p_2)}{q\cdot (p_1+p_2)}\bigg)\beta(u_1){M}_{\lambda_1\lambda_2}^-(s_2,t;u_1)\bigg].
\end{split}
\end{equation}
\end{widetext}
Details about the parameterization of the vertex $M_{\lambda_1\lambda_2}^\pm$ are given in~\cref{app:pin}.

\subsubsection{Polynomial backgrounds}
Empirically, it is found that the Deck mechanism is not sufficient to provide a completely satisfactory description of nonresonant background, and as a result a good description of the angular moments is not possible. The failure of the pure-Deck contribution to saturate the nonresonant background can be attributed to the presence of other nonresonant processes which can be expected to appear at these energies. For example, one expects $s$-channel contributions from nucleon resonances, or $t$-channel exchanges other than $\pi$, like the $\rho(770)$, which have not been explicitly considered here. Since the Deck model provides a parameter-free description of the leading background, it is possible to add an empirical polynomial background to absorb the remaining nonresonant production and achieve a reasonable description of the two-pion lineshape. In particular, for $J=0,1$ an additional amplitude with the same structure of Eq.~\eqref{eq:resonant_amplitude} is added to the model, where the unconstrained dynamical function $A^J(s_{12})$ is taken to be a polynomial of the form
\begin{equation}
A^J(s_{12})=(s_{12}-s_{12}^\text{min})(s_{12}-s_{12}^\text{max})
\end{equation}
where
\begin{align}
s_{12}^\text{min}&=4m_\pi^2,\\
s_{12}^\text{max}&=s+m_p^2-\frac{1}{2m_p^2}\bigg[(s+m_p^2)(2 m_p^2-t)\nonumber\\
&\quad-\lambda^{1/2}(s,m_p^2,0)\lambda^{1/2}(t,m_p^2,m_p^2)\bigg].
\end{align}
The constants, $s_{12}^\text{min}$ and $s_{12}^\text{max}$ were chosen to prevent an unnaturally large contribution from these non-resonant contributions, especially near threshold where the model is fit to data. Similar to the resonant amplitudes, complex, helicity-dependent couplings, $b_{M}^{\text{E},J}$ are determined by the data.

\section{Results}
\label{sec:results}

The model contains 30 free parameters, $\{a_{M}^{\text{E},\mathcal{R}}, b_{M}^{\text{E},J}\}$ which determine the relative strengths and phases of the production mechanisms. These parameters are fit to the experimental measurements of the angular moments reported in Ref.~\cite{CLAS:2009ngd}. Before comparing the model to data, details of the fit procedure are given. There exist four sets of moments measured by CLAS at photon laboratory frame energies 3.0--3.2 GeV, 3.2--3.4 GeV, 3.4--3.6 GeV, and 3.6--3.8 GeV. The Regge approach employed in this analysis is most suitable for large energies. Therefore the analysis is performed using data in the highest energy bin ($E_\gamma=3.6\text{--}3.8~\si{GeV}$). The model is evaluated at the central value $E_\gamma=3.7~\si{GeV}$.

\subsection{Fitting the Model to Angular Moments of Two-Pion Photoproduction}
\label{sec:model_fitting}
The results below show fits to experimental measurements of two-pion angular moments $\braket{Y_L^M}$ for $L=0,1,2$ and $M=0,\dots, L$. It is important to note that the $t$-dependence of two components of the model, namely the nonresonant Deck amplitude and the Pomeron-mediated $\rho$ photoproduction are not fit to data.

The $t$-dependence present in the experimental data reflects the complex interplay between different physical processes which contribute to two-pion photoproduction. Since  no strong physical constraints exist that allow determination of the $t-$dependence in the moderate $t$ region, here a data-driven or \mbox{\textit{bottom--up}}~\cite{JPAC:2021rxu} approach is taken by fitting each $t$ bin separately, effectively allowing the $t$-dependence to be entirely determined by the data. Each fit therefore takes into account 600 data points. Ascribing a physical mechanism to the effective $t$-dependence of the fitted parameters is left for future work.

\begin{figure*}
    \centering
    \includegraphics[scale=.53,clip]{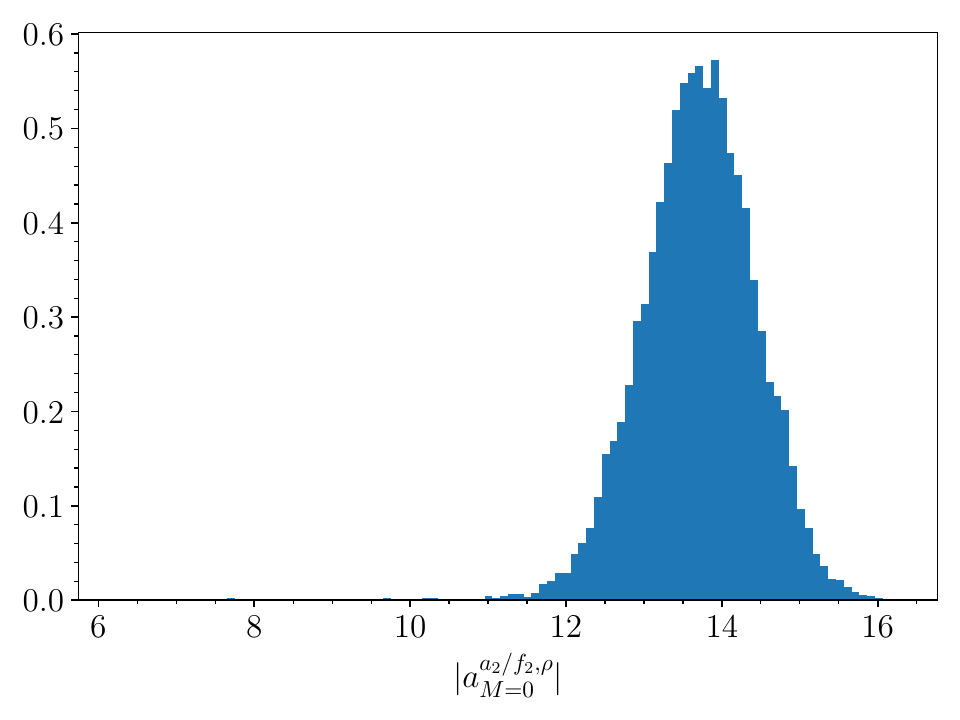}
    \includegraphics[scale=.53,clip]{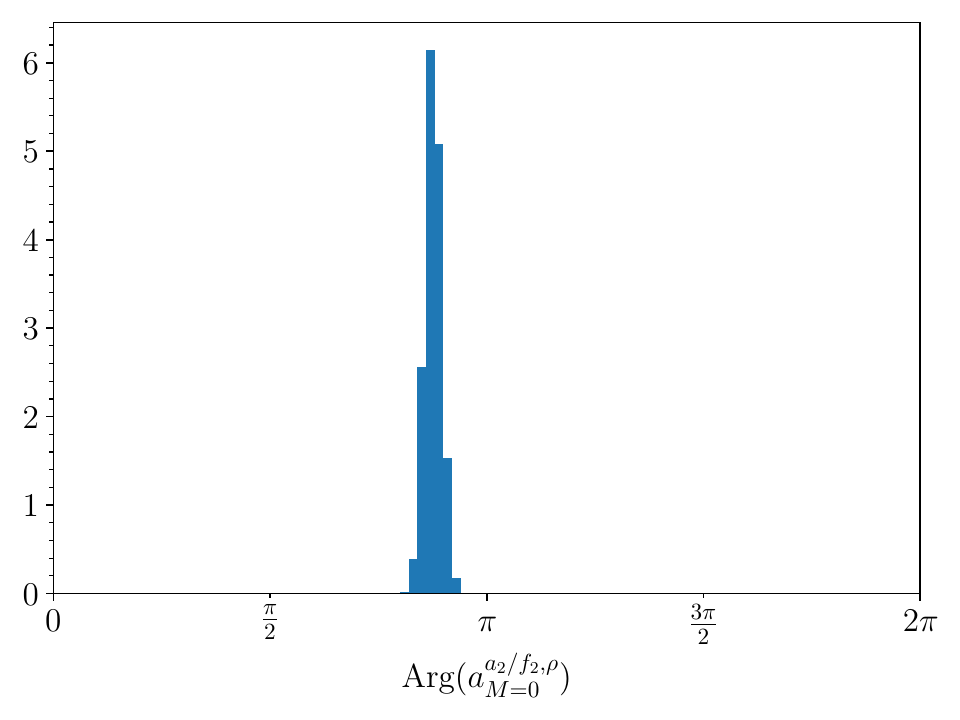}
    \caption{Normalized distribution of the absolute value (left) and the phase (right) of the $f_2$ exchange amplitude multiplier for the $\rho$ helicity $M=0$ fitted at $t=-0.45$ GeV$^2$.}
    \label{fig:f2_0}
\end{figure*}

\begin{figure*}[t]
\centering
\includegraphics[scale=0.45]{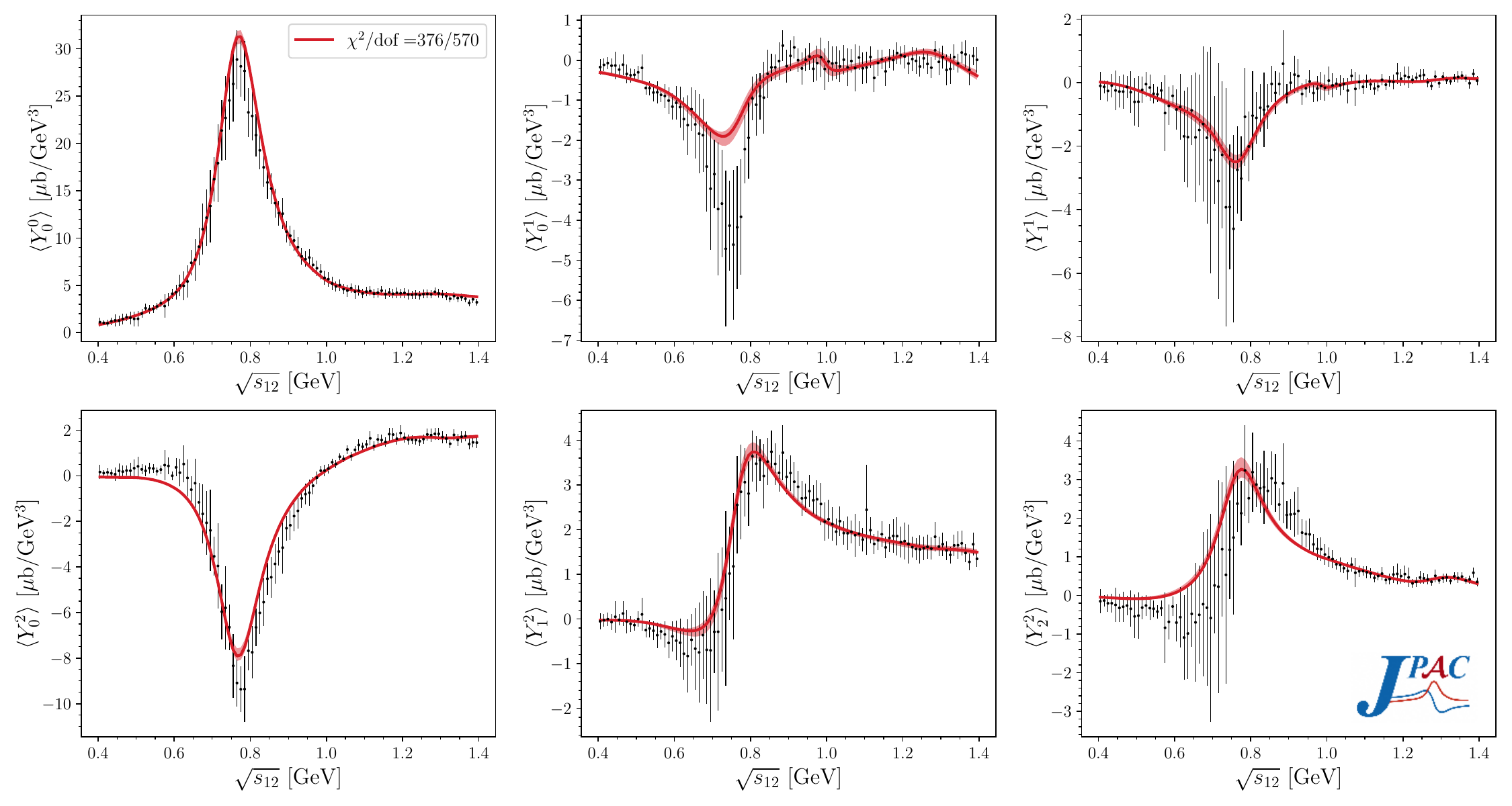}
\caption{Comparison of complete model fitted to exprimental measurements from Ref.~\cite{CLAS:2009ngd} of two-pion angular moments $\braket{Y^L_M}$ for $L=0,1,2$ and $M=0,\dots L$ for $E_\gamma=3.7~\si{GeV}$ and $t=-0.45~\si{GeV^2}$. Since all data shown here is fit simultaneously, this corresponds to $600-30=570$ degrees of freedom.}
\label{fig:YLM_t_0.45}
\end{figure*}

\begin{figure*}[t]
\centering
\includegraphics[scale=0.45]{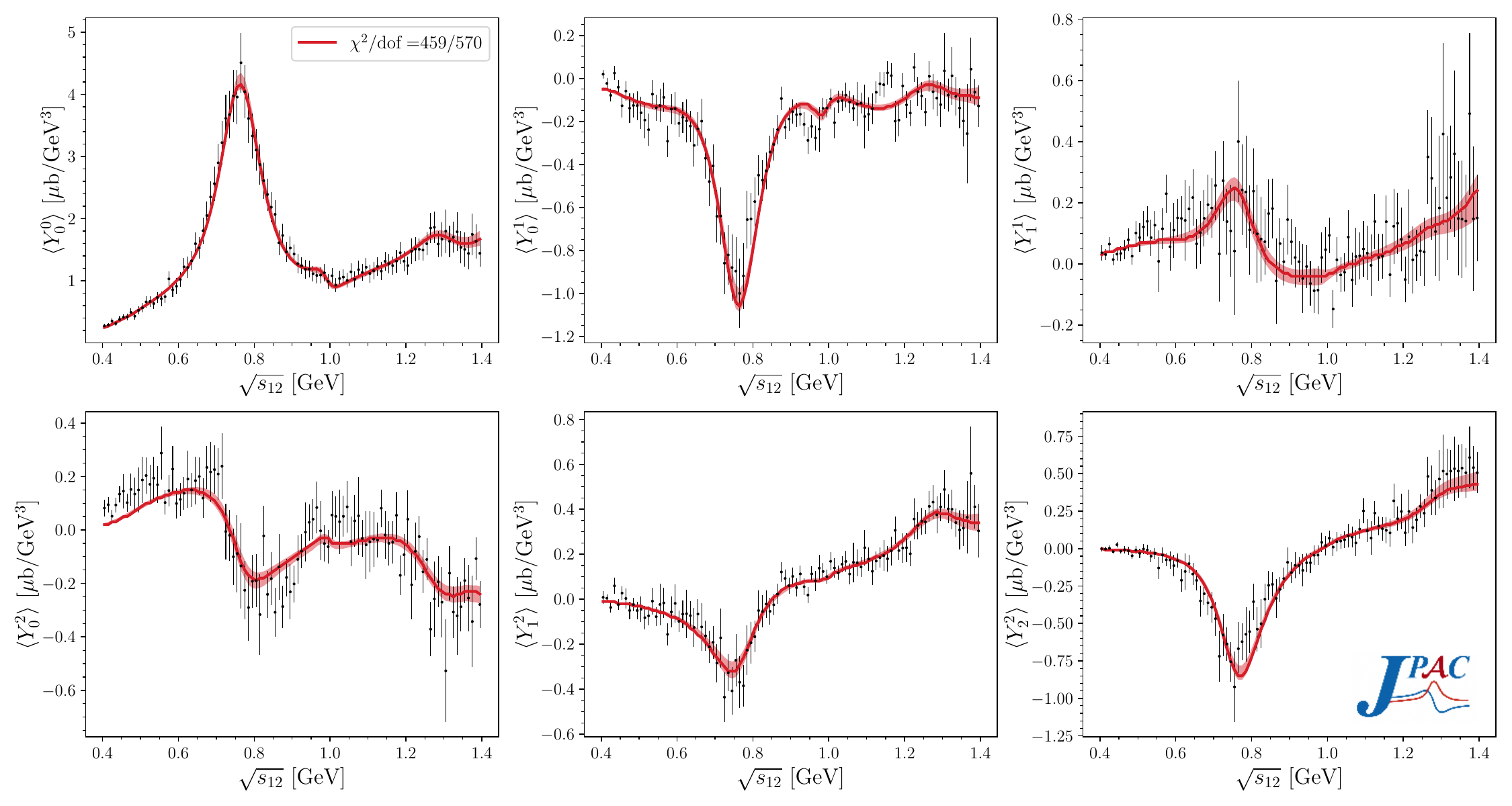}
\caption{Comparison of complete model fitted to exprimental measurements from Ref.~\cite{CLAS:2009ngd} of two-pion angular moments $\braket{Y^L_M}$ for $L=0,1,2$ and $M=0,\dots L$ for $E_\gamma=3.7~\si{GeV}$ and $t=-0.95~\si{GeV^2}$. Since all data shown here is fit simultaneously, this corresponds to $600-30=570$ degrees of freedom.}
\label{fig:YLM_t_0.95}
\end{figure*}

The results of these fits, as well as the associated statistical uncertainty bands obtained from the bootstrap method~\cite{JPAC:2021rxu} for representative $t$-bins, are shown in~\cref{fig:YLM_t_0.45,fig:YLM_t_0.95}. The rest of the fitted $t$-bins are shown in~\cref{app:comparisons_to_other_t}. The fit procedure consists of two steps. First, for each $t$-bin, $10^5$ randomly initialized fits are performed to the full set of moments. This procedure explores a large initialization region in the parameter space, providing confidence that the resulting best-fit values are near the global minimum of the $\chi^2$ function. Second, statistical uncertainties in the model parameters and moment values are determined by performing a bootstrap analysis with $10^4$ synthetic data samples. In this step, fits are initialized with the best-fit parameters determined in the first step. The resulting parameter distributions are Gaussian to good accuracy. A characteristic example of these parameter distributions is shown in~\cref{fig:f2_0}, where the absolute value and phase of the $a_{M=0}^{a_2/f_2,\rho}$ are shown. The distributions appear Gaussian and thus provide confidence that the parameter determinations are robust with no nearby minima that would suggest models with different physical interpretations. In addition, a good agreement between the model and data is obtained for all moments, leading to a global $\chi^2/\text{dof}\sim 1$ for each $t$-bin. The best-fit $\chi^2$ and $\chi^2/\text{dof}$ values are given in Table~\ref{tab:chi_squared}.\footnote{The $\chi^2$
function employed in these fits does not take into account correlations in data, which are expected to be large, since all moments are obtained from the same underlying set of measured events.} 
\begin{table}
\caption{$\chi^2$ and $\chi^2/\text{dof}$ for fitted data.}
\label{tab:chi_squared}
\renewcommand{\arraystretch}{1.2}
\begin{ruledtabular}
\begin{tabular}{c c c c c c c}
$|t|$ [\si{GeV^2}] & 0.45  & 0.55 & 0.65 & 0.75 & 0.85 & 0.95 \\ \hline
$\chi^2$ &            376  & 502  & 511  & 443  & 430  & 459 \\
$\chi^2/\text{dof}$ & 0.65 & 0.88 & 0.90 & 0.78 & 0.75 & 0.80 \\
\end{tabular}
\end{ruledtabular}
\end{table}
A slight increase in model uncertainty for masses around 1.4 GeV is because this region is expected to be dominated by the $D$-wave, due to the presence of the $f_2(1270)$ resonance. The only other $D$-wave contribution comes from the Deck amplitude, which, as mentioned already, is completely constrained, and contains no free parameters. As argued in \cite{Bibrzycki:2018pgu}, the $D$-wave Deck contribution is also considerably smaller than the $P$-wave.
Therefore the resonant $D$-wave helicity amplitudes do not have much to interfere with, and thus their relative strengths and phases are less constrained.

While the agreement between the model and data is generally good, some discrepancies can be observed. In particular, the dip predicted by the model for $\sqrt{s_{12}}$ values near the $\rho(770)$ mass and momentum transfers $|t|\in[0.45,0.75]~\si{GeV^2}$ for the $\braket{Y^1_{0}}$ moment is shallower than the experimental data. The fitted $\braket{Y^1_{1}}$ moment in turn exhibits slight phase discrepancies with the data. Both these points can be seen in~\cref{fig:YLM_t_0.45}.

These discrepancies could be attributed to the modeling of the nucleon-reggeon vertex, \mbox{$\mathcal{B}_{\lambda_1\lambda_2}^{\alpha}=\overline{u}_{\lambda_2}(p_2)\gamma^{\alpha} u_{\lambda_1}(p_1)$}, which is not the most general form. As a result, the model might exhibit certain relations between helicity amplitudes which are not due to physical requirements. Unfortunately, it is not possible to investigate this any further, as the information on individual nucleon helicity couplings is so far unaccessible experimentally. 

It is worth noting the increasing prominence of the resonances $f_0(980)$ and $f_2(1270)$ as the square of the four-momentum increases. These signals are not well established as they are typically at the 1 or 2$\sigma$ significance level, but mark regions of interest in $\pi\pi$ invariant mass for future experimental and theoretical studies. 

\subsection{Understanding the Angular Moments in Terms of the Underlying Production Mechanisms}
\begin{figure*}
    \includegraphics[scale=.5]{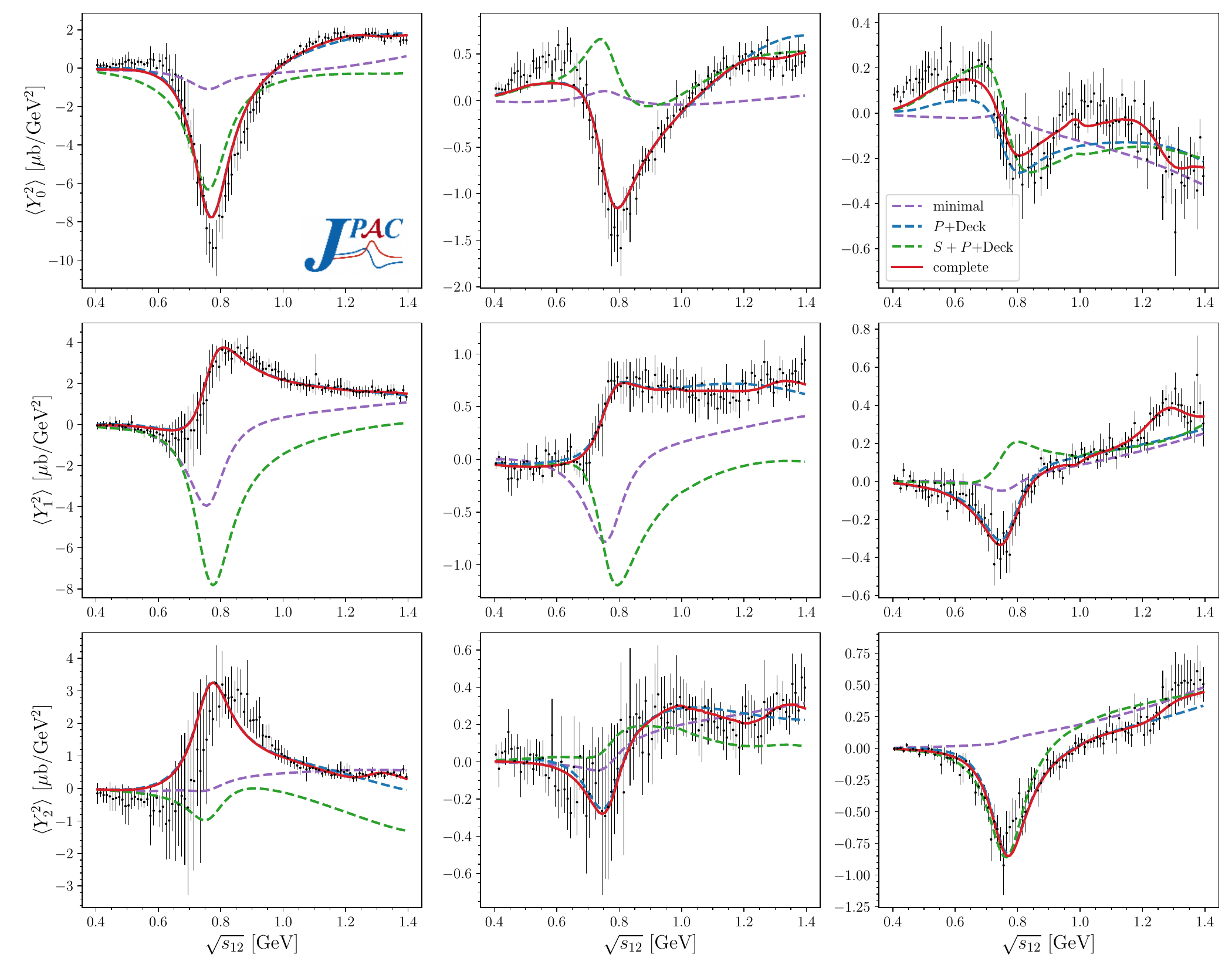}
    \caption{Studying the production amplitudes which contribute to $\braket{Y_M^2}$ at different momentum transfer. Results are shown for $t=-0.45~\si{GeV^2}$ (left) $t=-0.65~\si{GeV^2}$ (center) and $t=-0.95~\si{GeV^2}$ (right). Colors label the different model decompositions introduced in the text. Experimental data is taken from CLAS~\cite{CLAS:2009ngd}.}
    \label{fig:decompositions_20_21_22}
\end{figure*}

Physically motivated amplitude parameterizations offer the possibility of interpreting the relative contributions to the predicted angular moments in terms of the proposed production amplitudes. As already mentioned, physical constraints on the amplitudes in the moderate-$t$ region are rather mild. Therefore, apart from the Pomeron exchange and Deck amplitudes whose use can be treated as an extrapolation from the low-$t$ region, other amplitudes' parametrizations are kept as flexible as possible.
This section analyzes the model in relation to the included production mechanisms. It will be shown that each of the phenomena which were attributed to Pomeron exchange in the forward limit (the $\rho(770)$ lineshape, SCHC and the $t$-dependence),
cannot be interpreted in this way at the larger momentum transfers considered in this work. In order to do this, it is helpful to separate the various components of the model, and examine how they interfere to produce the resulting set of angular moments. Specifically, three decompositions plus the full model are defined. These are (in order of increasing complexity):
\begin{enumerate}
\item minimal,
\item $P+\text{Deck}$,
\item $S+P+\text{Deck}$,
\item complete.
\end{enumerate}
The first, labeled ``minimal", includes Pomeron-induced resonant $\rho(770)$ production and the Deck mechanism. Since no parameters in this decomposition are fit, it may be interpreted as a minimal two-pion photoproduction model. This model is similar in spirit to the models of S\"oding and Deck~\cite{Drell:1961zz,Soding:1965nh,Deck:1964hm,Pumplin:1970kp,PhysRevLett.25.485}, which describe the process as an interference between a nonresonant Deck amplitude and a resonant $\rho(770)$ photoproduction amplitude. Such descriptions have been shown to be suitable for a characterization of the angular moments at smaller momentum transfer. In particular, these models provide an explanation of the observed broadening and skewing of the $\rho(770)$ lineshape compared to the pion-initiated $\rho(770)$ production. As shall be seen, the minimal model generally does a poor job of describing the angular moments, motivating the necessity of incorporating additional amplitudes. Two amplitudes decompositions, ``$P+\text{Deck}$'' and ``$S+P+\text{Deck}$'' include the full set of the selected spin-$J$ resonances (e.g. the $\rho(770)$ for ``$P+\text{Deck}$''), plus the corresponding polynomial background. It is important to note that these split amplitudes cannot be interpreted as models in their own right, since parameters are not refit in these scenarios. Rather, they can be used to gauge the effects of including additional amplitudes to the model. Finally, ``complete'' is the fully developed model presented in this work and fitted to the data which includes all resonant amplitudes, plus the full non-resonant component arising from the sum of the Deck mechanism and the polynomial backgrounds. The improvement of this model over the ``minimal'' model can be seen clearly. Figures ~\ref{fig:decompositions_00_10_11} and \ref{fig:decompositions_20_21_22}
present the moment decompositions. 

\subsubsection{Analyzing $\braket{Y_0^0}$ and the $\rho(770)$ lineshape}
While information may be learned by studying each of the angular moments, it is useful to note that $\braket{Y_0^0}$ is the sum of squared partial-wave amplitudes (see~\cref{app:angular_moments_partial_waves}). Thus the analysis of this moment is simplified because interference between the partial waves does not occur. Note however, that since the Deck contributes to all two-pion partial waves, the decompositions are not bounded from above by the complete model. The leading contribution arises from the $\rho(770)$ resonance, which is known to be the dominant feature in this kinematic region. 

\begin{figure*}
    \centering
    \includegraphics[scale=.5]{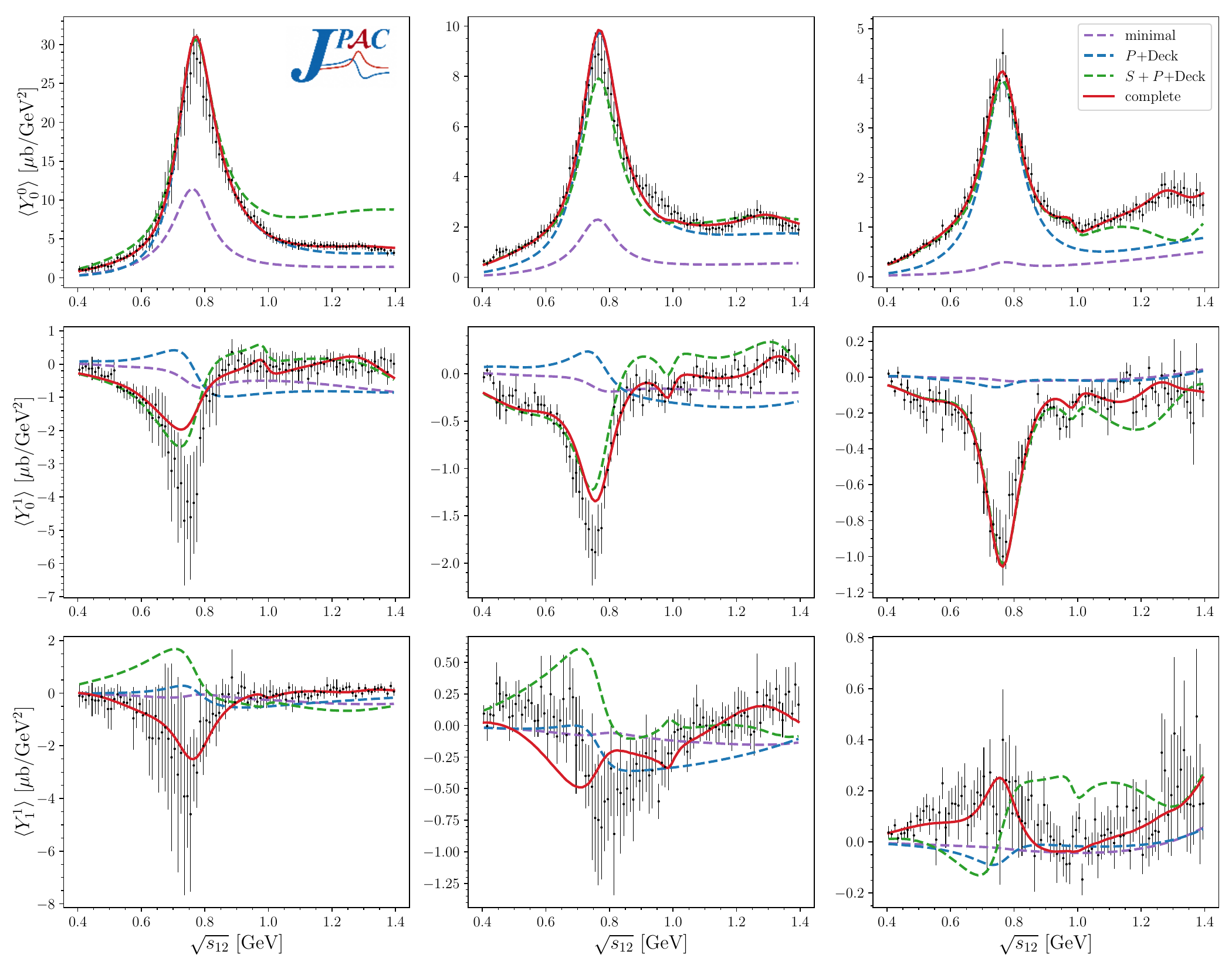}
    \caption{Studying the production amplitudes which contribute to $\braket{Y_M^L}$ for $L=0,1$ and all $M$ at different momentum transfer, $t$. Results are shown for $t=-0.45~\si{GeV^2}$ (left), $t=-0.65~\si{GeV^2}$ (center) and $t=-0.95~\si{GeV^2}$ (right). Colors label the different model decompositions introduced in the text. Experimental data is taken from CLAS~\cite{CLAS:2009ngd}.
    } 
    \label{fig:decompositions_00_10_11}
\end{figure*}

In the forward limit, standard Regge phenomenology implies that the dominant production mechanism for the $\rho(770)$ should be the Pomeron exchange. 
However, at the larger momentum transfers studied in this work, this is not the case. This may be seen by examining the prediction of the minimal model, which is shown in~\cref{fig:decompositions_00_10_11}, along with the other decompositions mentioned previously. Note that the minimal model seriously underestimates the angular moment $\braket{Y_0^0}$ for all $t$-bins, demonstrating clear evidence for physics beyond the Pomeron exchange models. 

Thus, it is important to identify the various physical mechanisms which contribute meaningfully to the angular moments in this kinematic region and study their $t$-dependence.
The model decomposition $P+\text{Deck}$ generally provides a reasonable description of $\braket{Y_0^0}$ in the vicinity of the $\rho(770)$ peak. However, note that this decomposition cannot reproduce both the magnitude of the data at the peak and the lineshape. In general, this decomposition always produces a lineshape which is narrower than the experimental data. At larger $t$ values, the magnitude of the $\rho(770)$ decreases. As a result, other discrepancies between this decomposition and the data become more pronounced, away from the $\rho(770)$ peak. The value of including $S$-wave resonances and the corresponding $S$-wave polynomial background in the decomposition $S+P+\text{Deck}$ can be particularly appreciated for $\sqrt{s_{12}}<m_\rho$. In this region, it is natural to expect large relative contributions from $S$-wave contributions, since other partial-waves are kinematically suppressed near threshold. 
For $t=-0.45~\si{GeV^2}$ (see~\cref{fig:decompositions_00_10_11}), the model exhibits strong $D$-wave cancellation between the resonant contribution due to the $f_2(1270)$ and the $D$-wave component of the nonresonant Deck amplitude. Large interference between resonant and nonresonant $D$-wave contributions are also seen at larger $t$. These large interference effects demonstrate the importance of including these subleading contributions to the amplitude. 

\subsubsection{Analyzing Angular Moments with $L=1,2$}
More information can be obtained by considering the higher angular moments. It is important to note, however, that these are sensitive to the interference between different partial waves. This makes the interpretation somewhat more complicated. For the full expressions of the angular moments in terms of partial waves, see~\cref{app:angular_moments_partial_waves}. In order to understand them, it is useful to note that the dominant feature present in the low angular moments is the $P$-wave $\rho(770)$ resonance. Since the angular moments are bilinear in the partial waves, this means that each term which contributes to a particular angular moment contains the product of zero, one or two $P$-wave amplitudes. Heuristically, terms which contain two $P$-wave amplitudes ($|P_0|^2$, say), can be expected to dominate relative to other contributions in the vicinity of the $\rho(770)$ peak. If an angular moment does not contain any terms which are the product of two $P$-wave amplitudes, it is natural to expect that the dominant contributions to the angular moments arise from the terms linear in the $P$-wave amplitudes. This general trend can be observed in the model decompositions shown below. 

The $S$-wave contributions can be most conveniently observed by studying the angular moments $\braket{Y^1_0}$ and $\braket{Y_1^1}$, since in these moments the $S$-wave interferes with the dominant $P$-wave amplitudes. In particular, it interferes with $P_0$ in $\braket{Y^1_0}$ and $(P_{+1}-P_{-1})$ in $\braket{Y_1^1}$, respectively. The only $S$-wave contribution in both the minimal model and $P+\text{Deck}$ decomposition arises from the Deck amplitude, which projects onto all partial waves. The minimal model predictions for the angular moments are shown in~\cref{fig:decompositions_00_10_11}. It is clear that the phase and magnitude of the $S$-wave amplitude are poorly described in these decompositions. This justifies the inclusion of other $S$-wave contributions beyond those from the Deck mechanism. The decomposition $S+P+\text{Deck}$ produces a much improved description of these angular moments allowing the global properties of these moments to be captured. For large invariant masses, the contribution of the resonant $D$-wave is also needed. Even though its global statistical significance is not very strong, there exists an interesting $t$ dependence of the interference pattern in the vicinity of the $f_0(980)$ mass. This effect is particularly visible in the $\braket{Y_0^{1}}$ moment. Namely, it starts as a maximum for $t=-0.45$ GeV$^2$ and continues as a minimum from $t=-0.65$ GeV$^2$ up to the $t=-0.95$ GeV$^2$ bin. This is suggestive of considerable relative phase motion between the resonant $S$-wave and the $P$-wave and is most likely due to the effects of different Regge exchanges and nonresonant contributions.

In previous $\pi^+\pi^-$ photoproduction analyses conducted at small momentum transfers, the shape of $\braket{Y_0^{2}}$ moment was interpreted as due to the dominant resonant $P_{+1}$ wave, see~\cref{eq:ylm20} in~\cref{app:angular_moments_partial_waves}. This interpretation is still applicable for the moment measured at $t=-0.45~\si{GeV^2}$. However, for larger $|t|$ bins, interference effects become increasingly important. 
Specifically, it is due to the combined interference of the $f_2$ exchange amplitude with Pomeron and $P$-wave nonresonant amplitudes that the model $\braket{Y^2_{0}}$ moment features the proper sign and magnitude at the $\rho(770)$ mass. The contribution of nonresonant amplitudes is also necessary to reproduce the plateau for masses above 1~\si{GeV}. With increasing $t$, this plateau is replaced by the interference of $S$- and $D$-waves.

If direct $\rho(770)$ production dominates other production mechanisms, the $\braket{Y^2_1}$ and $\braket{Y_{2}^2}$ moments are expected to exhibit activity only around the $\rho$ mass region, with interferences $(P_{+1}-P_{-1})P_0$ and $P_{+1}P_{-1}$, respectively. However, substantial activity above 1 GeV implies a prominent role of nonresonant effects. It is interesting to note that while the minimal model clearly fails to reproduce the data in the vicinity of the $\rho(770)$ region, as $|t|$ increases, the prediction of larger $\sqrt{s_{12}}\geq 1~\si{GeV}$ becomes increasingly well-described. This agreement between the model and the data in this kinematic region can be attributed to the Deck mechanism, which produces the approximately linearly growing component of the angular moment at large $\sqrt{s_{12}}$. The agreement for these moments at larger $\sqrt{s_{12}}$ and $|t|$, demonstrates the importance of the Deck mechanism.

Interference activity above 1 GeV in the $\braket{Y^2_{1}}$ moment is due to the interference of $D_{+1}$ and $D_{-1}$ waves with the $S$- wave and the $D_0$ wave. The activity in $\braket{Y^{2}_2}$ above 1 GeV is due to $D_{+1}D_{-1}$ interference. Contributions of the $P$- and $F$-waves in this region should amount to building a flat background, however resonant contributions of $\rho(1450)$ or the tentative $\rho(1250)$ from Ref.~\cite{PhysRevD.102.054029} cannot be excluded. 

From the study of these model decompositions, it is clear that the observed angular moments in this kinematic region cannot be interpreted in terms of Pomeron-induced $\rho(770)$ photoproduction alone. Rather, these decompositions demonstrate that the interference of different production mechanisms is required to reproduce the global features of the angular moments. Evidence for the influence of the $\rho(770)$, $f_0(980)$ and $f_2(1270)$ resonances on the angular moments are observed.

\subsection{Determination of the $\rho^0(770)$ photoproduction cross section}
\begin{figure}
    \centering
    \includegraphics[scale=0.5]{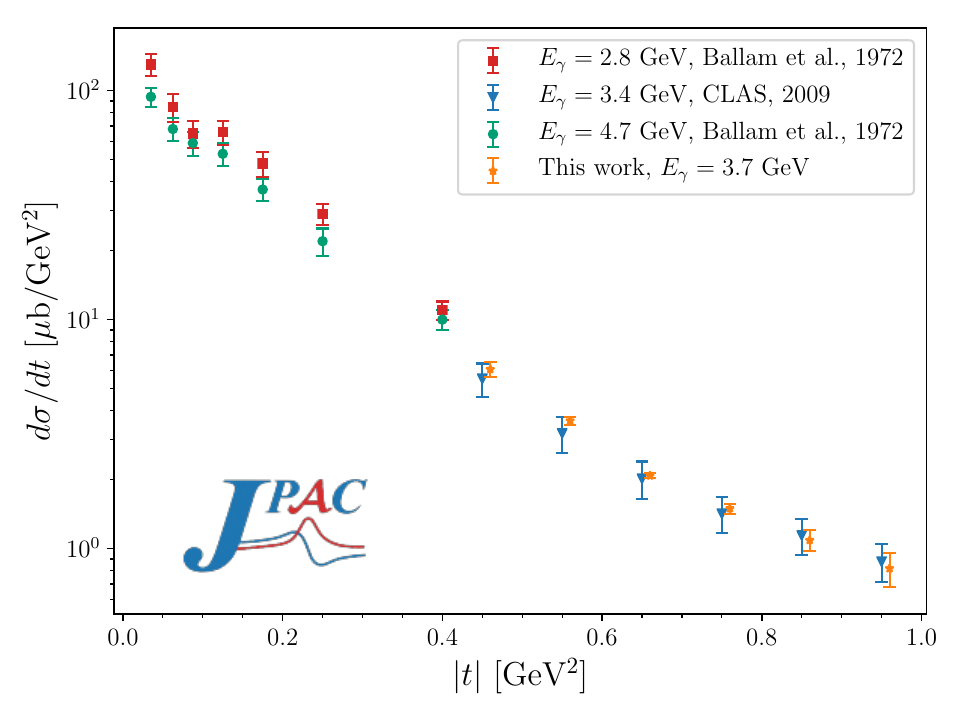}
    \caption{Comparison of differential cross section for process $\gamma p\to \rho^0(770) p$  computed in this work with previous measurements of this quantity. Uncertainties are estimated using bootstrap resampling. Ballam~{\it et al.} extract the $\rho$ contribution from $\pi^+\pi^-$ photoproduction data assuming a Breit-Wigner lineshape~\cite{Ballam:1972eq}. In CLAS, the full $P$-wave contribution is integrated over the $[0.4,~1.2]~\si{GeV}$ mass range and quoted~\cite{CLAS:2009ngd}. The cross section in this work is calculated by isolating the resonant $\rho(770)$ contribution and integrating over the same energy range. As the $\rho(770)$ dominates the $P$-wave, the difference with CLAS is negligible within errors. Note that values have been slightly offset horizontally for clarity.}
    \label{fig:rho_photoproduction}
\end{figure}

The model developed in this work contains two amplitudes which describe this process as proceeding by resonant production of the $\rho(770)$. By isolating just those amplitudes, and integrating over the lineshape of the $\rho(770)$, it is possible to extract the $\rho^0(770)$ photoproducton differential cross section. The contribution to $\expval{Y_0^0}$ from resonant $\rho(770)$ amplitudes is denoted $\expval{Y_0^0}_\rho$. Then, from Eq.~\eqref{eq:y00}, it is clear that the differential cross section can be obtained from computing 
\begin{equation}
\frac{d\sigma}{dt} = \frac{1}{\mathcal{B}}\int_{m_\text{min}}^{m_\text{max}} d \sqrt{s_{12}} \expval{Y_0^0}_\rho\,,
\end{equation}
where $\mathcal{B}\simeq 1$~\cite{ParticleDataGroup:2022pth} is the $\rho\to\pi\pi$ branching ratio. To compare with Ref.~\mbox{\cite{CLAS:2009ngd}}, the integration bounds are taken as $m_\text{min}=0.4~\si{GeV}$ and $m_\text{max}=1.2~\si{GeV}$. The resulting extracted $\rho(770)$ photoproduction cross section is shown in~\cref{fig:rho_photoproduction}. The explicit values are presented in Table~\ref{tab:rho_photoproduction}. This result can be compared with previous determinations of this process from Ref.~\cite{Ballam:1970qn} and the CLAS determination from Ref.~\cite{CLAS:2009ngd}. It is important to note that since the couplings are fit to each $t$-bin independently, they acquire and effective $t$-dependence. Thus the differential cross-section is not guaranteed to be a smooth function of $|t|$. The resulting differential cross section agrees within errors with the determination from Ref.~\cite{CLAS:2009ngd}. As shown in Fig.~\ref{fig:pomeron_model}, a Pomeron model would predict an exponentially decreasing differential cross section. However, at larger $|t|$ the non-Pomeron contributions enhance the differential cross section relative to the Pomeron model.

\begin{table}
\centering
\caption{Extracted $\rho^0(770)$ photoproduction cross section.}
\label{tab:rho_photoproduction}
\renewcommand{\arraystretch}{1.2}
\begin{ruledtabular}
\begin{tabular}{c c}
$|t|~[\si{GeV^2}]$  & $d\sigma/dt$ [$\mu\text{b}/\si{GeV^2}$] \\ \hline
0.45 & $6.074\pm0.446$\\
0.55 & $3.619\pm0.152$\\
0.65 & $2.080\pm0.051$\\
0.75 & $1.496\pm0.074$\\
0.85 & $1.089\pm0.112$\\
0.95 & $0.819\pm0.141$\\
\end{tabular}
\end{ruledtabular}
\end{table}

\subsection{$s$-channel helicity conservation and the partial waves}
SCHC expresses the observation that, at small $|t|$, the helicity amplitudes with $\lambda_\gamma \neq M$ are suppressed relative to the helicity-conserving amplitude. 
Physically, this means that the helicity of the produced $\rho(770)$ is strongly correlated with the helicity of the photon beam. 

\begin{figure*}
    \centering
    \subfloat[\label{subfig:partialwaves}]{
    \includegraphics[scale=0.35]{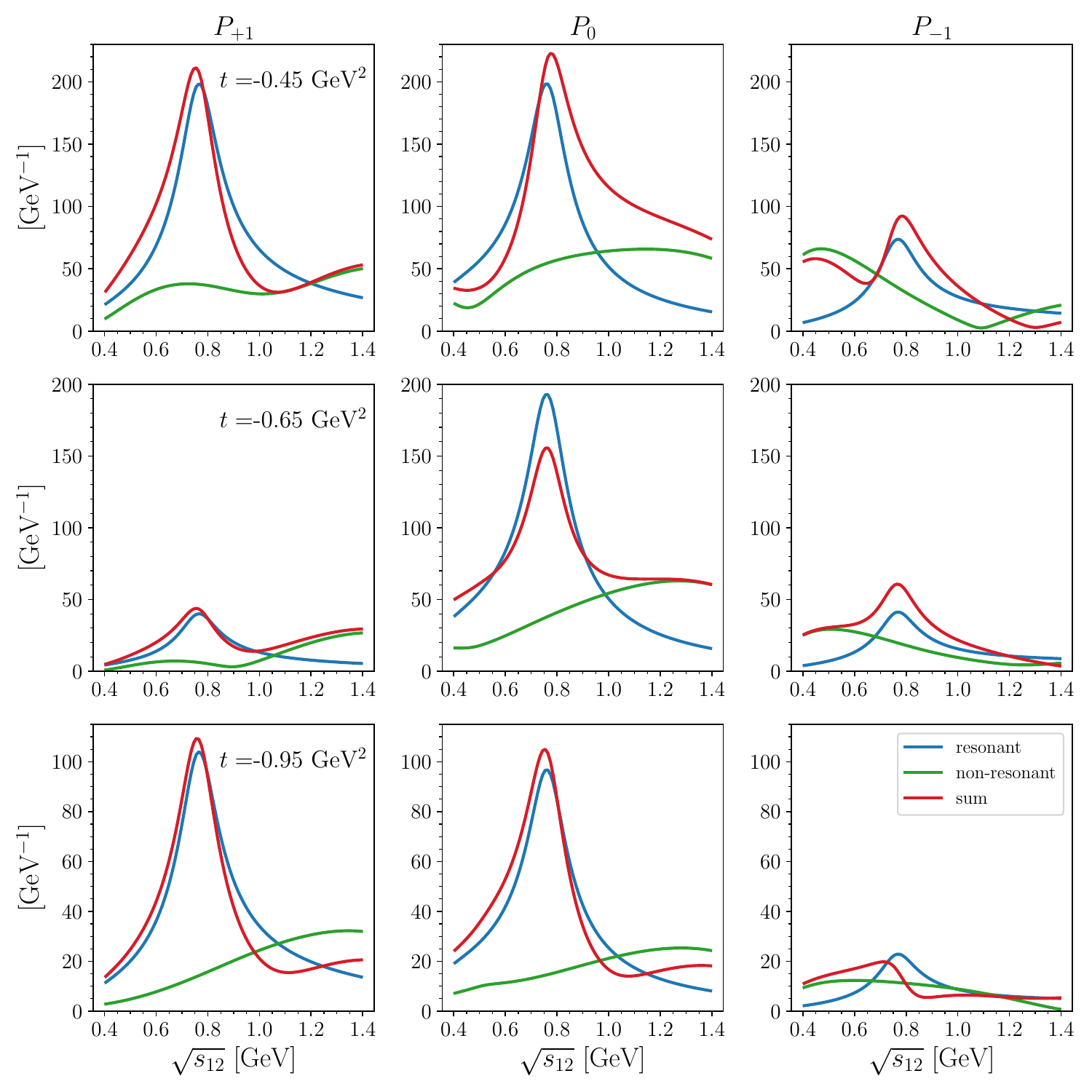}}
    \subfloat[\label{subfig:argandplus}]{
    \includegraphics[scale=0.35]{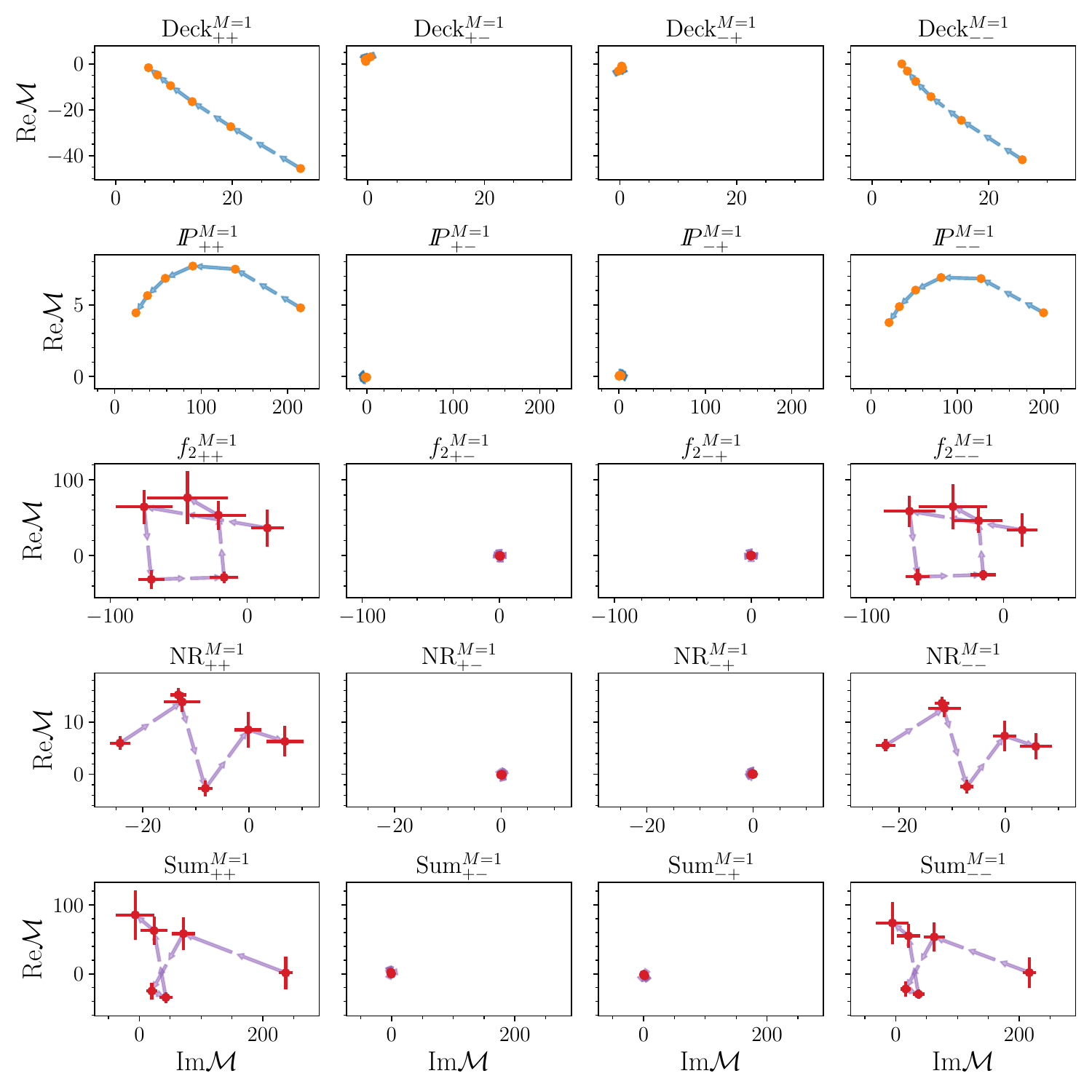}}
    \vspace{-10pt}

    \subfloat[\label{subfig:argand0}]{
    \includegraphics[scale=0.35]{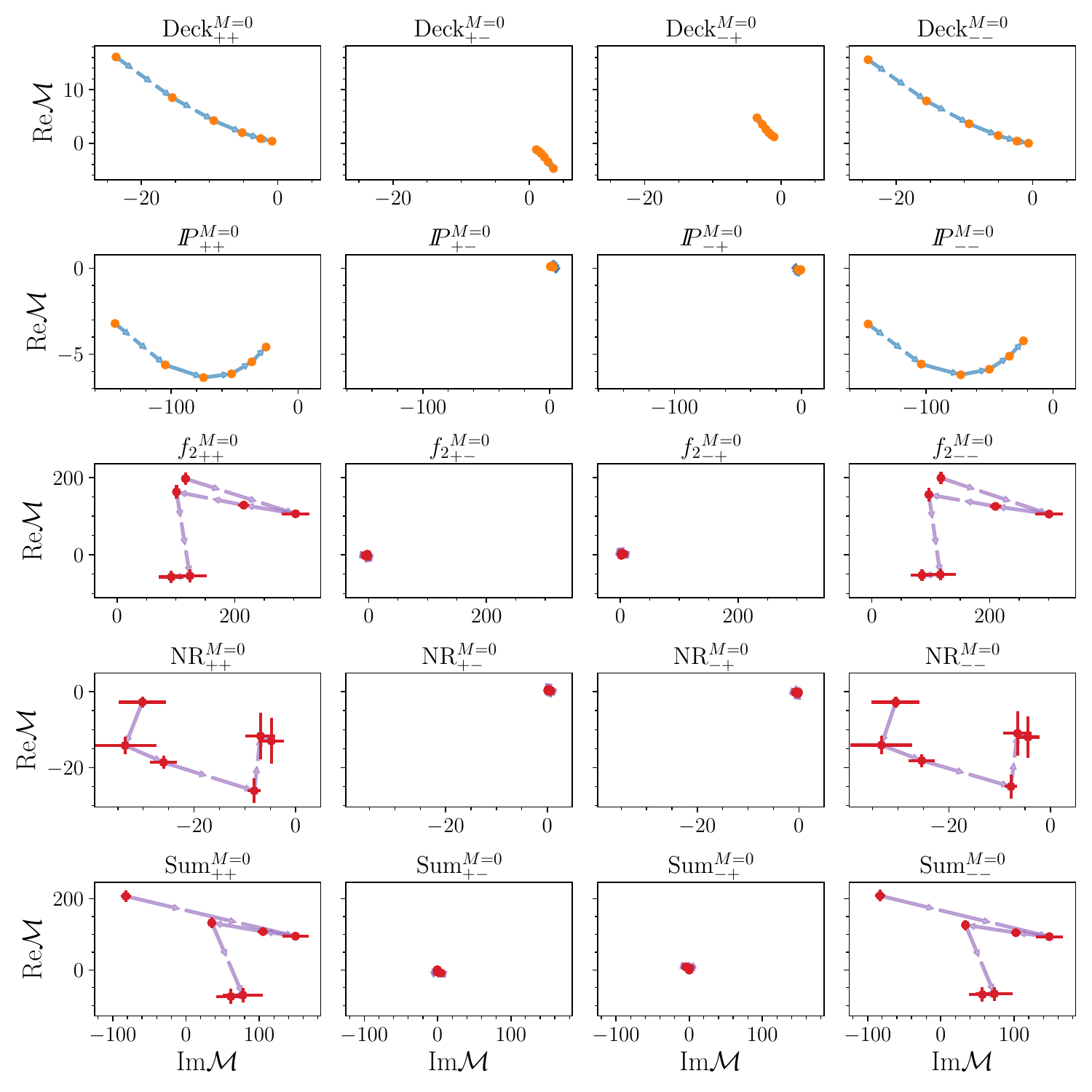}}
    \subfloat[\label{subfig:argandminus}]{
    \includegraphics[scale=0.35]{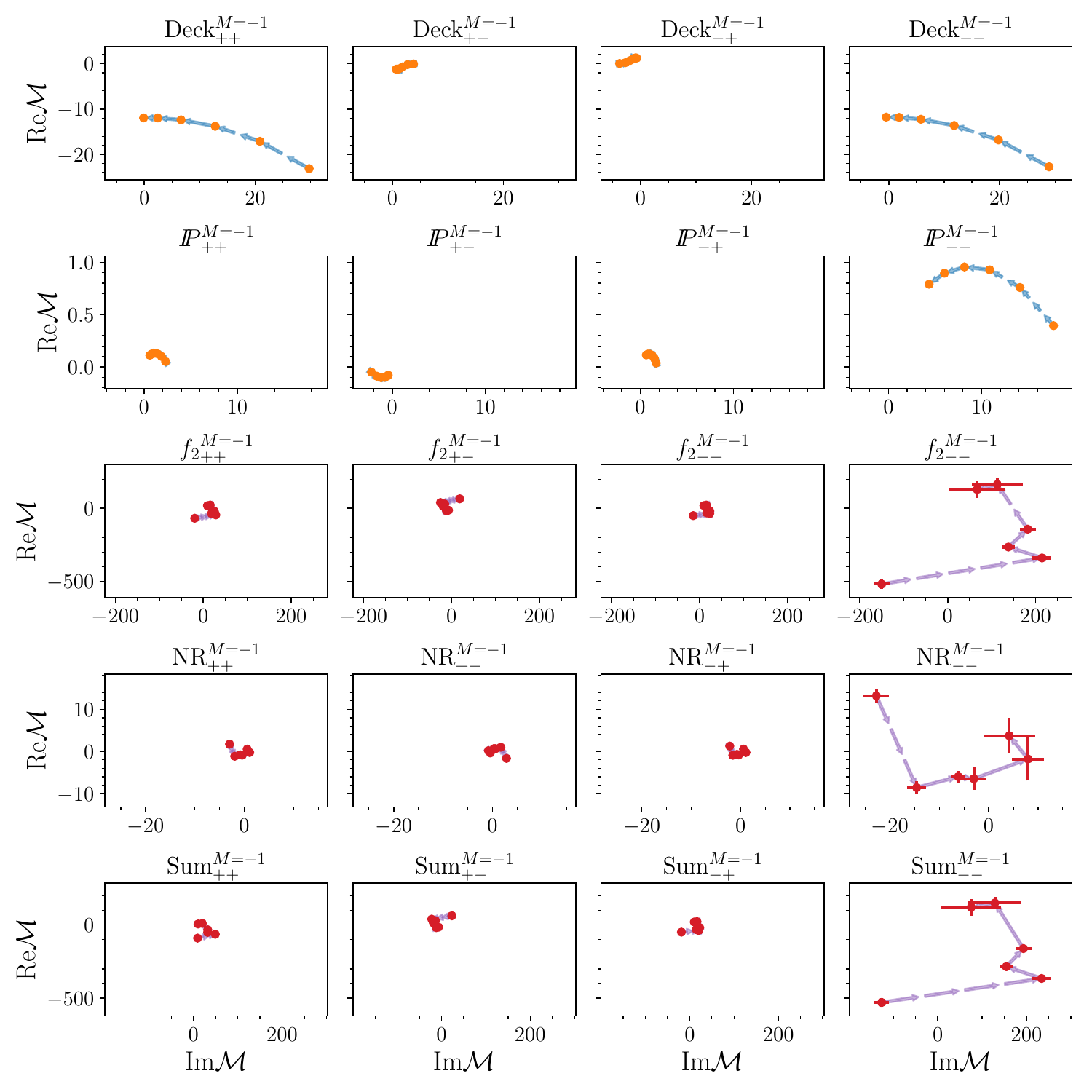}}

    \caption{Strengths of resonant and nonresonant $P$-waves for $M=+1,0,-1$ helicities for $|t|\in\{0.45~\si{GeV^2},0.65~\si{GeV^2},0.95~\si{GeV^2}\}$ (a). Real and imaginary parts of $P_{+1}$ (b), $P_0$ (c) and $P_{-1}$ (d) partial waves for all nucleon helicities. Arrows point in the direction of increasing $-t$. Orange data indicates that no parameters in this production mechanism were fit, while red data with statistical uncertainties implies that these production mechanisms were fit to experimental data.
    }
    \label{fig:ReImAmpP-wave}
\end{figure*}

Pomeron exchange provides a natural explanation for the empirically observed phenomenon of SCHC in the forward limit, since in Regge models helicity-flip amplitudes are suppressed by factors of $\sqrt{-t}$ relative to the leading helicity-conserving amplitude. However, as the momentum transfer increases, this mechanism predicts increasing contributions from these helicity-flip amplitudes, ultimately leading to an inversion of the hierarchy observed at small momentum transfers. It is important to note, however, that this prediction arises from extrapolating Regge amplitudes which are valid in the forward limit to larger $|t|$-values. It is unclear how appropriate this extrapolation is. However, there are few theoretical arguments that allow one to predict the dominant production mechanisms for particular resonances in the $\sqrt{s_{12}}$ and momentum transfer regions of interest.

The model developed here is more flexible than a Regge model, and thus it is interesting to examine to what extent the model supports the presence of SCHC in the experimental data with $|t|>0.4~\si{GeV^2}$. Evidence for SCHC may be obtained by studying the SDMEs, but this is beyond the scope of this work. Instead, the validity of SCHC can be inferred from a direct examination of the partial-wave amplitudes. In particular, SCHC implies the hierarchy of the $P$-waves,
\begin{equation}
|P_{+1}|\gg |P_0|, |P_{-1}|,
\end{equation}
in the vicinity of the $\rho(770)$ mass. Thus by examining the relative strengths of the three $P$-wave amplitudes in the model, evidence for or against SCHC can be obtained. Partial-wave projection is performed according to Eq.~\eqref{eq:partial_waves}. The real and imaginary parts of the $P$-wave amplitudes, as well as the magnitude of the dominant nucleon helicity-nonflip component, are shown in~\cref{fig:ReImAmpP-wave}. By considering only amplitudes that contribute to resonant $\rho(770)$ production, it is possible to examine the extent to which the helicity of the photon is conserved.~\cref{fig:ReImAmpP-wave}(d) shows that, even though the $P$-wave is dominated by resonant production of the $\rho(770)$, the hierarchy predicted by SCHC is not obeyed. In particular, it is clear that the $P_{0}$, $P_{-1}$ provide non-negligible contributions to the total $P$-wave intensity. 

It is possible to decompose these $P$-wave partial waves further into individual production mechanisms. Figs.~\ref{fig:ReImAmpP-wave}(b), (c) and (d) show the $t$-dependence of the real and imaginary parts of the $P$-wave components at a two-pion invariant mass of $\sqrt{s_{12}}=m_\rho$ for $P_{+1}$, $P_{0}$, and $P_{-1}$, respectively. As explained above, the model is fit to the low angular moments at fixed $t$. As a result, the free parameters acquire an effective $t$-dependence. It is known that no dynamical singularities can occur for $t<0$. Thus it is expected that the coupling dependence on $t$ should be analytical. However, no constraints were placed on the model to ensure that this occurred. 

The $t$-dependence of the $P$-wave amplitudes shown in Figs.~\ref{fig:ReImAmpP-wave}(b), (c) and (d) are either continuous, like for Deck mechanism and Pomeron exchange, or computed at selected $t$-values where fits were performed (connecting arrows are added to indicate the direction of $t$-variable evolution). Due to partial wave projection of the $\pi^+\pi^-$ system, the helicity structure of the upper vertex is encoded in the $t$-dependence of moments, and thus is self-analysing. Contrary to that, the information on the $t$-dependence of the lower vertex helicity couplings are lost in the experiment under analysis, and therefore could not be fitted. For the sake of simplicity, we assumed that the Lorentz structure of lower vertex is that of a vector particle exchange. This assumption entails the suppression of the helicity-flip amplitudes which is clearly visible in the plots. Otherwise, the $t$-dependence is rather smooth, albeit with nontrivial phase motion. The association of this phase dependence with particular Regge exchanges is left for a further work.

\subsection{Comparison of Model to Higher Moments}
\label{sec:higher_moments}
\begin{figure*}
\centering
\includegraphics[scale=0.45]{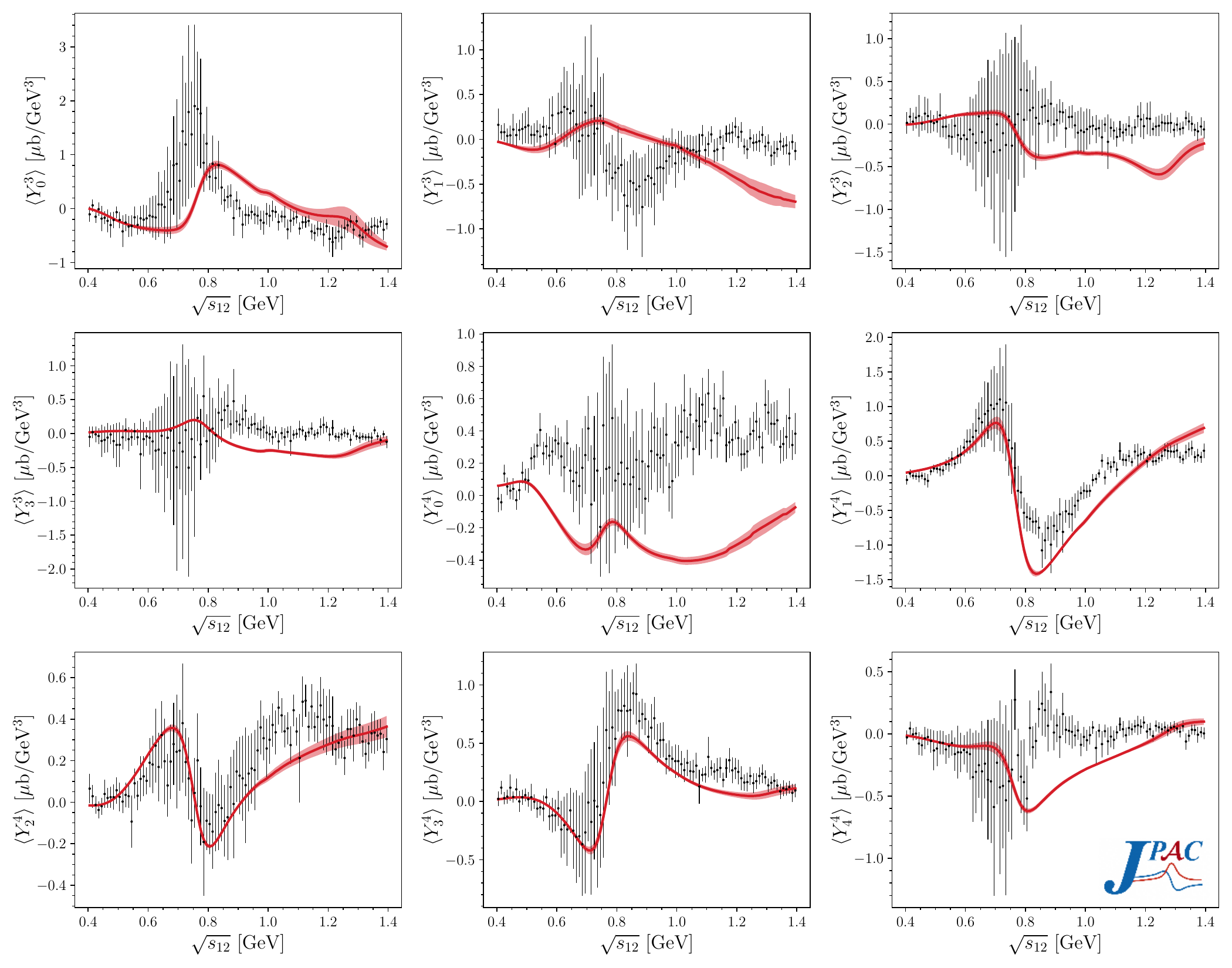}
\caption{Prediction of angular moments $\braket{Y^L_M}$ for $L=3,4$, $M=0,\dots, L$ and $t=-0.45~\si{GeV^2}$. The model curves are superimposed and not fitted to these data. Experimental data is taken from CLAS~\cite{CLAS:2009ngd}}
\label{fig:higher_moments_t_0.45}
\end{figure*}

While the model was fit to the lowest six moments ($L=0,1,2$, $M=0,\dots, L$), Ref.~\cite{CLAS:2009ngd} includes data for all moments up to $L=4$. As a result, it is possible to compare the model with these higher moments. Since these moments are sensitive to contributions from higher partial waves, it is to be expected that deficiencies may appear. However, if the main production mechanisms have been identified correctly, then reasonable descriptions of these higher moments should be obtained if no additional resonances with higher-angular moment are present. Results for these predictions of the model are shown in~\cref{fig:higher_moments_t_0.45} for $t=-0.45~\si{GeV^2}$. Comparisons between the model and higher moments for other $t$-bins are shown in~\cref{fig:higher_moment_t_0.55,fig:higher_moment_t_0.65,fig:higher_moment_t_0.75,fig:higher_moment_t_0.85,fig:higher_moment_t_0.95} in~\cref{app:comparisons_to_higher_moments}. While the description of these moments is not as good as for the lower angular moments, the overall magnitude and most characteristic patterns present in the data are in fair agreement with model predictions. Consider in particular the angular moments $\braket{Y^4_1}$, $\braket{Y^4_2}$ and $\braket{Y^4_3}$. From~\cref{app:comparisons_to_higher_moments}, it is possible to see that these moments contain $P$-$F$-wave interference. This interference produces the structures observed around the $\rho(770)$ mass. It is important to note that no resonant $F$-wave component is included in this model, because the lightest spin-3 $\pi\pi$ resonance peaks at 1.69 GeV, and was thus not considered in this study. Thus this structure is produced in the model due to the interference of the $P$-wave with the Deck amplitude, which projects onto all partial waves. The agreement here is suggestive that the relative magnitudes and phases of the respective partial waves are well constrained in the region of the $\rho(770)$. In this context it is important to stress the observation made in~\cite{Bibrzycki:2018pgu} that, for the Deck mechanism, the even partial waves are suppressed as compared to the odd ones. Thus, the moments where the interference with nonresonant even waves is essential (e.g. $\braket{Y_3^1}$) may be described inadequately by Deck amplitudes alone. While some of these angular moments appear to lack features present in the data, it is important to note that they are at least one order of magnitude smaller than the lower ones, and thus the remaining discrepancies are small in absolute terms.

\section{Conclusion}
\label{sec:conclusion}
Models of two-pion photoproduction which contain both a Pomeron-induced $\rho(770)$ resonant amplitude and a nonresonant $P$-wave amplitude can explain a number of features observed in two-pion photoproduction measurements at momentum transfers in the range \mbox{$|t|<0.4~\si{GeV^2}$.} In particular, such models can explain the $\rho(770)$
lineshape, SCHC and the $t$-dependence of the $\rho(770)$ photoproduction cross section. However, at larger momentum transfers, this model ceases to saturate the experimental data, and a more sophisticated model is required. In this work, a theoretical description of two-pion photoproduction is developed. The model incorporates the leading resonances which contribute to two-pion production, as well as the expected leading background from the Deck mechanism. The model contains a number of free parameters which are fit to data for the low angular moments. The resulting model provides a good global description of the low angular moments. By studying decompositions of the model, an understanding of the physical mechanisms which contribute to the $\rho(770)$
lineshape, SCHC and the $t$-dependence of the $\rho(770)$ photoproduction cross section can be identified. Rather than originating from a single production mechanism, data in this kinematic region can be understood as being due to several competing mechanisms. 

Further confidence is gained from a comparison of the fitted model to the higher angular moments. While the agreement with these moments is not as good as for the fitted ones, it is still reasonable. Angular moments with expected large $P$-wave components are well described by the model, while other angular moments likely suffer from the presence of more competing mechanisms which are not fully captured by the model.

This paper is primarily concerned with the theoretical development and fitting of the data for two-pion photoproduction. However, there exist several immediate applications of this model, which are left for further work. In particular, it can be applied to compute spin-density matrix elements for both two-pion resonances like the $\rho(770)$, as well as baryon resonances like the $\Delta$. Based on the $t$-dependence of the amplitudes, one can attempt to construct a microscopically motivated model of the resonance production not only in the $P$-wave but also in the $S$- and $D$-waves. In addition, it is interesting to examine whether such a model --suitably generalized-- would enable a prediction of two-pion electroproduction.

\section*{Acknowledgements}
VM and RJP have been supported by the projects \mbox{CEX2019-000918-M} (Unidad de Excelencia ``María de Maeztu''), \mbox{PID2020-118758GB-I00}, financed by \mbox{MICIU/AEI/10.13039/501100011033/} and FEDER, UE, as well as by the EU \mbox{STRONG-2020} project, under the program \mbox{H2020-INFRAIA-2018-1} Grant Agreement \mbox{No.~824093}. 
VM is a Serra H\'unter fellow and acknowledges support from \mbox{CNS2022-136085}. 
We gratefully acknowledge Polish high-performance computing infrastructure PLGrid (HPC Centers: ACK Cyfronet AGH) for providing computer facilities and support within computational grant No. \mbox{PLG/2023/016601}.
NH was supported by Polish research project Grant No. \mbox{2018/29/B/ST2/02576} (National Science Center).
CFR is supported by Spanish Ministerio de Ciencia, Innovación y Universidades \mbox{(MICIU)} under Grant \mbox{No.~BG20/00133}.
AA is supported by the U.S. National Science Foundation under Grant No. \mbox{PHY-2209183}.
This research was supported in part by Lilly Endowment, Inc., through its support for the Indiana University Pervasive Technology Institute.
This work was supported by the U.S. Department of Energy contract DE-AC05-06OR23177, under which Jefferson Science Associates, LLC operates Jefferson Lab, and DE-FG02-87ER40365, and it contributes to the aims of the U.S. Department of Energy ExoHad Topical Collaboration, contract DE-SC0023598.
DW is supported in part by DFG and NSFC through funds provided to the Sino-German CRC 110 ``Symmetries
and the Emergence of Structure in QCD'' (NSFC Grant No. 12070131001, DFG Project-ID 196253076).
\appendix
\section{Kinematics}
\label{app:kinematics}
Explicit definitions for the four-vectors of the particles in the $\pi^+\pi^-$ helicity rest frame are given here. All quantities may be described in terms of three invariants $s$, $t$ and $s_{12}$ and two angles $\Omega_\text{H}=(\theta_\text{H},\phi_\text{H})$ of the $\pi^+$ in this frame. The energies in this frame are
\begin{align}
E_{\mathbf{p}_{1}^\text{H}}&=\frac{s-m_N^2+t}{2\sqrt{s_{12}}},
\\
E_{\mathbf{q}^\text{H}}&=\frac{s_{12}-t}{2\sqrt{s_{12}}},
\\
E_{\mathbf{p}_{2}^\text{H}}&=\frac{s-s_{12}-m_N^2}{2\sqrt{s_{12}}},
\\
E_{\mathbf{k}_{1}^\text{H}}&=\frac{1}{2}\sqrt{s_{12}},
\\
E_{\mathbf{k}_{2}^\text{H}}&=\frac{1}{2}\sqrt{s_{12}}.
\end{align}
The direction of the recoiling proton ($p_2$) defines the negative $z$-axis. The production plane lies in the $x$-$z$ plane. These choices lead to the following set of three-vectors:
\begin{align}
\mathbf{p}_1^\text{H}&=|\mathbf{p}_1^\text{H}|(\sin\theta_1,0,\cos\theta_1),
\\
\mathbf{q}^\text{H}&=|\mathbf{q}^\text{H}|(-\sin\theta_q,0,\cos\theta_q),
\\
\mathbf{p}_2^\text{H}&=|\mathbf{p}_2^\text{H}|(0,0,-1),
\\
\mathbf{k}_1^\text{H}&=|\mathbf{k}_1^\text{H}|(\sin\theta^\text{H}\cos\phi^\text{H},\sin\theta^\text{H}\sin\phi^\text{H},\cos\theta^\text{H}),
\\
\mathbf{k}_2^\text{H}&=-|\mathbf{k}_2^\text{H}|(\sin\theta^\text{H}\cos\phi^\text{H},\sin\theta^\text{H}\sin\phi^\text{H},\cos\theta^\text{H}),
\end{align}
where the angles $\theta_1$ and $\theta_q$ are defined for convenience. They are given by
\begin{align}
\cos\theta_1&=\frac{2m_N^2-2E_{\mathbf{p}_1^\text{H}}E_{\mathbf{p}_2^\text{H}}-t}{2|\mathbf{p}_1^\text{H}||\mathbf{p}_2^\text{H}|},
\\
\cos\theta_q&=\frac{s+t-s_{12}-m_N^2-2E_{\mathbf{p}_1^\text{H}}E_{\mathbf{q}^\text{H}}}{2|\mathbf{p}_1^\text{H}||\mathbf{q}^\text{H}|}.
\end{align}

\section{$\pi N$ Amplitude Parameterization}
\label{app:pin}
It is necessary to define the quasi pion-nucleon amplitude ${M}_{\lambda_1\lambda_2}^\pm(s_i,t,u_j)$, since it depends on the pion virtuality, $u_j$ and is thus unphysical. Without loss of generality, this amplitude may be decomposed as
\begin{equation}
\begin{split}
{M}_{\lambda_1\lambda_2}^\pm(s_i,t,u_j)&=\overline{u}({p}_2,\lambda_2)\bigg[\tilde{A}^\pm(s_i,t,u_j)
\\
&+\frac{1}{2}(\slashed{{K}}_i+\slashed{{k}}_j)\tilde{B}^\pm(s_i,t,u_j)\bigg]u({p}_1,\lambda_1),
\end{split}
\end{equation}
where $K_i=q-k_i$ and $\tilde{A}^\pm(s_i,t,u_j)$ and $\tilde{B}^\pm(s_i,t,u_j)$ are scalar functions of the relevant Lorentz invariants.

In the limit that $u_j\to m_\pi^2$, this amplitude reduces to elastic $\pi N $ scattering. In order to propose a suitable form for the scalar amplitudes, $\tilde{A}^\pm$ and $\tilde{B}^\pm$, it is useful to recall some facts about the on-shell process. The scattering amplitude admits the standard decomposition~\cite{Chew:1957zz}:
\begin{equation}
\begin{split}
\mathcal{M}_{\lambda_1\lambda_2}^\pm(s,t)&=\overline{u}({p}_2,\lambda_2)\bigg[A^\pm(s,t)
\\
&+\frac{1}{2}(\slashed{k}_1+\slashed{k}_2)B^\pm(s,t)\bigg]u({p}_1,\lambda_1),
\end{split}
\end{equation}
where $s=(q_a+q_b)^2=(q_1+q_2)^2$, $t=(q_a-q_1)^2=(q_b-q_2)^2$ and $2m_N^2+2m_\pi^2=s+t+u$. In the $s$-channel CM frame, it is possible to show that the energies are
\begin{align}
E_a&=\frac{s-m_N^2+m_\pi^2}{2\sqrt{s}},
\\
E_b&=\frac{s+m_N^2-m_\pi^2}{2\sqrt{s}},
\\
E_1&=\frac{s-m_N^2+m_\pi^2}{2\sqrt{s}},
\\
E_2&=\frac{s+m_N^2-m_\pi^2}{2\sqrt{s}},
\end{align}
and the corresponding three-momenta are
\begin{equation}
|\mathbf{q}_a|=\frac{\lambda^{1/2}(s,m_N^2,m_\pi^2)}{2\sqrt{s}}=|\mathbf{q}_b|\equiv |\mathbf{q}|.
\end{equation}
It is possible to express the cosine of the scattering angle, $ z_s = \cos\theta_s$, in terms of $s$ and $t$ via
\begin{equation}
z_s=\frac{t-2m_N^2+2E_bE_2}{2q_b q_2}.
\end{equation}
The amplitudes themselves may be written in terms of reduced helicity amplitudes,
\begin{align}
\frac{1}{4\pi}\tilde{A}^\pm(s,t)&=\frac{\sqrt{s}+m_N}{Z_1^+ Z_2^+}f_1^\pm(s,z_{s})-\frac{\sqrt{s}-m_N}{Z_1^- Z_2^-}f_2^\pm(s,z_{s}),
\\
\frac{1}{4\pi}\tilde{B}^\pm(s,t)&=\frac{1}{Z_1^+ Z_2^+}f_1^\pm(s,z_{s})-\frac{1}{Z_1^- Z_2^-}f_2^\pm(s,z_{s}),
\end{align}
where $Z_i^\pm=\sqrt{E_i \pm m_N}$. These reduced helicity amplitudes may be partial-wave expanded as
\begin{align}
f_1^\pm(s_i,z_{s})&=\frac{1}{\sqrt{q_b q_2}}\sum_{l=0}^\infty f_{l+}^\pm(s) P_{l+1}^\p(z_{s}) \nonumber
\\
&-\frac{1}{\sqrt{q_b q_2}}\sum_{l=2}^\infty f_{l-}^\pm(s) P_{l-1}^\p(z_{s}),
\\
f_2^\pm(s,z_{s})&=\frac{1}{\sqrt{q_b q_2}}\sum_{l=1}^\infty \bigg[f_{l-}^\pm(s)-f_{l+}^\pm(s)\bigg] P_{l}^\p(z_{s}),
\end{align}
where $f_{l}^\pm(s)$ are the $\pi N$ partial waves. At low energies, it suffices to truncate the partial wave expansion at some $l_\text{max}$, while at large $s$, all partial waves contribute, and a description in terms of crossed-channel Reggeon exchanges is more appropriate. 

In this work, the initial pion is off-shell. Since such an amplitude is unphysical, the off-shell extension of this model is defined by a model-dependent prescription. The virtuality of the pion enters through the modified kinematics. In the unphysical CM frame for the virtual pion-nucleon system, denoted by the asterisk ($^*$), the energies are
\begin{align}
E_a^*&=\frac{s_i-m_N^2+u_j}{2\sqrt{s_i}},
\\
E_b^*&=\frac{s_i+m_N^2-u_j}{2\sqrt{s_i}},
\\
E_1^*&=\frac{s_i-m_N^2+m_\pi^2}{2\sqrt{s_i}},
\\
E_2^*&=\frac{s_i+m_N^2-m_\pi^2}{2\sqrt{s_i}},
\end{align}
and the corresponding magnitudes of the three-momenta of the incident pion and nucleon are
\begin{align}
q_a^*=\frac{\lambda^{1/2}(s_i,m_N^2,u_j)}{2\sqrt{s_i}}=q_b^*,
\\
q_1^*=\frac{\lambda^{1/2}(s_i,m_N^2,m_\pi^2)}{2\sqrt{s_i}}=q_2^*.
\end{align}
Note the presence of $u_j$ in the energies and magnitudes of three-momenta in the incoming hadrons.
The resulting cosine of the scattering angle is
\begin{equation}
z_{s_i}^*=\frac{t-2m_N^2+2E_b^*E_2^*}{2q_b^* q_2^*}.
\end{equation}
Using these kinematics, the scalar amplitudes $\tilde{A}^\pm(s_i,t,u_j)$ and $\tilde{B}^\pm(s_i,t,u_j)$ are \textit{defined} as
\begin{align}
\frac{1}{4\pi}\tilde{A}^\pm(s_i,t,u_j)&\equiv\frac{\sqrt{s}+m_N}{Z_1^+ Z_2^+}f_1^\pm(s_i,z_{s_i}^*) \nonumber
\\
&-\frac{\sqrt{s}-m_N}{Z_1^- Z_2^-}f_2^\pm(s_i,z_{s_i}^*),
\\
\frac{1}{4\pi}\tilde{B}^\pm(s_i,t,u_j)&\equiv\frac{1}{Z_1^+ Z_2^+}f_1^\pm(s_i,z_{s_i}^*) \nonumber \\
&-\frac{1}{Z_1^- Z_2^-}f_2^\pm(s_i,z_{s_i}^*),
\end{align}
where $Z_i^\pm=\sqrt{E_i^* \pm m_N}$. These reduced helicity amplitudes may be partial-wave expanded as
\begin{align}
f_1^\pm(s_i,z_{s_i}^*)&=\frac{1}{\sqrt{q_b^* q_2^*}}\sum_{l=0}^\infty f_{l+}^\pm(s_i) P_{l+1}^\p(z_{s_i}^*) \nonumber
\\
&-\frac{1}{\sqrt{q_b^* q_2^*}}\sum_{l=2}^\infty f_{l-}^\pm(s_i) P_{l-1}^\p(z_{s_i}^*),
\\
f_2^\pm(s_i,z_{s_i}^*)&=\frac{1}{\sqrt{q_b^* q_2^*}}\sum_{l=1}^\infty \bigg[f_{l-}^\pm(s_i)-f_{l+}^\pm(s_i)\bigg] P_{l}^\p(z_{s_i}^*).
\end{align}
In this work, the pion-nucleon amplitudes from Ref.~\cite{Mathieu:2018xyc} are employed. These amplitudes interpolate between the SAID parameterization~\cite{Arndt:1994bu,Arndt:1995bj,Arndt:2003if} at low-energies and a Regge parameterization at high-energies using rigorous constraints from finite-energy sum rules~\cite{Dolen:1967zz}. 

\onecolumngrid

\newpage
\section{Angular Moments in terms of Partial Waves}
\label{app:angular_moments_partial_waves}
This appendix explores the complex correlation between angular moments and partial waves, as outlined by Eqs.~\eqref{angular-moments} and \eqref{ang.mom} in~\cref{sec:kinematics}. These moments are expressed as a summation over partial waves, each characterized by spectroscopic notation truncated to $L = 4$ (G-wave). The adoption of spectroscopic notation, wherein $l$ is substituted by $S$, $P$, $D$, etc., facilitates a clearer representation of the partial waves. In the following expressions, the prefactor \(\kappa\), defined in Eq.~\eqref{kappa}, includes a \(1/2^2\) factor to accommodate the averaging over the spins of the photon and nucleon. Consequently, an additional factor of 2 was introduced to eliminate the sum over \(\lambda_\gamma\).
{\allowdisplaybreaks
\begin{align}
\braket{Y_0^0}&=2\kappa \sum_{\lambda_1\lambda_2}\Bigg(\left|S\right|^2+\left|P_{-1}\right|^2+\left|P_0\right|^2+\left|P_{1}\right|^2+\left|D_{-2}\right|^2+\left|D_{-1}\right|^2+\left|D_{0}\right|^2+\left|D_{1}\right|^2+\left|D_{2}\right|^2+\left|F_{-3}\right|^2+\left|F_{-2}\right|^2+\left|F_{-1}\right|^2\\&+\left|F_{0}\right|^2+\left|F_{1}\right|^2+\left|F_{2}\right|^2+\left|F_{3}\right|^2+\left|G_{-4}\right|^2+\left|G_{-3}\right|^2+\left|G_{-2}\right|^2+\left|G_{-1}\right|^2+\left|G_{0}\right|^2+\left|G_{1}\right|^2+\left|G_{2}\right|^2+\left|G_{3}\right|^2+\left|G_{4}\right|^2\Bigg),\nonumber
\end{align}
\begin{align}
\braket{Y_0^1}&=2\kappa\sum_{\lambda_1\lambda_2}\Re\Bigg(2\big(S^*P_0\big)+\frac{2}{\sqrt{5}}\Big(\sqrt{3}\big(P_{-1}^*D_{-1}+P_{1}^*D_{1}\big)+2\big(P_0^*D_{0}\big)\Big)+2\sqrt{\frac{3}{7}}\big(D_{-2}^*F_{-2}+D_{2}^*F_{2}\big)+ 2\sqrt{\frac{3}{35}}\Big(2\sqrt{2}\big(D_{-1}^*F_{-1}\\&+D_{1}^*F_{1}\big)+3D_{0}^*F_0\Big) +\frac{2}{\sqrt{3}}\big(F_{-3}^*G_{-3}+F_{3}^*G_{3} \big)+\frac{4}{\sqrt{7}}\big(F_{-2}^*G_{-2}+F_{2}^*G_{2}\big)+2\sqrt{\frac{5}{7}}\big(F_{-1}^*G_{-1}+ F_{1}^*G_{1}\big)+\frac{8}{\sqrt{21}}\big(F_{0}^*G_0\big)\Bigg),\nonumber
\end{align}
\begin{align}
\braket{Y_1^1}&=2\kappa\sum_{\lambda_1\lambda_2}\Re\Bigg(-S^*P_{-1}+S^*P_{1}+\frac{1}{\sqrt{5}}\big(P_{-1}^*D_{0}-P_{1}^*D_{0}\big)+\sqrt{\frac{6}{5}}\big(P_{1}^*D_{2} - P_{-1}^*D_{-2}\big)-\sqrt{\frac{3}{5}}\big(P_{0}^*D_{-1}+ P_{0}^*D_{1}+D_{0}^*F_{1}\big)\\&+\sqrt{\frac{3}{35}}\big(D_{-2}^*F_{-1}-D_{2}^*F_{1}+\sqrt{3}D_{-1}^*F_{0}\big)+\sqrt{\frac{6}{7}}\big(D_{1}^*F_{2}-D_{-1}^*F_{-2}\big)-\sqrt{\frac{2}{35}}\Big(2\big(D_{0}^*F_{-1}\big)+ 3\big(D_{1}^*F_{0}\big)\Big)+\frac{3\sqrt{7}}{70}\big(D_{2}^*F_{3}\big)\nonumber\\&+\frac{1}{\sqrt{21}}\big(F_{-3}^*G_{-2}-F_{3}^*G_{2}\big)+\frac{1}{\sqrt{7}}\Big(F_{-2}^*G_{-1}+F_{2}^*G_{1}-3\big(D_{-2}^*F_{-3}\big)\Big)+\sqrt{\frac{5}{7}}\big(F_{1}^*G_{2}- F_{-1}^*G_{-2}\big)+\sqrt{\frac{2}{7}}\big(F_{-1}^*G_{0}-F_{1}^*G_{0}\big)\nonumber\\&+\sqrt{\frac{10}{21}}\big(-F_{0}^*G_{-1} +F_{0}^*G_{1}\big) -\frac{12\sqrt{3}}{7}\big(F_{-3}^*G_{-4}\big)- F_{-2}^*G_{-3} +F_{2}^*G_{3}+\frac{2}{\sqrt{3}}\big(F_{3}^*G_{4}\big)\Bigg),\nonumber
\end{align}
\begin{align}
\label{eq:ylm20}
\braket{Y_0^2} = 2\kappa & \sum_{\lambda_1\lambda_2} \Bigg(\sqrt{\frac{1}{5}}\big(
-\left|P_{-1}\right|^2 + 2\left|P_{0}\right|^2- \left|P_{1}\right|^2+\left|F_{-1}\right|^2+\frac{4\left|F_{0}\right|^2}{3}+\left|F_{1}\right|^2\big)
+ \frac{1}{7}\Big(2\sqrt{5}\big(\left|D_{0}\right|^2-\left|D_{-2}\right|^2-\left|D_{2}\right|^2 \big)\\& +\sqrt{5}\big(\left|D_{-1}\right|^2 +\left|D_{1}\right|^2 \big)\Big)- \frac{\sqrt{5}}{3}\big(\left|F_{-3}\right|^2+ \left|F_{3}\right|^2 \big)- \frac{1\sqrt{5}}{11}\Big(4\big(\left|G_{-4}\right|^2 +\left|G_{4}\right|^2 \big) - \left|G_{-3}\right|^2-\left|G_{3}\right|^2\Big) \nonumber\\&  \frac{\sqrt{5}}{77}\Big(8\big(\left|G_{-2}\right|^2 +\left|G_{2}\right|^2\big) + 17\big(\left|G_{-1}\right|^2 +\left|G_{1} \right|^2\big) + 20\left|G_{0}\right|^2\Big)  +\Re\Big(2\big(S^*D_{0}\big)+ 6\sqrt{\frac{1}{35}}\Big(\sqrt{2}\big(P_{-1}^*F_{-1}+P_{1}^*F_{1}\big) \nonumber\\& + \sqrt{3}\big(P_{0}^*F_{0}\Big) + \frac{2}{7}\sqrt{15}\big(D_{-2}^*G_{-2}+D_{2}^*G_{2}\big)+ \frac{2}{7}\sqrt{30}\big (D_{-1}^*G_{-1}+ D_{1}^*G_{1}\big)+ \frac{12}{7}\big(D_{0}^*G_{0}\big) \Bigg),\nonumber
\end{align}
\begin{align}
 \braket{Y_1^2} &=2\kappa \sum_{\lambda_1\lambda_2}\Re\Bigg(  -S^* D_{-1}+S^* D_{1}- \sqrt{\frac{3}{5}}\big(P_{-1}^*P_{0}+P_{1}^* P_{0}\big)  +\sqrt{\frac{6}{7}}\big(P_{1}^* F_{2}- P_{-1}^* F_{-2}\big)+ \frac{3}{\sqrt{35}}\big( P_{-1}^*F_{0}-P_{1}^* F_{0}\big)- \frac{4\sqrt{3}}{\sqrt{70}}\big(P_{0}^* F_{-1}\big)\\&+ \frac{4\sqrt{210}}{77}\big(P_{0}^*F_{1}\big) - \frac{\sqrt{30}}{7}\big( D_{-2}^* D_{-1}+D_{0}^* G_{-1}-D_{2}^* D_{1}-D_{0}^*G_{1}\big)  - \frac{\sqrt{35}}{7}\big( D_{-2}^* G_{-3}\big) + \frac{\sqrt{5}}{7}\big(D_{-2}^*G_{-1}-D_{2}^* G_{1}
 +D_{0}^* D_{1} \nonumber\\&- D_{0}^* D_{-1}\big)+ \frac{2\sqrt{10}}{7}\big(D_{1}^*G_{2} -D_{-1}^* G_{-2}\big)  +\frac{4}{7} \big( D_{-1}^*G_{0} -D_{1}^*G_{0}\big) - \frac{1}{3}\sqrt{\frac{5}{11}}\big( F_{-2}^* F_{-3}\big) + \frac{1}{\sqrt{3}}\big(F_{2}^* F_{1}-F_{-1}^* F_{-2}\big) \nonumber\\&+\frac{\sqrt{10}}{15}\big(F_{1}^* F_{0}-F_{0}^* F_{-1}\big) + \frac{\sqrt{5}}{3}\big(F_{3}^* F_{2}\big)+\frac{2 \sqrt{15}}{11}\big(G_{4}^* G_{3} -G_{-3}^* G_{-4}\big)  +\frac{\sqrt{105}}{77} \big(5G_{3}^* G_{2}-2G_{-2}^* G_{-3}\big) \nonumber\\& -\frac{9\sqrt{15}}{77}\big(G_{-1}^* G_{-2}-G_{2}^* G_{1}\big) - \frac{5\sqrt{6}}{77} \big(G_{0}^* G_{-1}- G_{1}^* G_{0}\big) + \sqrt{\frac{5}{7}} \big(G_{3}^* D_{2}\big)\Bigg),\nonumber
\end{align}
\begin{align}
 \braket{Y_2^2} &=2\kappa \sum_{\lambda_1\lambda_2}\Re\Bigg( S^* D_{2} + S^* D_{-2}+ \frac{3}{\sqrt{7}} \big(P_{-1}^* F_{-3}+F_{3}^* P_1\big)  - \sqrt{\frac{6}{5}}\big(P_{-1}^*P_1\big) + \frac{\sqrt{105}}{35} \big(F_{1}^* P_{-1}+P_1^* F_{-1}\big) + \sqrt{\frac{3}{7}}\big(P_0^* F_{-2} + F_{2}^* P_0\big) \\& + \sqrt{\frac{10}{7}} \big( D_{-2}^*G_{-4}+G_{4}^* D_{2}\big) + \sqrt{\frac{5}{7}} \big(D_{-1}^* G_{-3}\big)+ \frac{\sqrt{5}}{7}\Big(D_{1}^* G_{-1}+G_0^* D_{2}+G_{1}^* D_{-1}-2\big(D_{0}^* D_{-2}+D_{0}^* D_{2}\big)\Big) + \frac{1}{7}\big(G_0^* D_{-2}\big)\nonumber\\&- \frac{\sqrt{30}}{7}\big(D_{1}^* D_{-1}\big) + \frac{\sqrt{15}}{7}\big(D_{0}^* G_{-2}+ G_{2}^* D_{0}\big)+ \frac{\sqrt{35}}{7}\big(G_{3}^* D_{1}\big)- \frac{\sqrt{2}}{3} \big(F_{-1}^* F_{-3}+F_{3}^* F_{1}\big)- \frac{2}{3}\big(F_{0}^* F_{-2}+F_{2}^* F_{0}\big)\nonumber\\&  - \frac{\sqrt{30}}{77}\big(2F_{1}^* F_{-1}+10G_{1}^* G_{-1}\big)-\frac{\sqrt{210}}{77}\Big(2\big(G_{-2}^* G_{-4}+ G_{4}^* G_{2}\big)+ 3\big(G_{-1}^* G_{-3}+G_{3}^* G_{1}\big)\Big)- \frac{3\sqrt{3}}{7}\big(G_0^* G_{-2}\big)  - \frac{30\sqrt{3}}{77}\big(G_{2}^* G_0\big) \Bigg),\nonumber
\end{align}
\begin{align}
\braket{Y_0^3} &= 2\kappa \sum_{\lambda_1\lambda_2}\Re\Bigg(2\big(S^* F_{0}\big) -\frac{3\sqrt{35}}{7}\big(P_{-1}^* D_{-1}+P_{1}^* D_{1}\big)+ 2\sqrt{\frac{10}{21}}\big(P_{-1}^* G_{-1}+P_{1}^*G_{1}\big)+\frac{8}{\sqrt{21}}\big(P_{0}^* G_{0}\big)+\frac{3\sqrt{105}}{7}\big(P_{0}^* D_{0}\big)\\& -\frac{3}{4}\big(D_{-2}^*F_{-2}\big)+\frac{2\sqrt{10}}{15}\big(D_{-1}^*F_{-1}+D_{1}^* F_{1}\big)+\frac{8\sqrt{5}}{15}\big(D_{0}^*F_{0}\big) 
-\frac{4}{3}\Re\big(D_{2}^* F_{2}\big)
-\frac{6\sqrt{7}}{11}\big(F_{-3}^*G_{-3}+F_{3}^*G_{3}\big)+\frac{12}{11}\big(F_{0}^*G_{0}\big)\nonumber\\&-\frac{2\sqrt{3}}{11}\big(F_{-2}^*G_{-2}+F_{2}^*G_{2}\big)
+\frac{2\sqrt{15}}{11}\big(F_{-1}^*G_{-1}+F_{1}^* G_{1}\big)\Bigg),\nonumber
\end{align}
\begin{align}
 \braket{Y_1^3} &= 2\kappa \sum_{\lambda_1\lambda_2}\Re\Bigg(S^* F_{1}- S^* F_{-1}+\frac{\sqrt{105}}{35}\big(D_{-2}^* P_{-1}-D_{2}^* P_{1}\big) - \frac{3\sqrt{70}}{35}\big( D_{0}^* P_{-1}-D_{0}^* P_{1}\big) - \frac{\sqrt{35}}{7}\big( G_{-2}^* P_{-1}+G_{-1}^* P_{0}\big)\\& + \frac{\sqrt{14}}{7}\big( G_{0}^* P_{-1}-G_{0}^* P_{1}\big) - \frac{2\sqrt{210}}{35}\big(D_{-1}^* P_{0} -  D_{1}^* P_{0}\big) +\frac{\sqrt{35}}{7}\big(G_{1}^* P_{0}+  G_{2}^* P_{1}\big)  + \frac{\sqrt{2}}{3} \big(F_{-3}^* D_{-2}-F_{3}^* D_{2}\big) \nonumber\\& -\frac{2\sqrt{30}}{15}\big(D_{-2}^* F_{-1}- D_{2}^* F_{1}\big) - \frac{1}{\sqrt{3}} \big(F_{-2}^* D_{-1}-F_{2}^* D_{1}\big) - \frac{\sqrt{10}}{6}\big( F_{0}^* D_{-1}\big)  +\frac{1}{\sqrt{5}} \big(D_{0}^* F_{1} - D_{0}^* F_{-1}\big)+\frac{\sqrt{10}}{15}\big( F_{0}^* D_{1}\big)\nonumber\\&    +\frac{\sqrt{42}}{11}\big(G_{-4}^* F_{-3}- G_{4}^* F_{3}\big) -\frac{\sqrt{6}}{3} \big(G_{-2}^* F_{-3}-G_{2}^* F_{3}\big)   -\frac{\sqrt{14}}{11}\big( G_{-3}^* F_{-2}-G_{3}^* F_{2}\big)  -\frac{4\sqrt{2}}{11} \big(G_{-1}^* F_{-2}-G_{1}^* F_{2}\big)  \nonumber\\& -\frac{\sqrt{15}}{11}\big(G_{-1}^* F_{0}- G_{1}^* F_{0}\big) +\frac{1}{11}\big( G_{0}^* F_{1}-F_{-1}^*G_{0}\big)+ \frac{2\sqrt{10}}{11}\big( G_{2}^* F_{1}-G_{-2}^* F_{-1}\big) \Bigg),\nonumber
\end{align}
\begin{align}
\braket{Y_2^3} &= 2\kappa \sum_{\lambda_1\lambda_2}
\Re\Bigg(S^*F_{-2}+S^*F_2+\frac{2}{11}\Big(P_{-1}^*G_{-3}-\sqrt{\frac{15}{2}}\big(F_{-3}^*G_{-1}+F_3^*G_1\big)
+\sqrt{\frac{7}{2}}\big(F_1^*G_3+F_{-1}^*G_{-3}\big)\Big)\\&
-2\sqrt{\frac{3}{14}}\big(P_{-1}^*D_1\big)+\frac{1}{\sqrt{7}}\big(P_{-1}^*G_1+P_1^*G_{-1}\big)+\sqrt{\frac{3}{7}}\big(P_0^*D_{-2}+P_0^*D_2\big)+\frac{2}{\sqrt{7}}\big(P_0^*G_{-2}+P_0^*G_2\big)-\sqrt{\frac{1}{7}}\big(P_1^*D_{-1}\big)\nonumber\\&+\big(P_1^*G_3\big)-\frac{2}{3}\big(D_{-2}^*F_0\big)-\sqrt{\frac{1}{3}}\big(D_{-1}^*F_1+D_1^*F_{-1}\big)+\frac{\sqrt{5}}{3}\big(D_{-1}^*F_{-3}+D_1^*F_3\big)+\sqrt{\frac{35}{2}}\big(F_{-2}^*G_{-4}+\frac{2}{11}F_2^*G_4\big)\nonumber\\&-\frac{7}{11}\big(F_{-2}^*G_0+F_2^*G_0+F_2^*G_0\big)-\frac{4\sqrt{2}}{11}\big(F_{-1}^*G_1+F_1^*G_{-1}\big)-\frac{\sqrt{3}}{11}\big(F_0^*G_{-2}+ F_0^*G_2\big)\Bigg),\nonumber
\end{align}
\begin{align}
\braket{Y_3^3} &=2\kappa \sum_{\lambda_1\lambda_2}\Re \Bigg(S^* F_{3}-  S^* F_{-3}-\frac{3}{\sqrt{7}}\big(D_{2}^* P_{-1}-D_{-2}^* P_{1}\big) - \frac{2}{\sqrt{3}}\big(G_{-4}^* P_{-1}-G_{4}^* P_{1}\big)   + \frac{1}{\sqrt{21}}\big(G_{2} P_{-1}-G_{-2}^* P_{1}\big)\\&-
 \frac{1}{\sqrt{3}}\big(G_{-3}^* P_{0}- G_{3}^* P_{0}\big) - \frac{\sqrt{2}}{3}\big(F_{1}^* D_{-2}-F_{-1}^* D_{2}\big)- \frac{\sqrt{5}}{3}\big(F_{2}^* D_{-1}- D_{0}^* F_{-3} + F_{ -2}^* D_{1}+D_{0}^*  F_{3}\big)  - \frac{3}{11}\big(G_{ 0}^* F_{-3}-G_{0}^* F_{3}\big) \nonumber\\& - \frac{\sqrt{30}}{11}\big(G_{1}^* F_{-2}-G_{-1}^*F_{2}\big) +\frac{\sqrt{42}}{11}\big(G_{-4}^*  F_{-1}-G_{4}^*  F_{1}\big) -\frac{3\sqrt{6}}{11}\big(G_{2}^*  F_{-1}-G_{-2}^*  F_{1}\big) +\frac{3\sqrt{7}}{11}\big(G_{-3}^*  F_{0} - G_{3}^*  F_{0}\big) \Bigg),\nonumber
\end{align}
\begin{align}
  \braket{Y_0^4} &=2\kappa \sum_{\lambda_1\lambda_2}\Bigg( \frac{1}{7}\big( \left| D_{-2}\right|^2+ 6\left| D_{0}\right| ^2 +\left| D_{2}\right|^2\big)-\frac{52}{91}\big( \left| D_{-1}\right|^2 + \left| D_{1}\right|^2\big)+\frac{1}{11}\Big(3\big(\left| F_{-3}\right|^2+\left| F_{3}\right|^2\big) -7\big(\left| F_{-2}\right|^2+\left| F_{2}\right|^2\big)+ \left| F_{-1}\right|^2\\&+\left| F_{1}\right|^2 +6 \left| F_{0}\right|^2\Big)-\frac{27}{99} \left| G_{-2}\right|^2+\frac{243}{1001}\big(\left| G_{-1}\right|^2+\left| G_{1}\right|^2\big)  -\frac{27}{91} \left| G_{2}\right|^2 +\frac{81}{143}\big(6\left| G_{0}\right|^2-2\left| G_{3}\right|^2-\left| G_{-3}\right|^2\big)+\frac{36}{47} \left| G_{4}\right|^2 \nonumber\\& +\Re\Big(2\big(S^* G_{0}\big)-\frac{26\sqrt{14}}{91}\big(F_{-1}^*P_{-1}\big)+\frac{8}{\sqrt{21}}\big(P_{0}^* F_{0}\big)-\frac{60\sqrt{3}}{77} \big(G_{-2}^*D_{-2}+D_{2}^*G_{2}\big)+\frac{10\sqrt{6}}{77}\big(D_{-1}^* G_{-1}\big)  +\frac{40\sqrt{5}}{77}\big(D_{0}^* G_{0}\big)\Big)\Bigg),\nonumber
\end{align}
\begin{align}
\braket{Y_1^4} &=2\kappa \sum_{\lambda_1\lambda_2}\Re\Bigg(-S^* G_{-1}+S^* G_{1} +\sqrt{\frac{1}{7}}\big( P_{-1}^* F_{-2}-F_{2}^* P_{1}\big) -\frac{\sqrt{210}}{21}\big(F_{0}^* P_{-1}-P_{1}^* F_{0}\big) -\frac{\sqrt{35}}{7}\big(P_{0}^* F_{-1} -  F_{1}^* P_{0}\big)\\&
+\frac{3\sqrt{210}}{77}\big(D_{-2}^* G_{-3}-G_{3}^* D_{2}\big)-\frac{10\sqrt{30}}{77}\big(D_{-2}^* G_{-1}-D_{2}^* G_{1}\big)+\frac{\sqrt{5}}{7}\big( D_{-1}^* D_{-2}-D_{2}^* D_{1}\big) -\frac{\sqrt{30}}{7}\big(D_{0}^* D_{-1}-D_{1}^* D_{0}\big)\nonumber\\&
-\frac{5\sqrt{6}}{77}\big(G_{0}^* D_{-1}-D_{1}^* G_{0}\big)-\frac{9\sqrt{15}}{77}\big(D_{-1}^* G_{-2}-D_{1}^* G_{2}\big)
 -\frac{17\sqrt{5}}{77}\big(D_{0}^* G_{-1}- D_{0}^* G_{1}\big) + \frac{\sqrt{30}}{11}\big(F_{-2}^* F_{-3}-F_{3}^* F_{2}\big)\nonumber\\& 
-\frac{4 \sqrt{2}}{11}\big(F_{-1}^* F_{-2}-F_{2}^* F_{1}\big)-\frac{\sqrt{15}}{11}\big(F_{0}^* F_{-1}-F_{1}^* F_{0}\big) +\frac{27\sqrt{10}}{143}\big(G_{-3}^* G_{-4}-_{4}^*G_{3}\big)-\frac{9\sqrt{70}}{334}\big(G_{-2}^* G_{-3}-G_{3}^* G_{2}\big)\nonumber\\&-\frac{243}{1001}\big(G_{0}^* G_{-1}-G_{0}^* G_{1}\big)+\frac{162\sqrt{10}}{1001}\big(G_{2}^* G_{1}-G_{-2}^* G_{-1}\big)\Bigg),\nonumber
\end{align}
\begin{align}
\braket{Y_2^4} &=2\kappa \sum_{\lambda_1\lambda_2}\Re\Bigg(S^* G_{-2}+S^* G_{2}-\sqrt{\frac{1}{21}}\big(P_{-1}^* F_{-3}-F_{3}^* P_{1}\big) -\frac{\sqrt{35}}{7}\big(F_{1}^* P_{-1}+P_{1}^* F_{-1}\big) +\frac{2}{\sqrt{7}}\big(P_{0}^* F_{-2}+ F_{2}^* P_{0}\big)\\& -\frac{2\sqrt{210}}{77}\big(D_{-2}^* G_{-4}+G_{4}^* D_{2}\big) +\frac{\sqrt{15}}{7}\big(D_{0}^* D_{-2}+ D_{-2}* D_{0}\big)-\frac{30\sqrt{3}}{77} \big(G_{0}^* D_{-2}+D_{2}^* G_{0}\big)
+\frac{5\sqrt{105}}{77}\big(D_{-1}^* G_{-3}+G_{3}^* D_{1}\big) \nonumber\\& -2\sqrt{10} \big(\frac{1}{7}D_{1}^* D_{-1}+\frac{1}{11}F_{1}^* F_{-1}\big)-\frac{9\sqrt{15}}{77}\big(G_{1}^* D_{-1}+D_{1}^* G_{-1}\big)+\frac{8\sqrt{5}}{77}\big(D_{0}^* G_{-2}+ D_{0}^* G_{2}\big)+\frac{\sqrt{6}}{11}\big( F_{-1}^* F_{-3}+ F_{3}^* F_{1}\big)\nonumber\\&-\frac{\sqrt{3}}{11}\big(F_{0}^* F_{-2}+F_{2}^* F_{0}\big)+\frac{81\sqrt{70}}{1001}\big(G_{-4}^* G_{-2}+ G_{4}^* G_{2}\big)+\frac{27\sqrt{70}}{1001}\big(G_{-1}^* G_{-3}+G_{3}^* G_{1}\big) -\frac{27}{91}\big(G_{0}^* G_{-2}+G_{0}^* G_{2}\big)\nonumber\\& -\frac{162\sqrt{10}}{1001}\big(G_{1}^* G_{-1}\big)  \Bigg),\nonumber
\end{align}
\begin{align}
\braket{Y_3^4} &=2\kappa \sum_{\lambda_1\lambda_2} \Re\Bigg(-S^*G_{-3}+ S^* G_{3}- P_{-1}^* F_{2}  + P_{1}^* F_{-2}-\sqrt{\frac{1}{3}}\big(F_{-3}^* P_{0}- P_{0}^* F_{3}\big) - \frac{3\sqrt{210}}{77}\big(D_{-2}^* G_{1}-G_{-1}^* D_{2}\big)\\&+ \frac{\sqrt{35}}{7}\big(D_{1}^* D_{-2}-D_{-1}^* D_{2}\big) +\frac{14\sqrt{15}}{77}\big(D_{1}^* G_{4}- D_{-1}^*G_{-4}\big)- \frac{5\sqrt{105}}{77}\big( D_{-1}^* G_{2}+D_{1}^*G_{-2}\big)+ \frac{27\sqrt{10}}{143}\big(G_{-4}^* G_{-1}- G_{4}^*G_{1}\big) \nonumber\\&
+ \frac{3\sqrt{7}}{11}\big(F_{-3}^* F_{0}-F_{0}^* F_{3}\big)  +\frac{\sqrt{5}}{11}\big(D_{0}^*G_{-3}-D_{0}^*G_{3} \big)+ \frac{\sqrt{14}}{11} \big(F_{-2}^* F_{1}- F_{-1}^* F_{2}\big)   + \frac{81}{143}\big(G_{0}^*G_{-3}-G_{0}^* G_{3}\big) \nonumber\\& +\frac{27\sqrt{70}}{1001}\big(G_{-2}^* G_{1}-G_{-1}^* G_{2}\big) \Bigg),\nonumber
\end{align}
\begin{align}
   \braket{Y_4^4} &=2\kappa \sum_{\lambda_1\lambda_2}\Re\Bigg( S^*G_{-4}+ S^*G_{4}-\frac{2}{\sqrt{3}}\big(P_{-1}^* F_{3} + F_{-3}^* P_{1}\big)  +\frac{\sqrt{70}}{7}\big( D_{-2}^* D_{2}\big)-\frac{2\sqrt{210}}{77}\big(D_{-2}^* G_{2}+G_{-2}^* D_{2}\big)\\&-\frac{2\sqrt{15}}{11}\big(D_{-1}^*G_{3}+G_{-3}^* D_{1}\big) -\frac{4\sqrt{5}}{11}\big( D_{0}^*G_{-4}\big) -\frac{4\sqrt{5}}{11}\big( D_{0}^*G_{4}\big)+\frac{\sqrt{42}}{11}\big(F_{-3}^* F_{1}+F_{-1}^* F_{3}\big)+\frac{\sqrt{70}}{11}\big(F_{-2}^* F_{2} +\frac{81}{91}G_{-2}^*G_{2}\big)\nonumber\\&+\frac{54}{143}\big(G_{0}^* G_{-4}+G_{0}^* G_{4}\big)+\frac{27\sqrt{10}}{143}\big(G_{-3}^* G_{1}+ G_{-1}^*G_{3}\big)  \Bigg).\nonumber
\end{align}
}

\FloatBarrier
\section{Angular Moments at other $t$}
\label{app:comparisons_to_other_t}
In section ~\ref{sec:model_fitting} the fitting procedure was discussed. The resulting quality of fit was demonstrated with the fits from two characteristic $t$-bins. Other $t$-bins show a qualitatively similar goodness of fit. These confirm that the overall fit quality is good, the salient features present in the data are properly accounted for by the model, while model uncertainty derived from bootstrap is under control. Also the increasing role of the $S$- and $D$-wave effects as $t$ increases is clearly seen in the data and the model.

Finally, the changing interference pattern in the $P$-wave dominated moments $\langle Y^2_1 \rangle$  and $\langle Y^2_2\rangle$ implies nontrivial physics beyond the Pomeron exchange dominated small $t$ region.
\begin{figure*}
    \centering
    \includegraphics[scale=0.4]{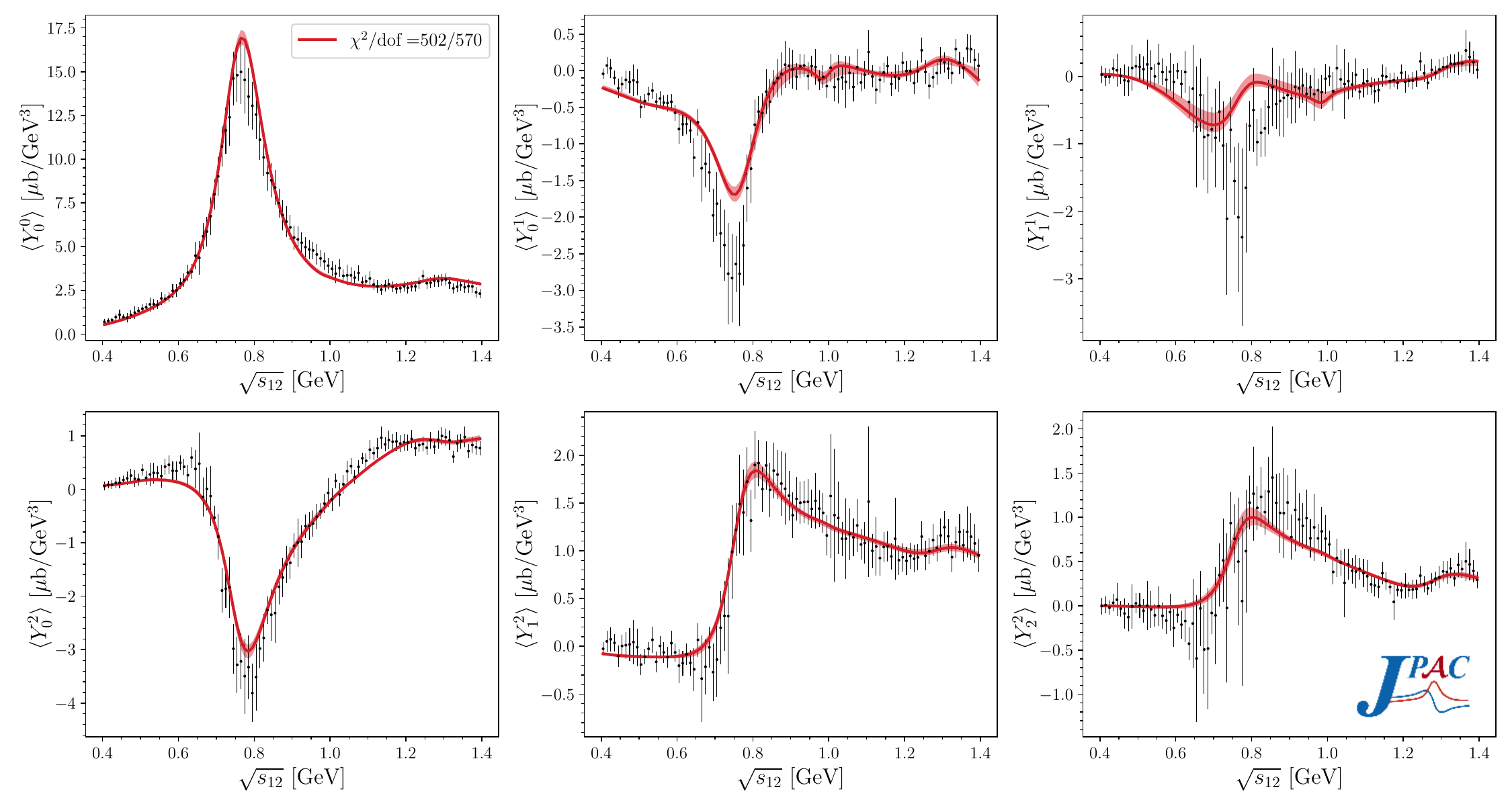}
    \caption{Comparison of complete model fitted to exprimental measurements from Ref.~\cite{CLAS:2009ngd} of two-pion angular moments $\braket{Y^L_M}$ for $L=0,1,2$ and $M=0,\dots L$ for $E_\gamma=3.7~\si{GeV}$ and $t=-0.55~\si{GeV^2}$.}
    \label{fig:YLM_t_0.55}
\end{figure*}
\begin{figure*}
    \centering
    \includegraphics[scale=0.4]{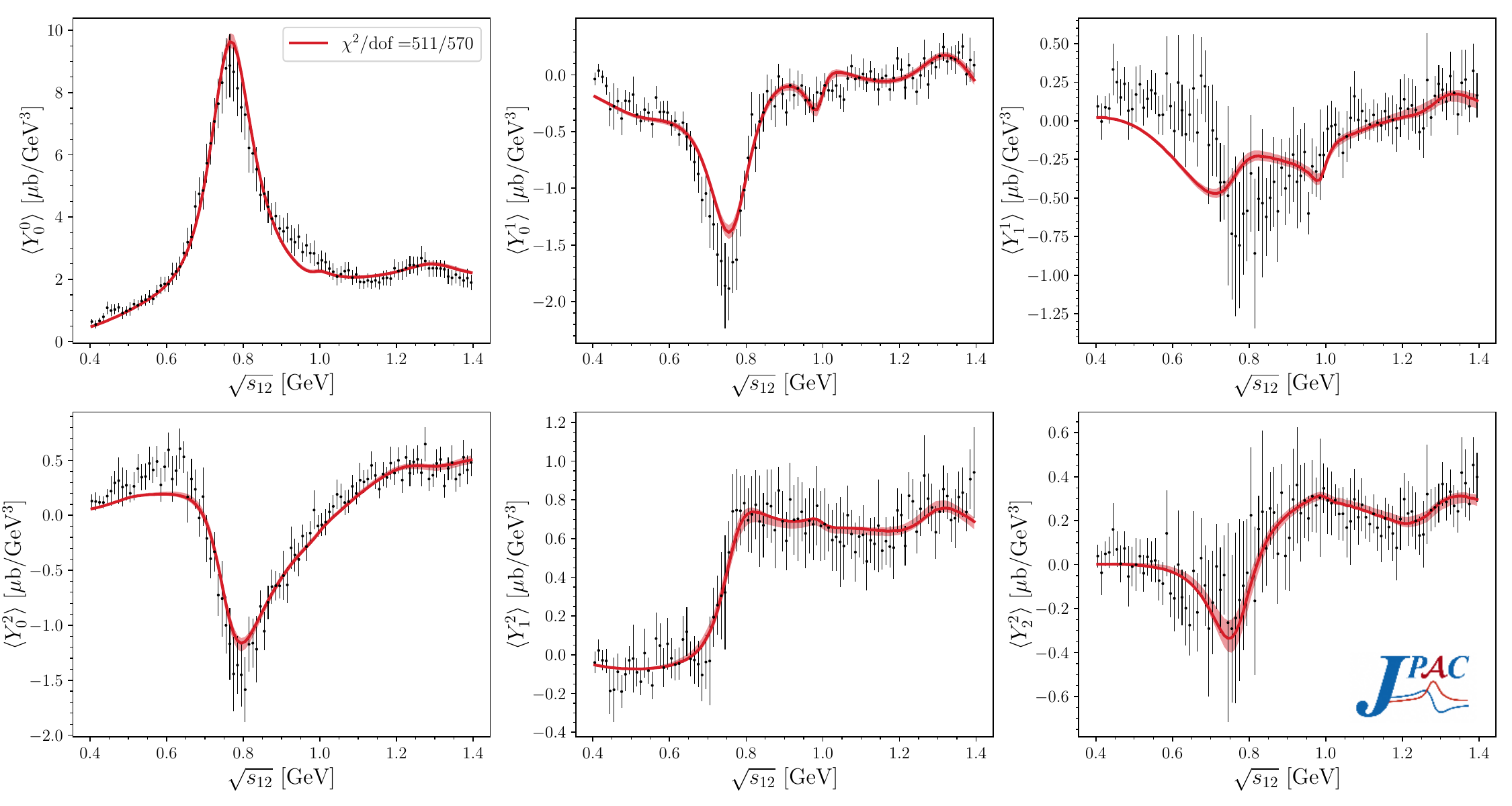}
    \caption{Comparison of complete model fitted to exprimental measurements from Ref.~\cite{CLAS:2009ngd} of two-pion angular moments $\braket{Y^L_M}$ for $L=0,1,2$ and $M=0,\dots L$ for $E_\gamma=3.7~\si{GeV}$ and $t=-0.65~\si{GeV^2}$.}
    \label{fig:YLM_t_0.65}
\end{figure*}
\begin{figure*}
    \centering
    \includegraphics[scale=0.4]{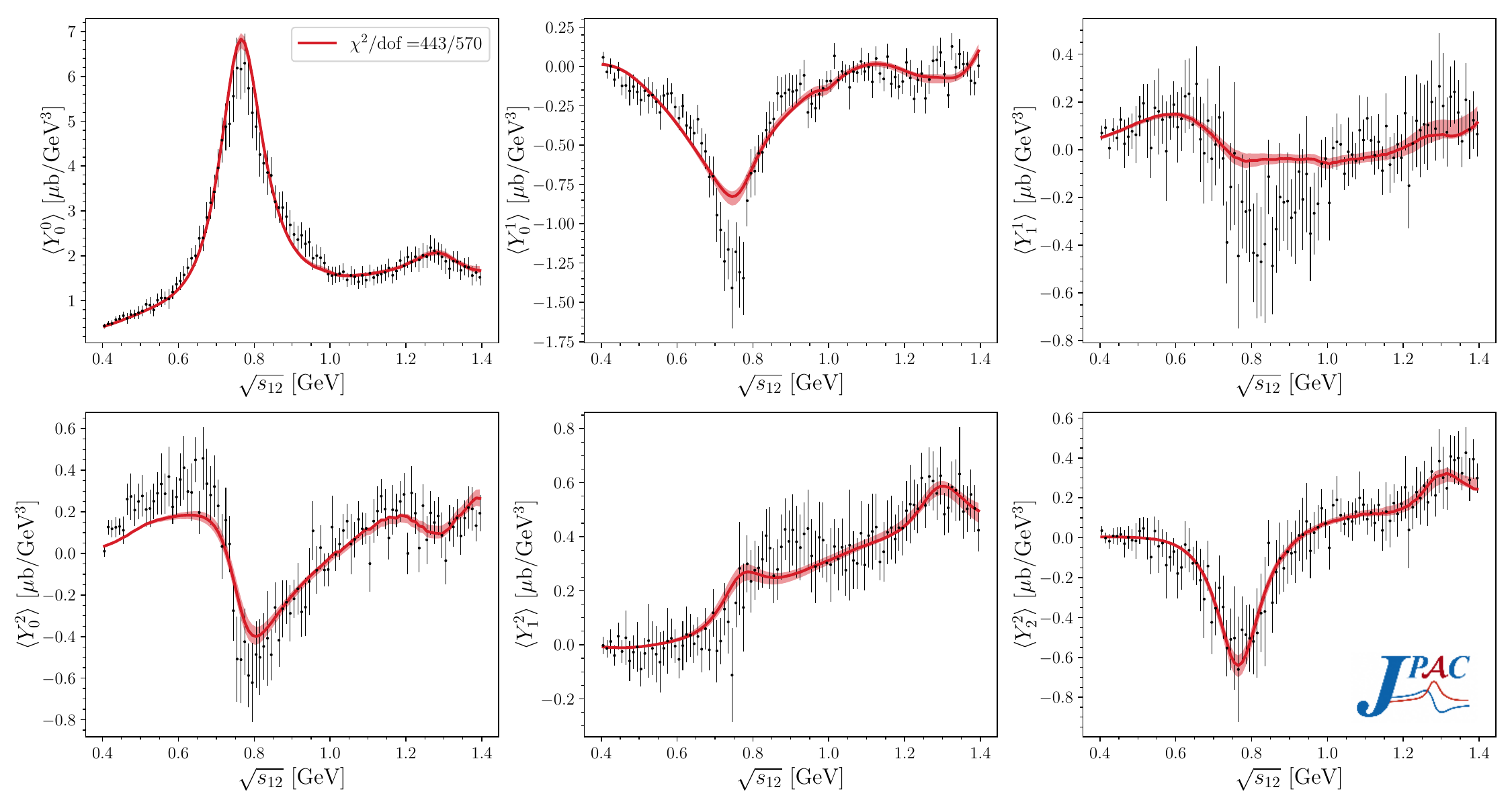}
    \caption{Comparison of complete model fitted to exprimental measurements from Ref.~\cite{CLAS:2009ngd} of two-pion angular moments $\braket{Y^L_M}$ for $L=0,1,2$ and $M=0,\dots L$ for $E_\gamma=3.7~\si{GeV}$ and $t=-0.75~\si{GeV^2}$.}
    \label{fig:YLM_t_0.75}
\end{figure*}

\begin{figure*}
    \centering
    \includegraphics[scale=0.4]{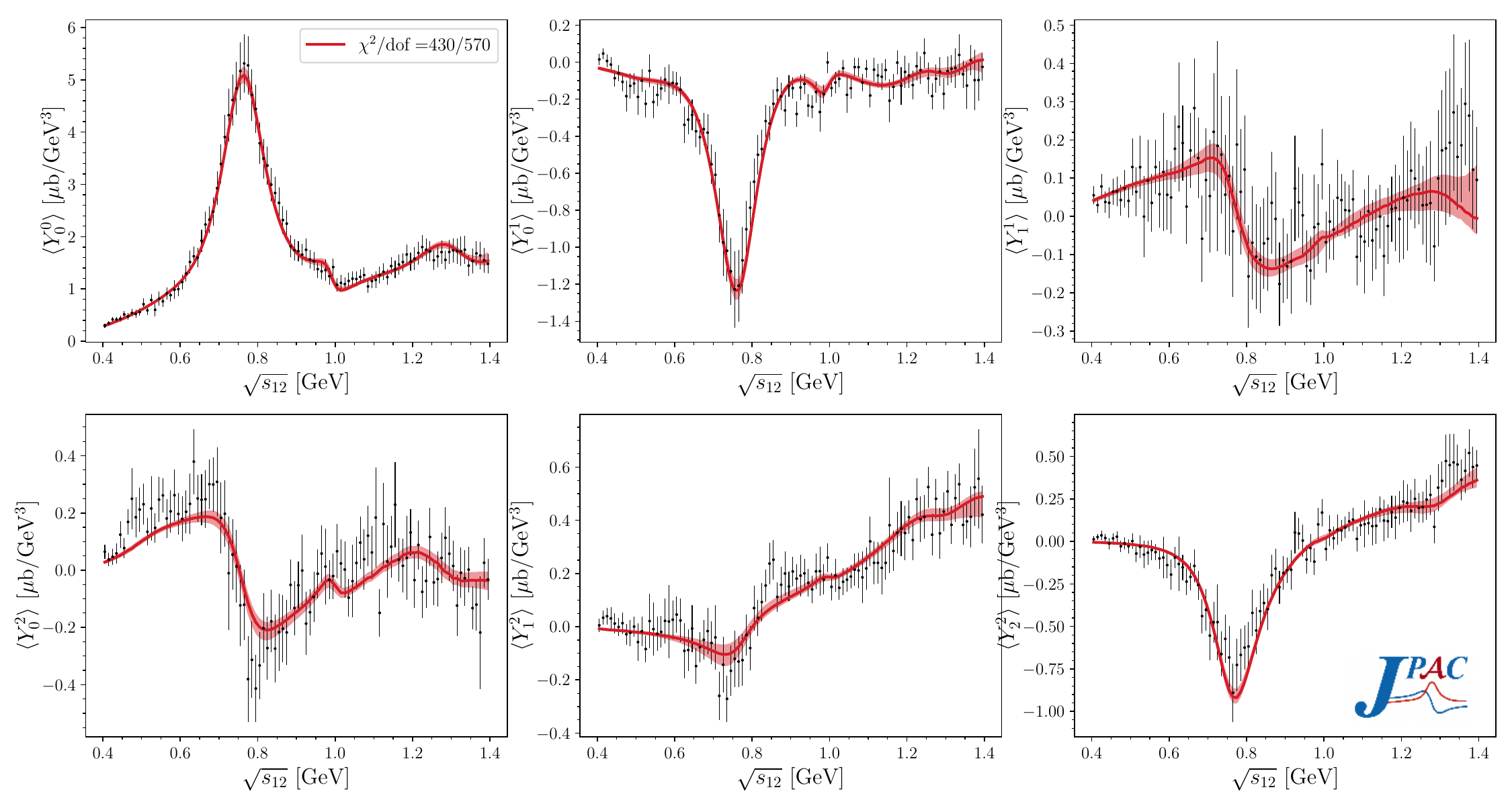}
    \caption{Comparison of complete model fitted to exprimental measurements from Ref.~\cite{CLAS:2009ngd} of two-pion angular moments $\braket{Y^L_M}$ for $L=0,1,2$ and $M=0,\dots L$ for $E_\gamma=3.7~\si{GeV}$ and $t=-0.85~\si{GeV^2}$.}
    \label{fig:YLM_t_0.85}
\end{figure*}

\begin{figure*}
        \centering
    \includegraphics[scale=0.39]{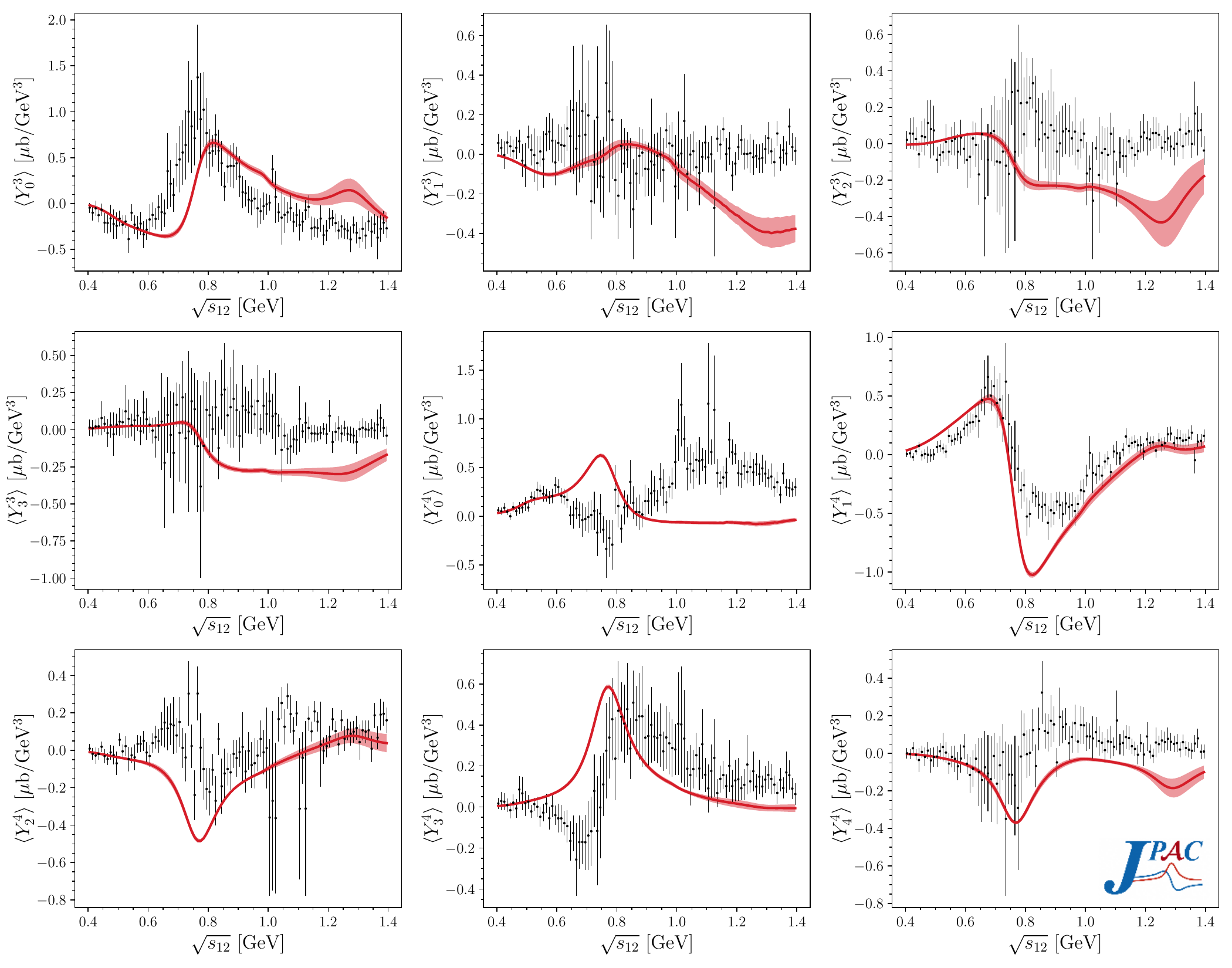}
        \caption{Prediction of angular moments $\braket{Y^L_M}$ for $L=3,4$, $M=0,\dots, L$ and $t=-0.55~\si{GeV^2}$. The model curves are superimposed and not fitted to these data. Experimental data is taken from CLAS~\cite{CLAS:2009ngd}}
    \label{fig:higher_moment_t_0.55}
    \end{figure*}

\FloatBarrier
\section{Comparisons to Higher Moments}
\label{app:comparisons_to_higher_moments}
Even though moments contain contributions from all partial waves, the resonant contributions in the two-pion energy region of interest are saturated by resonances in the $S$-, $P$-, and $D$-waves. All higher waves result from nonresonant contributions like the Deck mechanism. Thus, a compaarison of the model with these higher moments serves as a consistency check of the model. It can be seen that while the model properly accounts for moments $\langle Y^4_1\rangle$, $\langle Y^4_2\rangle$ and $\langle Y^4_3\rangle$ which are dominated by the $P$-$F$ interference the contribution of other background sources is required in other moments.

\begin{figure*}
    \centering
    \includegraphics[scale=0.35]{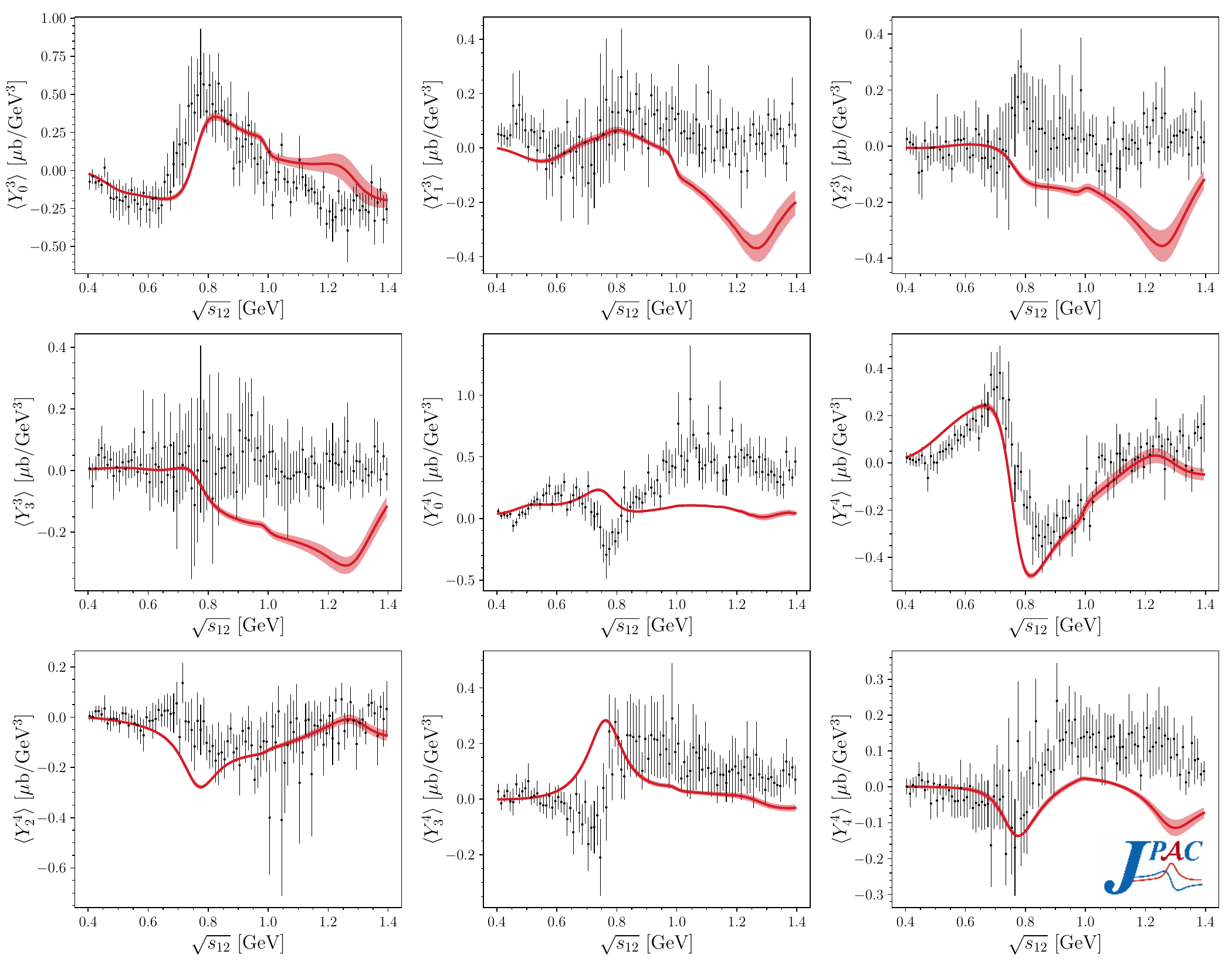}
    \caption{Prediction of angular moments $\braket{Y^L_M}$ for $L=3,4$, $M=0,\dots, L$ and $t=-0.65~\si{GeV^2}$. The model curves are superimposed and not fitted to these data. Experimental data is taken from CLAS~\cite{CLAS:2009ngd}}
    \label{fig:higher_moment_t_0.65}
\end{figure*}

\begin{figure*}
    \centering
    \includegraphics[scale=0.35]{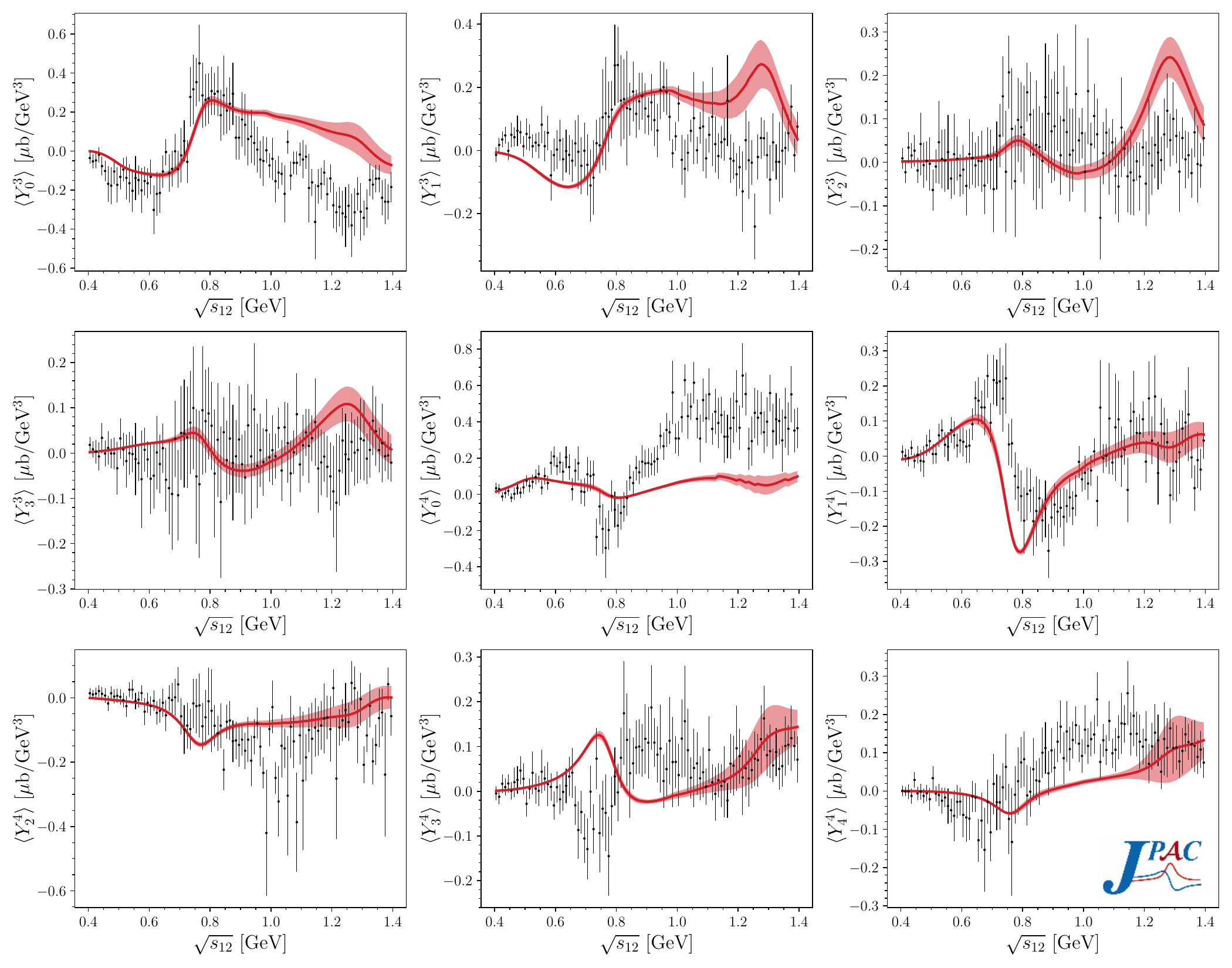}
    \caption{Prediction of angular moments $\braket{Y^L_M}$ for $L=3,4$, $M=0,\dots, L$ and $t=-0.75~\si{GeV^2}$. The model curves are superimposed and not fitted to these data. Experimental data is taken from CLAS~\cite{CLAS:2009ngd}}
    \label{fig:higher_moment_t_0.75}
\end{figure*}

\FloatBarrier
\begin{figure*}
    \centering
    \includegraphics[scale=0.35]{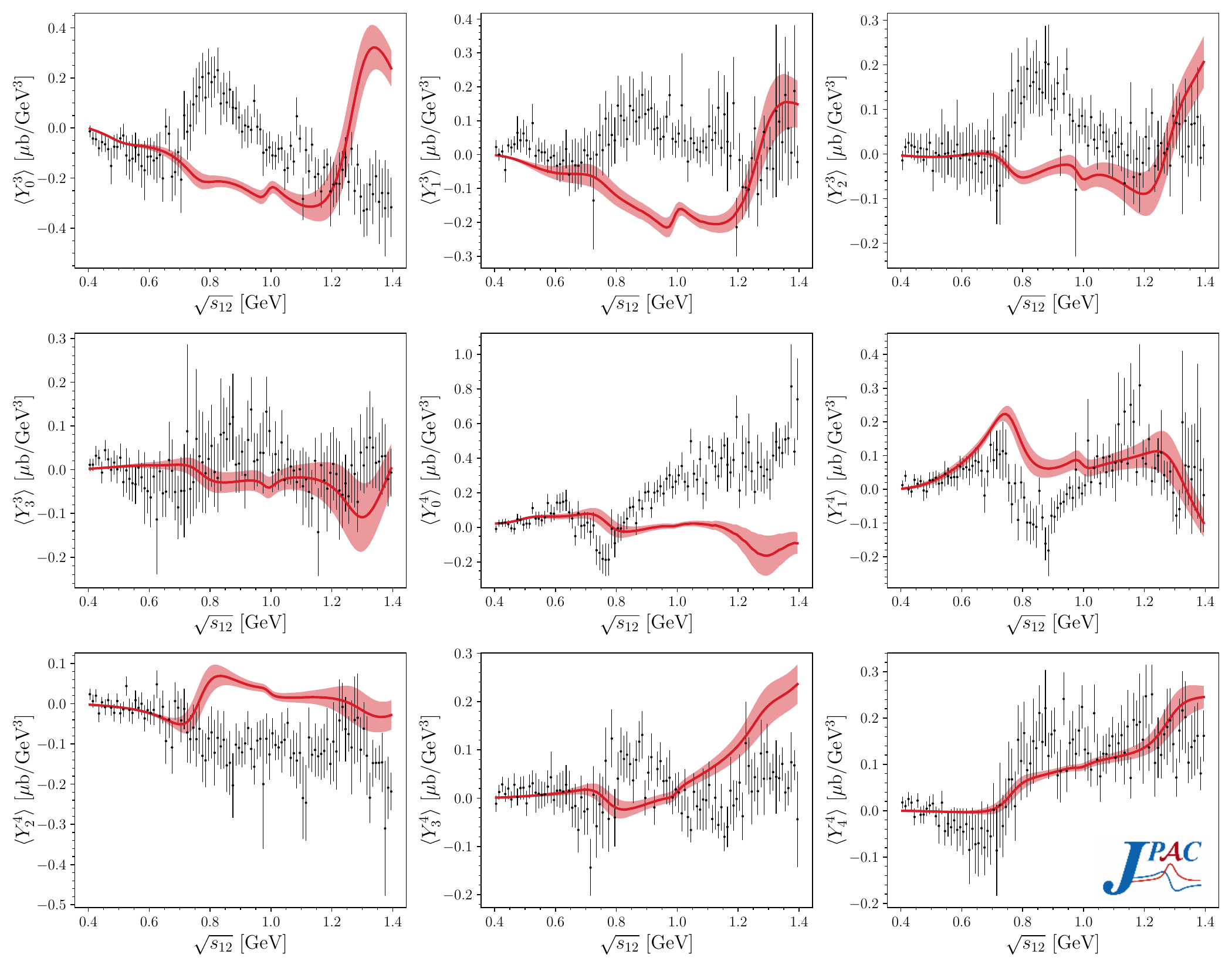}
    \caption{Prediction of angular moments $\braket{Y^L_M}$ for $L=3,4$, $M=0,\dots, L$ and $t=-0.85~\si{GeV^2}$. The model curves are superimposed and not fitted to these data. Experimental data is taken from CLAS~\cite{CLAS:2009ngd}}
    \label{fig:higher_moment_t_0.85}
\end{figure*}

\FloatBarrier
\begin{figure*}
    \centering
    \includegraphics[scale=0.35]{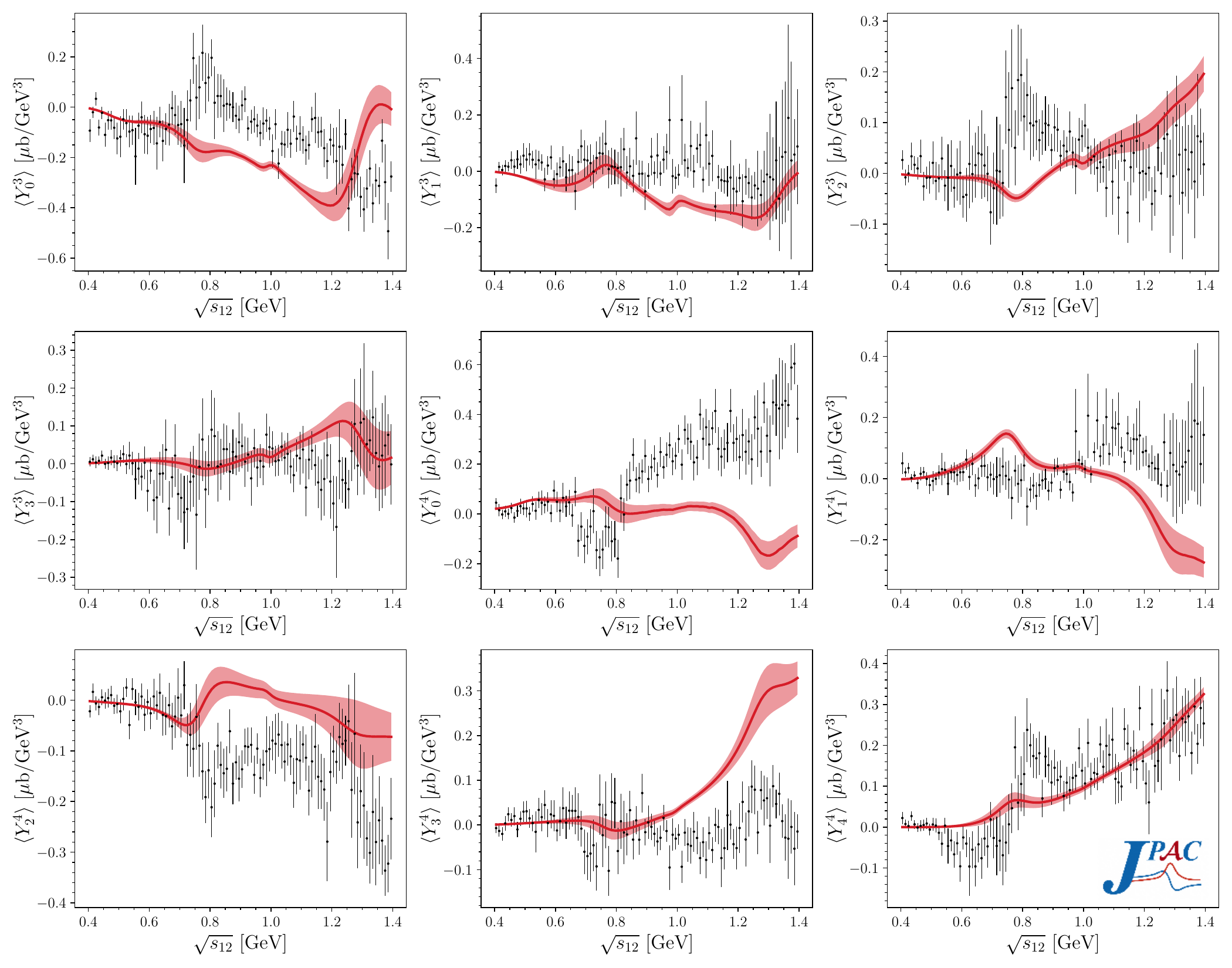}
    \caption{Prediction of angular moments $\braket{Y^L_M}$ for $L=3,4$, $M=0,\dots, L$ and $t=-0.95~\si{GeV^2}$. The model curves are superimposed and not fitted to these data. Experimental data is taken from CLAS~\cite{CLAS:2009ngd}}
    \label{fig:higher_moment_t_0.95}
\end{figure*}

\section{Parameters of the model}
\label{app:parameter_values}
In this appendix a complete list of model parameter values is provided. Recall that these values were obtained in a two-stage procedure. First, $10^5$-fold fitting of the model predictions to experimentally measured moments of angular distribution was performed for each $t$-bin. In this way, the obtained parameter values that should lie very close to the global $\chi^2$ minima for all sets of momets. Next, starting from these best fit values, fits were performed on $10^4$ synthetic datasets to obtain the central values and uncertainties of moments and parameters for all $t$-bins. The distributions of parameter and observable values were Gaussian to very good accuracy.
\begin{table}[th]
    \centering
\caption{Average values and standard deviations of model parameters obtained with bootstrap method for all $t$-bins under study. Parameters for the $\rho(770)$ photoproduction with Pomeron exchange were fixed to values
\mbox{$g_{\gamma \mathbb{P}\rho}=5.96$}, \mbox{$g_{\rho\pi\pi}= 2.506$} and \mbox{$\beta_\mathbb{P}(t)=\exp(bt)$} with $b=3.6~\si{GeV^{-2}}$ taken from Ref.~[22].}
\begin{tabular}{l|c|c|c|c|c|c}%
Parameter&t=-0.45 GeV$^2$&t=-0.55 GeV$^2$&t=-0.65 GeV$^2$&t=-0.75 GeV$^2$&t=-0.85 GeV$^2$&t=-0.95 GeV$^2$\\%
\hline%
$|b^\text{Deck}|$&1.0&1.0&1.0&1.0&1.0&1.0\\%
$\text{Arg}(b^\text{Deck})$&0.0&0.0&0.0&0.0&0.0&0.0\\%
$|a_{M=0,\pm 1}^{\mathbb{P},\rho}|$&\mbox{$g_{\gamma \mathbb{P}\rho} g_{\rho\pi\pi}\beta_\mathbb{P}(t)$}&\mbox{$g_{\gamma \mathbb{P}\rho} g_{\rho\pi\pi}\beta_\mathbb{P}(t)$}&\mbox{$g_{\gamma \mathbb{P}\rho} g_{\rho\pi\pi}\beta_\mathbb{P}(t)$}&\mbox{$g_{\gamma \mathbb{P}\rho} g_{\rho\pi\pi}\beta_\mathbb{P}(t)$}&\mbox{$g_{\gamma \mathbb{P}\rho} g_{\rho\pi\pi}\beta_\mathbb{P}(t)$}&\mbox{$g_{\gamma \mathbb{P}\rho} g_{\rho\pi\pi}\beta_\mathbb{P}(t)$}\\%
$\text{Arg}(a_{M=0,\pm 1}^{\mathbb{P},\rho})$&0.0&0.0&0.0&0.0&0.0&0.0\\%
$|a^{a_2/f_2,\rho}_{M=+1}|$&1.81$\pm$0.61&4.99$\pm$0.37&4.63$\pm$0.89&2.52$\pm$0.87&6.04$\pm$1.15&12.23$\pm$2.56\\%
$\text{Arg}(a^{a_2/f_2,\rho}_{M=+1})$&3.99$\pm$2.55&0.92$\pm$0.23&1.84$\pm$0.08&2.33$\pm$0.25&3.69$\pm$2.84&3.59$\pm$2.80\\%
$|a^{a_2/f_2,\rho}_{M=0}|$&13.70$\pm$0.73&20.72$\pm$0.32&18.07$\pm$0.25&15.98$\pm$0.52&13.71$\pm$0.53&13.56$\pm$0.55\\%
$\text{Arg}(a^{a_2/f_2,\rho}_{M=0})$&2.75$\pm$0.06&1.92$\pm$0.05&1.97$\pm$0.05&2.31$\pm$0.09&0.73$\pm$0.32&0.54$\pm$0.75\\%
$|a^{a_2/f_2,\rho}_{M=-1}|$&264.32$\pm$10.13&193.44$\pm$4.91&148.54$\pm$2.31&124.15$\pm$3.26&126.79$\pm$7.29&118.00$\pm$11.65\\%
$\text{Arg}(a^{a_2/f_2,\rho}_{M=-1})$&3.00$\pm$0.05&3.71$\pm$0.07&3.47$\pm$0.10&3.76$\pm$0.09&5.23$\pm$0.35&5.08$\pm$0.84\\%
$|a^{\rho/\omega,f_0(500)}|$&29.51$\pm$1.44&33.46$\pm$3.13&152.47$\pm$7.59&109.51$\pm$4.82&24.45$\pm$1.83&21.22$\pm$1.60\\%
$\text{Arg}(a^{\rho/\omega,f_0(500)})$&4.09$\pm$0.12&3.53$\pm$0.16&3.72$\pm$0.07&5.10$\pm$0.10&4.32$\pm$2.72&5.59$\pm$0.98\\%
$|a^{\rho/\omega,f_0(980)}|$&0.71$\pm$0.29&1.20$\pm$0.50&8.03$\pm$1.72&1.45$\pm$0.72&3.08$\pm$0.55&1.35$\pm$0.28\\%
$\text{Arg}(a^{\rho/\omega,f_0(980)})$&4.26$\pm$0.65&2.83$\pm$2.62&0.89$\pm$1.71&2.75$\pm$0.91&2.47$\pm$2.78&4.56$\pm$2.38\\%
$|a^{\rho/\omega,f_0(1370)}|$&10.47$\pm$3.21&24.44$\pm$6.97&165.17$\pm$17.65&182.57$\pm$15.14&107.09$\pm$13.31&45.64$\pm$10.01\\%
$\text{Arg}(a^{\rho/\omega,f_0(1370)})$&4.04$\pm$0.41&3.79$\pm$0.29&3.95$\pm$0.11&4.28$\pm$0.11&0.72$\pm$0.49&0.75$\pm$0.35\\%
$|a^{\rho/\omega,f_2}_{M=+2}|$&4.75$\pm$1.55&11.55$\pm$1.19&10.47$\pm$1.40&8.41$\pm$2.20&6.73$\pm$2.58&9.95$\pm$2.50\\%
$\text{Arg}(a^{\rho/\omega,f_2}_{M=+2})$&0.80$\pm$0.73&0.88$\pm$0.17&0.81$\pm$0.13&3.61$\pm$0.47&3.30$\pm$0.75&4.87$\pm$2.02\\%
$|a^{\rho/\omega,f_2}_{M=+1}|$&4.14$\pm$2.14&5.87$\pm$1.75&5.62$\pm$1.43&13.00$\pm$1.89&18.94$\pm$3.44&6.11$\pm$2.38\\%
$\text{Arg}(a^{\rho/\omega,f_2}_{M=+1})$&3.13$\pm$0.77&3.07$\pm$0.42&3.84$\pm$0.28&0.92$\pm$0.20&1.06$\pm$1.91&4.67$\pm$0.72\\%
$|a^{\rho/\omega,f_2}_{M=0}|$&23.07$\pm$8.34&16.95$\pm$5.53&6.64$\pm$2.88&5.87$\pm$3.11&20.03$\pm$6.52&17.86$\pm$3.28\\%
$\text{Arg}(a^{\rho/\omega,f_2}_{M=0})$&3.78$\pm$0.48&3.19$\pm$0.32&4.43$\pm$0.72&2.70$\pm$0.78&3.62$\pm$0.35&3.32$\pm$0.34\\%
$|a^{\rho/\omega,f_2}_{M=-1}|$&67.07$\pm$19.57&79.09$\pm$9.85&90.67$\pm$11.71&46.52$\pm$22.34&93.56$\pm$25.98&129.39$\pm$10.75\\%
$\text{Arg}(a^{\rho/\omega,f_2}_{M=-1})$&4.35$\pm$2.64&0.62$\pm$1.38&0.41$\pm$0.27&4.03$\pm$2.50&4.79$\pm$0.92&5.18$\pm$1.85\\%
$|a^{\rho/\omega,f_2}_{M=-2}|$&301.94$\pm$78.71&253.50$\pm$61.51&228.06$\pm$48.92&351.98$\pm$47.81&240.02$\pm$55.91&139.73$\pm$64.65\\%
$\text{Arg}(a^{\rho/\omega,f_2}_{M=-2})$&4.32$\pm$0.33&1.27$\pm$0.37&1.09$\pm$0.18&1.62$\pm$0.78&1.68$\pm$2.11&3.92$\pm$2.27\\%
$|b^{\rho/\omega,J=0}_{M=0}|$&29.26$\pm$2.68&35.46$\pm$7.62&155.55$\pm$11.81&116.40$\pm$5.86&71.66$\pm$6.53&51.00$\pm$4.94\\%
$\text{Arg}(b^{\rho/\omega,J=0}_{M=0})$&0.48$\pm$0.11&2.26$\pm$2.93&5.88$\pm$1.24&5.96$\pm$0.80&2.23$\pm$0.30&2.18$\pm$0.29\\%
$|b^{a_2/f_2,J=1}_{M=+1}|$&5.83$\pm$0.31&5.43$\pm$0.63&6.99$\pm$0.47&3.78$\pm$0.64&5.11$\pm$1.22&7.41$\pm$1.26\\%
$\text{Arg}(b^{a_2/f_2,J=1}_{M=+1})$&5.55$\pm$1.85&5.44$\pm$0.12&5.29$\pm$0.07&2.48$\pm$2.94&4.25$\pm$0.43&3.30$\pm$0.48\\%
$|b^{a_2/f_2,J=1}_{M=0}|$&10.63$\pm$0.50&13.76$\pm$0.32&13.38$\pm$0.31&13.23$\pm$0.46&8.63$\pm$1.23&10.69$\pm$1.05\\%
$\text{Arg}(b^{a_2/f_2,J=1}_{M=0})$&3.39$\pm$0.15&3.58$\pm$0.13&3.63$\pm$0.10&4.14$\pm$0.13&3.76$\pm$0.43&3.80$\pm$0.42\\%
$|b^{a_2/f_2,J=1}_{M=-1}|$&74.51$\pm$5.96&47.30$\pm$4.08&24.96$\pm$3.75&23.84$\pm$5.95&34.05$\pm$10.54&31.93$\pm$13.36\\%
$\text{Arg}(b^{a_2/f_2,J=1}_{M=-1})$&5.91$\pm$0.13&0.54$\pm$0.13&0.63$\pm$0.24&0.94$\pm$0.76&2.50$\pm$0.74&3.20$\pm$0.69\\%
\hline%
\end{tabular}%
\end{table}

\end{fmffile}
\twocolumngrid
\FloatBarrier
\bibliography{bibliography}
\end{document}